\documentclass{article}
\usepackage[english]{babel}
\usepackage{csquotes}
\usepackage{amssymb}
\usepackage{lipsum}
\usepackage{authblk}
\usepackage{mathtools}
\usepackage{booktabs}
\usepackage{siunitx}
\usepackage{bm}
\usepackage{comment}

\usepackage[letterpaper,top=2cm,bottom=2cm,left=3cm,right=3cm,marginparwidth=1.75cm]{geometry}

\usepackage{caption}
\usepackage{subcaption}
\usepackage{amsmath}
\usepackage{graphicx}
\usepackage[colorlinks=true, allcolors=blue]{hyperref}
\usepackage{cleveref}
\crefname{figure}{Fig.}{Figs.}
\crefname{equation}{Eq.}{Eqs.}

\usepackage[square, numbers, comma, sort&compress]{natbib}

\usepackage{tikz}

\title{Environment heterogeneity creates fast amplifiers of natural selection in graph-structured populations}
\date{}
\author{Cecilia Fruet\textsuperscript{1,2}, Arthur Alexandre\textsuperscript{1,2}, Alia Abbara\textsuperscript{1,2}, Claude Loverdo\textsuperscript{3,4}, \\Anne-Florence Bitbol\textsuperscript{1,2,*}}
\affil{
\textbf{1} Institute of Bioengineering, School of Life Sciences, École Polytechnique Fédérale  de Lausanne (EPFL), Lausanne, Switzerland\\
\textbf{2} SIB Swiss Institute of Bioinformatics, Lausanne, Switzerland\\
\textbf{3} Sorbonne Université, CNRS, Laboratoire Jean Perrin, LJP, Paris, France\\
\textbf{4} Sorbonne Université, CNRS, Inserm, Institut de
Biologie Paris-Seine, IBPS, Paris, France\\
* Corresponding author. Email: \texttt{anne-florence.bitbol@epfl.ch}
}

\begin{document}

\maketitle

\begin{abstract}
Complex spatial structure, with partially isolated subpopulations, and environment heterogeneity, such as gradients in nutrients, oxygen, and drugs, both shape the evolution of natural populations. We investigate the impact of environment heterogeneity on mutant fixation in spatially structured populations with demes on the nodes of a graph. When migrations between demes are frequent, we find that environment heterogeneity can amplify natural selection and simultaneously accelerate mutant fixation and extinction, thereby fostering the quick fixation of beneficial mutants. We demonstrate this effect in the star graph, and more strongly in the line graph. We show that amplification requires mutants to have a stronger fitness advantage in demes with stronger migration outflow, and that this condition allows amplification in more general graphs. As a baseline, we consider circulation graphs, where migration inflow and outflow are equal in each deme. In this case, environment heterogeneity has no impact to first order, but increases the fixation probability of beneficial mutants to second order. Finally, when migrations between demes are rare, we show that environment heterogeneity can also foster amplification of selection, by allowing demes with sufficient mutant advantage to become refugia for mutants. 
\end{abstract}

\section*{Introduction}
Natural microbial populations often possess complex spatial structures, with partially isolated subpopulations, e.g.\ in soil-associated or host-associated microbiota~\cite{fierer_embracing_2017, donaldson_gut_2016, wu_modulation_2022, edwards_structure_2015, bickel_soil_2020}. In addition, the environment in which natural microbial populations live is generally heterogeneous,  featuring spatial variability in nutrient availability, temperature, pH, and concentrations of drugs or toxins. For instance, in soils, nutrient levels vary with depth, proximity to plant roots, and localized microbial metabolic activity~\cite{jobbagy_thedistribution_2001, naylor_trends_2022}. This variability creates microhabitats that shape microbial community composition and function~\cite{torsvik_prokaryotic_2002}. Aquatic environments are heterogeneous too~\cite{stocker_marine_2012}, influenced by factors such as oxygen gradients~\cite{fenchel_oxygen_2008}. 
The human body also constitutes a spatially structured and heterogeneous environment for its microbiota~\cite{costello_bacterial_2009}. In particular, the gut features gradients of oxygen, pH, bile salts, and antibiotic drugs when they are taken. 
This environment heterogeneity gives rise to heterogeneous selection pressures on microorganisms, leading to complex ecological and evolutionary dynamics~\cite{mccallum_gut_2023, donaldson_gut_2016,Verdon25}. To understand how natural microbial populations evolve, it is thus important to investigate the joint impact of spatial structure and of environment heterogeneity on population genetics. 

The vast majority of theoretical studies investigating the evolution of spatially structured populations assume that the environment is homogeneous. They show interesting impacts of spatial structure on evolution, in particular on the fundamental process of mutant fixation, whereby a mutant type takes over the population. 
Considering well-mixed subpopulations, known as demes or patches, connected by migrations~\cite{Wright31,Kimura64}, it was shown that the probability that a mutant fixes (i.e.\ takes over) in the population is not impacted by spatial structure, if migrations are sufficiently symmetric~\cite{maruyama70, maruyama74}, and demes do not get extinct~\cite{Barton93}. Models on graphs where each node comprises one individual have allowed to consider more complex spatial structures~\cite{lieberman_evolutionary_2005}. In this framework, known as evolutionary graph theory, it was shown that specific graphs, under specific update rules defining the dynamics, can amplify or suppress natural selection~\cite{lieberman_evolutionary_2005,Kaveh15,Hindersin15,Pattni15,tkadlec_limits_2020}. Amplifiers of selection increase the fixation probability of a beneficial mutant compared to a well-mixed population with the same size, and decrease the one of deleterious mutants. Meanwhile, suppressors of selection decrease the fixation probability of beneficial mutants and increase that of deleterious ones. Amplifiers of natural selection have been the subject of sustained attention~\cite{lieberman_evolutionary_2005,Kaveh15,Hindersin15,Pattni15,tkadlec_limits_2020,adlam_amplifiers_2015,pavlogiannis_amplification_2017}, given their potential to enhance adaptation, which might be useful e.g.\ in directed evolution. However, they generally slow down mutant fixation~\cite{frean_effect_2013, hindersin_exact_2016, tkadlec_population_2019,tkadlec_fast_2021}, which limits their impact. Evolutionary graph theory models have been generalized by placing well-mixed demes with fixed size on graph nodes, also using specific update rules~\cite{Houchmandzadeh11,Houchmandzadeh13,Constable14,yagoobi2021fixation,Yagoobi23}. Recently, more general models, which do not assume strictly fixed deme sizes or update rules, were proposed by some of us~\cite{marrec_toward_2021,Abbara__Frequent}. Under rare migrations, these models generalize the findings of evolutionary graph theory, and show that whether a graph amplifies or suppresses natural selection strongly depends on the asymmetry of migration between demes~\cite{marrec_toward_2021}. Moreover, under frequent asymmetric migrations, suppression of selection was found to be pervasive, and associated to an acceleration of mutant fixation or extinction~\cite{Abbara__Frequent}. 

How does environment heterogeneity across demes impact mutant fixation in spatially structured populations? In population genetics, environment heterogeneity has been studied in continuous spatially extended populations, but in the context of species range limits and local adaptation rather than mutant fixation in the whole population~\cite{haldane_relation_1956,slatkin_gene_1973,kirkpatrick_evolution_1997,lenormand_gene_2002,kawecki_conceptual_2004,polechova_limits_2015,kottler_draining_2021}. Importantly, these studies typically consider antagonistic effects of selection across space, while here, our main focus is on mutants that are beneficial throughout the structure, with the strength of their beneficial effect varying across demes. In evolutionary graph theory models with one individual per node, graphs with nodes falling into two environment types have been considered~\cite{maciejewski_environmental_2014, kamran_environmental_2019, kaveh_themoran_2020, brendborg_fixation_2022, svoboda_coexistence_2023, nemati_counterintuitive_2023}, mainly focusing on the optimal repartition of these two types of nodes to foster mutant fixation~\cite{kaveh_themoran_2020, brendborg_fixation_2022, nemati_counterintuitive_2023} and on their impact on evolutionary timescales~\cite{maciejewski_environmental_2014, svoboda_coexistence_2023}. Here, we consider a more general model with demes on nodes of the graph, where mutant fitness advantage can be different in each deme. We investigate how environment heterogeneity affects mutant fixation in deme-structured populations on graphs, comparing heterogeneous and homogeneous environments.

We generalize the spatially structured population model introduced in Ref.~\cite{Abbara__Frequent} to heterogeneous environments, modeled via a deme-dependent mutant fitness advantage. We first focus on the frequent migration regime. In circulation graphs, where migration inflow and outflow are equal in each deme, we find that environment heterogeneity has no impact to first order, but increases mutant fixation probability to second order. This simplest category of graphs serves as a baseline for the rest of our study. In the star and in the line, we show that environment heterogeneity substantially impacts mutant fixation probability, and that it can lead to amplification of selection, together with acceleration of mutant fixation and extinction. This stands in contrast with the usual tradeoff between amplification and slower dynamics. 
We determine analytically the conditions for amplification of selection in the star and in the line, using a branching process approach. We show that a key ingredient for amplification is that mutants should be strongly favored in demes with strong migration outflow. We generalize these findings to fully connected graphs with one special deme that differs from others both by mutant advantage and by migration outflow. 
Furthermore, we show that amplification can exist in all connected graphs with five nodes that are not circulations, with strong migration outflow from demes where mutants are advantaged. Finally, we turn to the rare migration regime, and show that environment heterogeneity can turn suppressors of selection into amplifiers in this regime too, but for a different reason: demes with sufficient mutant advantage become refugia for mutants.

\section*{Results}

\begin{figure}[htbp]
    \centering
    \includegraphics[width=\textwidth]{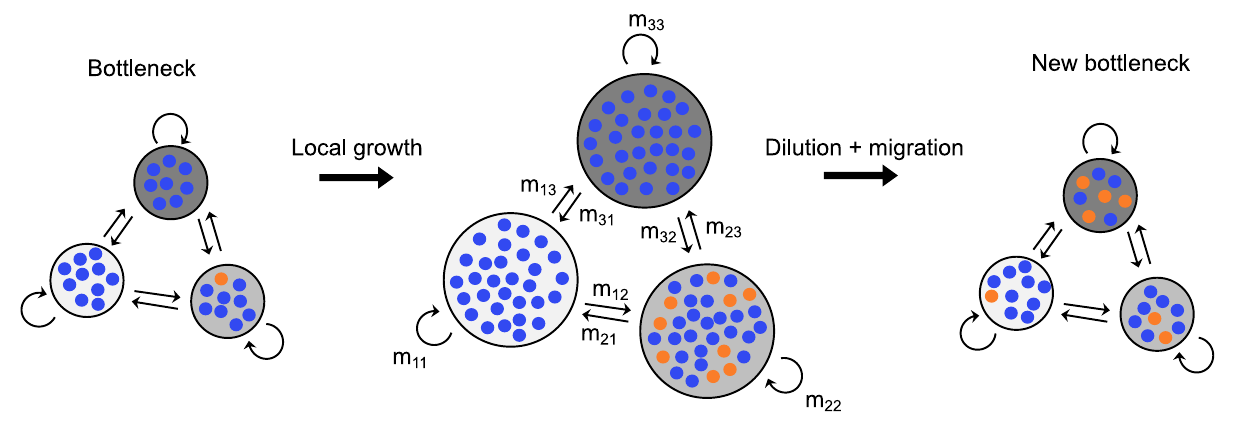}
    \caption{\textbf{Serial dilution model for spatially structured populations with environmental heterogeneities.} Schematic of an elementary step of the serial dilution model. Starting from a bottleneck, each deme undergoes a deterministic local growth step followed by a dilution and migration step, leading to a new bottleneck. 
    Blue markers represent wild-type individuals and orange markers represent mutants. Gray backgrounds with different darkness represent different environments.}
    \label{fig:methods}
\end{figure}

\paragraph{Including environment heterogeneity in deme-structured populations on graphs.} The impact of spatial structure on mutant fixation has traditionally been studied assuming homogeneous environments, where the fitness difference between mutants and wild-types is the same in the whole structured population~\cite{Wright31,Kimura64,maruyama70,maruyama74,Nagylaki80,Slatkin81,Barton93,Whitlock97,Nordborg02,whitlock2003,Sjodin05, lieberman_evolutionary_2005, Houchmandzadeh11,Houchmandzadeh13,Constable14,hauert_fixation_2014,allen_evolutionary_2017, marrec_toward_2021,yagoobi2021fixation,Yagoobi23,Abbara__Frequent,kreger_role_2023}. 
Here, we use our model of deme-structured population on graphs~\cite{Abbara__Frequent} and generalize it to include heterogeneous environments: each deme, located on a node of the graph, 
may have a different environment that can modulate the mutant fitness advantage. 
We denote the baseline mutant fitness advantage by $s$, and the local modulation in deme $i$ by the prefactor $\delta_i$, so that the mutant fitness advantage in that deme is $s\delta_i$.
In our serial dilution model, illustrated in Figure~\ref{fig:methods}, each deme starts at bottleneck size $K$, grows exponentially for a fixed time $t$, and then undergoes a dilution and migration step, returning to the original bottleneck size $K$~\cite{Abbara__Frequent}. These steps are then iterated. This description generalizes over structured Wright-Fisher models~\cite{Lessard07,Burden18} and allows explicit modeling of experiments with serial transfers~\cite{chakraborty_experimental_2023,abbara_mutant_2024}. The fitness advantage of mutants matters during the growth phase, and results in an increase of the fraction of mutants in demes where mutants were present initially. This variation of mutant fraction only depends on $st$, which we thus refer to as the baseline effective fitness advantage (see Methods). After a growth phase, migrations take place with probability $m_{ij}$ from deme $i$ to deme $j$ during each migration-dilution step, modeled using binomial samplings (see Methods).

Throughout, we focus on the fixation of a single mutant appearing uniformly at random in the structure, reflecting biologically realistic scenarios such as DNA replication errors, or mutations due to external stress affecting all cells. We mainly address the frequent migration regime~\cite{Abbara__Frequent}, 
which is the most relevant experimentally~\cite{Kryazhimskiy12,Nahum15,france_relationship_2019,Chakraborty23,abbara_mutant_2024}.
In the final section, we turn to the rare migration regime, where mutant fixation or extinction in a deme occurs much faster than migration~\cite{Slatkin81,marrec_toward_2021}.

Before investigating the role of environment heterogeneity, we briefly recall how spatial structure impacts mutant fixation in deme-structured populations on graphs with homogeneous environments. We showed in Ref.~\cite{Abbara__Frequent} that circulation graphs, where each node has the same total incoming and outgoing migration flow, do not impact fixation probability. This extends Maruyama's theorem~\cite{maruyama70,maruyama74} and the circulation theorem of evolutionary graph theory models with one individual per node of a graph~\cite{lieberman_evolutionary_2005}. Furthermore, we showed that all other graphs suppress natural selection, and we found that this is associated to an acceleration of mutant fixation or extinction~\cite{Abbara__Frequent}. These results were obtained under frequent migrations using the branching process approximation. 

\paragraph{Environment heterogeneity in circulation graphs: a baseline.}
We first consider circulation graphs, which do not impact fixation probability under homogeneous environments. For frequent migrations, using a branching process approximation under the assumptions that $s>0$, $K\gg 1$ and $1/K\ll st\ll 1$, the fixation probability of a mutant appearing uniformly at random can be expressed as 
\begin{equation}
    \rho=a st-\frac{b}{2}(st)^2\,,
    \label{expand}
\end{equation}
to second order in the baseline effective fitness advantage $st$ (see SI Section~\ref{SI-expand}). This expansion holds for any connected graph, but the coefficients $a$ and $b$ depend, \textit{a priori}, on the graph and on the environment heterogeneity.

For circulation graphs, we find that $a=2\langle\delta\rangle$ (see SI Section~\ref{subs:circ-fo}). Hence, to first order in $st$, the fixation probability $\rho=2\langle\delta\rangle st$ only depends on the mean effective fitness advantage $\langle\delta\rangle st$ of the mutant across the structure, and is not affected by environment heterogeneity. We also find that the fixation probability does not depend on where the mutant starts. Furthermore, it coincides with the fixation probability in a well-mixed population with the same mean effective fitness advantage, matching Haldane's classical result~\cite{haldane_amathematical_1927}. 
Thus, to first order in $st$, the circulation theorem extends to heterogeneous environments: in circulation graphs, neither spatial structure nor environment heterogeneity impact mutant fixation probability. 

By contrast, the second-order coefficient $-b$ in the expansion of the fixation probability in Eq.~\ref{expand} depends on environment heterogeneity even in circulation graphs (see SI Section~\ref{sec:circ-SO}). For the clique (fully connected graph), corresponding to Wright's island model~\cite{Wright31}, our second-order calculation shows that environment heterogeneity slightly increases the fixation probability of beneficial mutants. For a specific cycle, corresponding to the circular stepping-stone model~\cite{Kimura64,maruyama70,maruyama_stepping_1970}, we obtain similar, though more complex, results (see SI Section~\ref{sec:circ-SO}). These findings demonstrate that, beyond first order, environment heterogeneity has non-trivial theoretical effects even in circulation graphs, leading to a breakdown of the circulation theorem to second order. However, this effect remains quantitatively small, as it scales as $(st)^2$. This motivates turning to non-circulation graphs to investigate whether environment heterogeneity has stronger effects there. Importantly, circulation graphs, and in particular the clique, will serve as a baseline independent of environment heterogeneity to first order in $st$.

\begin{figure}[htb!]
    \centering
    \includegraphics[width=\linewidth]{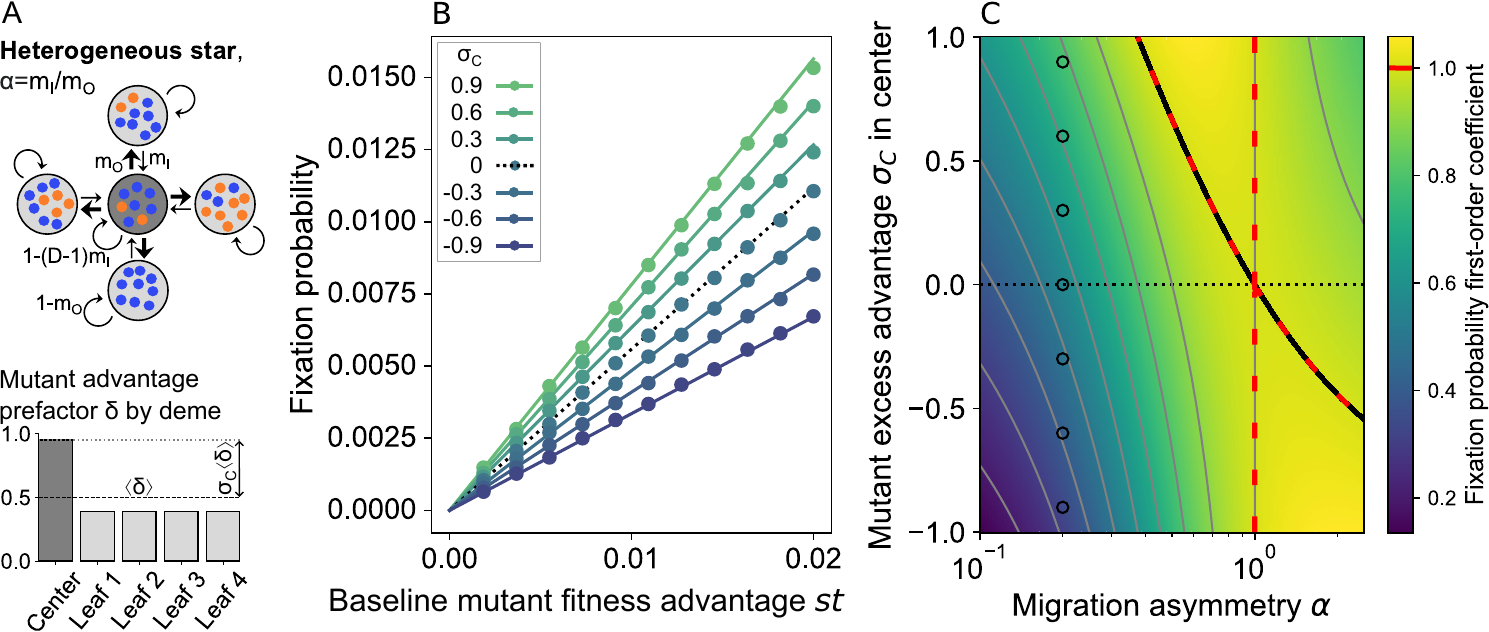} 
    \caption{\textbf{Impact of environmental heterogeneity on mutant fixation in the star.} Panel A: Schematic of the heterogeneous star, with migration asymmetry $\alpha=m_I/m_O$, and deme-dependent mutant advantage prefactor $\delta$ shown in the histogram below. The relative mutant fitness excess in the center is defined as $\sigma_C = (\delta_C - \langle \delta \rangle) / \langle \delta \rangle$. Panel B: Mutant fixation probability versus the baseline mutant fitness advantage $st$, for different relative mutant fitness excess $\sigma_C$ in the center, with migration asymmetry $\alpha=0.2$. Markers: stochastic simulation results; lines: analytical predictions (Eq.~\ref{eq:fo-star}). Panel C: Heatmap of the first-order coefficient in the expansion of the mutant fixation probability in the baseline effective mutant fitness advantage $st$ (denoted by $a$ in Eq.~\ref{expand}), versus the migration asymmetry $\alpha$ and the relative mutant fitness excess $\sigma_C$ in the center of the star.
    Red dashed lines: numerically-determined boundaries of the region where the star amplifies selection; solid black line: analytical prediction for this boundary, given by $f_{\textrm{star}}(\alpha,D)$ in Eq.~\ref{eq:fstar}. Horizontal dotted line: homogeneous environment case ($\sigma_C=0$). Markers: parameter values considered in panel B. Parameter values: in panels B and C, $D=5$; $K=1000$; $\langle\delta \rangle=0.5$. In Panel B: $m_O=0.6$; $\alpha=0.2$; each marker comes from $5\times 10^5$ stochastic simulation realizations.}
    \label{fig:fig4-star}
\end{figure}

\paragraph{Environment heterogeneity strongly impacts mutant fixation probability in a star.}
\label{sec:star1}
We next consider the star graph, which comprises a central node connected to $D-1$ leaf nodes by migrations (see Figure~\ref{fig:fig4-star}A). The star has been the focus of many studies on the impact of spatial structure on mutant fixation under homogeneous environments. In evolutionary graph theory models with one individual per node, the star either amplifies or suppresses natural selection depending on update rules ~\cite{lieberman_evolutionary_2005,Kaveh15,Hindersin15,Pattni15}. With demes on each node, in the rare migration regime, whether it amplifies or suppresses natural selection depends on migration asymmetry ~\cite{marrec_toward_2021}. By contrast, for more frequent migrations, it can only suppress selection~\cite{Abbara__Frequent}. 

How does environment heterogeneity impact mutant fixation in the star with a deme on each node? To address this question, we compare a star with heterogeneous environments (see Figure~\ref{fig:fig4-star}A) to a star with homogeneous environment, where the mean mutant fitness advantage is the same, i.e.\ where the mutant fitness advantage is $\langle\delta\rangle s$ in each deme. In the branching process approach, we show that, to first order in $s t$, the mutant fixation probability in the heterogeneous star only depends on selection through the mean mutant fitness effect, via $\langle\delta\rangle$, and through the relative fitness excess in the center, defined as $\sigma_C = (\delta_C - \langle \delta \rangle)/\langle \delta \rangle$ (see SI Section~\ref{sec:fo-highsym}, Eq.~\ref{eq:fo-star}). Fixation probability is also impacted by migration asymmetry $\alpha=m_I/m_O$, where $m_I$ is the probability of incoming migrations to the center from a leaf, while $m_O$ is the probability of outgoing migrations from the center to a leaf. When there is more migration outflow from the center than inflow to the center ($\alpha<1$), we find that environment heterogeneity increases mutant fixation probability if the center features a larger mutant fitness advantage than the leaves ($\sigma_C > 0$). Conversely, if there is more inflow to the center than outflow ($\alpha>1$), heterogeneity increases mutant fixation probability if the center features a smaller mutant fitness advantage than the leaves ($\sigma_C < 0$), see SI Section~\ref{subs:homo-hetero-star}. This suggests that having a stronger mutant advantage in locations with more migration outflow enhances mutant fixation. Finally, if $\alpha=1$, to first order in $st$, environment has no impact on fixation probability. This is consistent with our results above, as in this specific case the star is a circulation graph.

Figure~\ref{fig:fig4-star}B shows the impact of environment heterogeneity, via $\sigma_C$, on the mutant fixation probability in the star. We observe that the farther $\sigma_C$ is from $0$, the more the fixation probability deviates from the baseline of the homogeneous star. We observe that environment heterogeneity yields an increase of the fixation probability when $\sigma_C>0$, and a decrease of it when $\sigma_C<0$. As $\alpha<1$ in this figure, this is fully consistent with our analytical predictions. In fact, for each value of $\sigma_C$ considered, Figure~\ref{fig:fig4-star}B shows excellent agreement between our stochastic simulations and our analytical predictions. 

Figure~\ref{fig:fig4-star}C further illustrates how migration asymmetry $\alpha$ and environmental heterogeneity, via $\sigma_C$, impact the first-order coefficient $a$ in Eq.~\ref{expand} in the expansion in $st$ of the mutant fixation probability in the star. This heatmap highlights the strong impact of both migration asymmetry and environmental heterogeneity on fixation probability in the star. Notably, environmental heterogeneity can enhance fixation probability in a star beyond that of an equivalent well-mixed population or circulation with the same average mutant fitness advantage $\langle\delta\rangle s$. This reference fixation probability is $2\langle \delta \rangle st$ to first order, as shown above. Hence, a heterogeneous star can amplify natural selection. This effect is further illustrated in Figure~\ref{fig:faD}. We find an analytical condition on $\sigma_C$ for amplification to occur, see SI Section \ref{sub:sup-star}. Figure~\ref{fig:fig4-star}C shows that it matches the numerical finding for the boundary of the area where the fixation probability in the heterogeneous star exceeds that of an equivalent well-mixed population. This result is particularly striking because the star with frequent asymmetric migrations is known to suppress natural selection in a homogeneous environment~\cite{Abbara__Frequent}. Even with rare migrations, no amplification is possible in the homogeneous star with $\alpha<1$~\cite{marrec_toward_2021}, while here, amplification exists for $\alpha<1$, for sufficiently large $\sigma_C$. Therefore, our results demonstrate that the impact of environmental heterogeneity is strong enough to counteract this suppression, and lead to amplification of selection.

\begin{figure}[htb!]
    \centering
    \includegraphics[width=\linewidth]{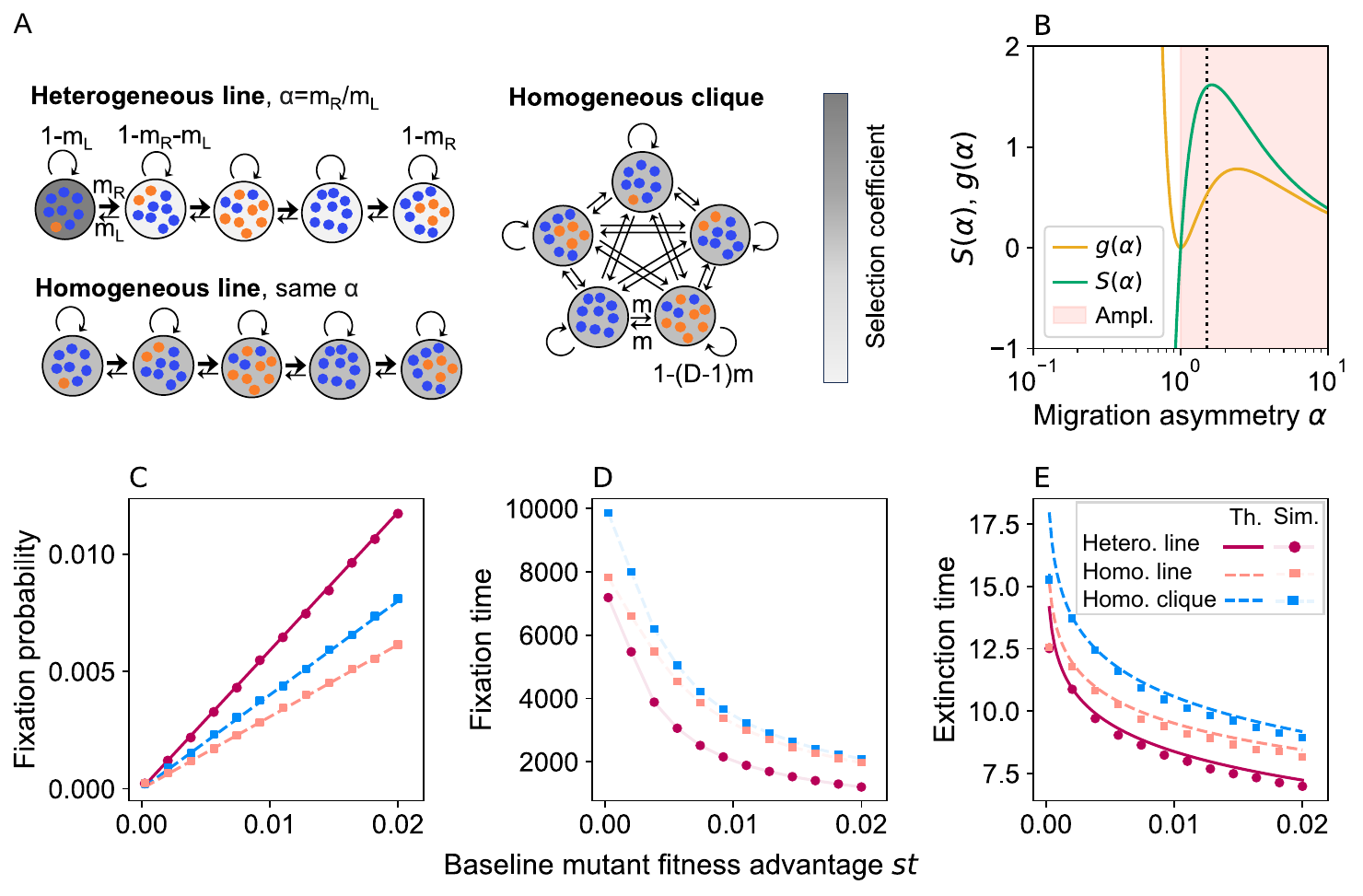}
    \caption{\textbf{Amplification of selection and accelerated dynamics in the heterogeneous line.} Panel A: Schematics of the structures considered: heterogeneous line with step environmental profile and $\alpha=1.5$; homogeneous line with same $\alpha$ and $\langle\delta\rangle$; homogeneous clique with same $\langle\delta\rangle$. Panel B: Condition for amplification of natural selection in the heterogeneous line shown in the top schematic. The functions $S$ and $g$, defined respectively in Eqs.~\ref{eq:cond-line1} and \ref{eq:cond-line3}, are plotted versus migration asymmetry $\alpha$. Pink-shaded region: range of $\alpha$ where $S(\alpha)>g(\alpha)$, i.e.\ where the heterogeneous line amplifies natural selection. Dotted line: migration asymmetry ($\alpha=1.5$) chosen in panels C-E. Panel C: Mutant fixation probability versus baseline mutant fitness advantage $st$, in the different spatial structures shown in panel A. Predictions from the branching process theory (``Th.''), specifically Eq.~\ref{eq:aline}, and results from stochastic simulations (``Sim.''), are shown. 
    Panel D: Mutant fixation time, in number of bottlenecks, in the different structures considered, versus baseline mutant fitness advantage $st$. Panel E: Same as in D, but for mutant extinction time. Theoretical predictions are from Eq.~\ref{eq:text}. Parameter values for all structures: $D=5$, $K=1000$; for all lines: $\alpha = 1.5$, $m_L = 0.3$; for the clique: $m=0.15$; for the line with step profile: $\delta=1$ in the leftmost deme, $\delta=0$ in other ones. Each marker comes from $5\times 10^5$ stochastic simulation realizations.} 
    \label{fig:line-proba-time}
\end{figure}

\paragraph{The heterogeneous line can substantially amplify selection and accelerate mutant fixation.} Another important spatial structure is the line of demes, or linear stepping-stone model~\cite{Kimura64,maruyama70,maruyama_stepping_1970}, see schematics in Figure~\ref{fig:line-proba-time}A. How does the line impact mutant fixation probability? In SI Section~\ref{sec:fo-highsym}, we derive the fixation probability of a mutant in a line of demes with frequent migrations in the branching process regime, both in the homogeneous and in the heterogeneous case, see Eq.~\ref{eq:aline}. Note that the line was not addressed earlier in this regime, e.g.\ not in Ref.~\cite{Abbara__Frequent}. In the homogeneous case, as the star and all other graphs~\cite{Abbara__Frequent}, the line suppresses natural selection as soon as it has some migration asymmetry, i.e.\ $\alpha\neq 1$, where $\alpha = m_R/m_L$ is the ratio of migration probabilities to the right and to the left. This is explicitly shown in SI Section \ref{sec:SI-homline-homclique} and illustrated in Figure~\ref{fig:cond-aD}. Note that for $\alpha=1$, the line is a circulation.

How does environment heterogeneity impact mutant fixation in the line? To address this question, we compare a line with an environment that can differ in each node $i$, through the mutant fitness advantage $\delta_is$, to a homogeneous line with the same mean mutant fitness advantage $\langle\delta\rangle s$ (see Figure~\ref{fig:line-proba-time}A). Let us denote by 
$\sigma_i=(\delta_i -\langle \delta \rangle)/ \langle \delta \rangle$ the relative mutant fitness excess in deme $i$. The fixation probability in the line with heterogeneous fitness is higher than in the homogeneous line if and only if:
\begin{equation}
   S(\alpha) \equiv \sum_{i=1}^D \sigma_i  \alpha^{-i}>0\,,
    \label{eq:cond-line1}
\end{equation}
see SI Section~\ref{subs:homo-hetero-line}. This entails that, in a heterogeneous line with monotonically decreasing mutant fitness advantage from left to right, the mutant fixation probability is higher than in the homogeneous line for $\alpha>1$, see SI Section~\ref{subs:homo-hetero-line}. The opposite holds for $\alpha<1$, and there is no impact of heterogeneity for $\alpha=1$ (when the line becomes a circulation), to first order in the baseline effective mutant fitness advantage $st$. Therefore, in the line, mutant spread and fixation are facilitated when the demes conferring the highest selective advantage to mutants are upstream of the overall migration flow.

We have shown that environment heterogeneity can increase fixation probability in the line. Can this result in amplification of selection? To address this question, we consider the first-order expansion in $st$ of the fixation probability, and we compare the line to a well-mixed population with the same average mutant fitness advantage, i.e.\ the same $\langle\delta\rangle$, or equivalently, to a clique with the same $\langle\delta\rangle$ (see Figure~\ref{fig:line-proba-time}A). 
Comparing the fixation probabilities to first order, we show that a heterogeneous line has a higher fixation probability than a clique if and only if:
\begin{equation}
   S(\alpha) > g(\alpha)\equiv \frac{(D-1)\alpha -(D+1) + (D+1)\alpha^{1-D} -(D-1)\alpha^{-D}}{\alpha^2-1}\,, 
    \label{eq:cond-line3}
\end{equation}
see SI Section~\ref{sec:heteroC-L}. 

Figure~\ref{fig:line-proba-time}B shows $S$ and $g$ as a function of $\alpha$ for the heterogeneous line shown in Figure~\ref{fig:line-proba-time}A, which has $D=5$ demes and features a step profile of mutant fitness advantages, with $\delta_i=1$ for $i=1$ and $\delta_i=0$ for $i> 1$. For such a decreasing profile of $\delta_i$, amplification of selection cannot exist for $\alpha<1$ (see SI Section \ref{sec:heteroC-L}). We observe on Figure~\ref{fig:line-proba-time}B that amplification of selection exists when $\alpha>1$ (Eq.~\ref{eq:cond-line3} is satisfied). Furthermore, for this particular environmental profile, $S(\alpha)>g(\alpha)$ in asymptotic behavior for large $\alpha$, see SI Section \ref{sec:heteroC-L}. Note that for other environmental profiles, we observed that $g$ and $S$ cross for an $\alpha^*>1$, and that amplification exists in a finite region from $\alpha=1$ to $\alpha=\alpha^*$.

Figure~\ref{fig:line-proba-time}C shows the mutant fixation probability in the above-described heterogeneous line with step environmental profile for $\alpha=1.5$, and evidences a strong amplification of selection compared to a homogeneous clique with the same average fitness. This stands in contrast to the homogeneous line with the same $\alpha$ and $\langle\delta\rangle$, which suppresses selection, see Figure~\ref{fig:line-proba-time}C, and with a heterogeneous line with an opposite environmental profile but the same $\alpha$, which further suppresses selection, see Figure~\ref{fig:grads-1}. Moreover, Figure~\ref{fig:line-proba-time}D-E shows that the average mutant fixation and extinction times are both faster in our heterogeneous line than in the homogeneous clique with the same $\langle\delta\rangle$. Hence, the heterogeneous line is able to both amplify selection and accelerate fixation at the same time. Importantly, this differs from the usual behavior of amplifiers in evolutionary graph theory, which slow down fixation~\cite{frean_effect_2013, hindersin_exact_2016, tkadlec_population_2019}. This joint amplification and acceleration is observed for diverse environmental profiles, as illustrated in Figure~\ref{fig:grads-2}. The results in Figure~\ref{fig:grads-2} suggest that larger environmental contrast between demes leads to more amplification of selection and to more acceleration of mutant fixation and extinction. Note that the same effect of joint amplification and acceleration, albeit less strong, exists for the heterogeneous star, see SI Section \ref{sub:sup-star} and Figure~\ref{fig:faD}.

\paragraph{Locally increasing mutant fitness advantage and migration outflow yield amplification of selection and acceleration of fixation.} In the star and the line, we showed that environment heterogeneity can amplify natural selection and accelerate fixation when mutants have a stronger fitness advantage upstream of the overall migration flow. Do these findings hold beyond the star and the line, when migration asymmetries arise solely from differences in migration probabilities, rather than from the graph connection pattern? To address this question, we consider a graph where all nodes are connected by migrations, but where one special deme (called deme number 1) differs from all others both by migration outflow and by environment. The prefactor of mutant fitness advantage is $\delta_1$ in the special deme and $\delta_2$ in other demes. 

\begin{figure}[htbp]
    \centering
    \includegraphics[width=\linewidth]{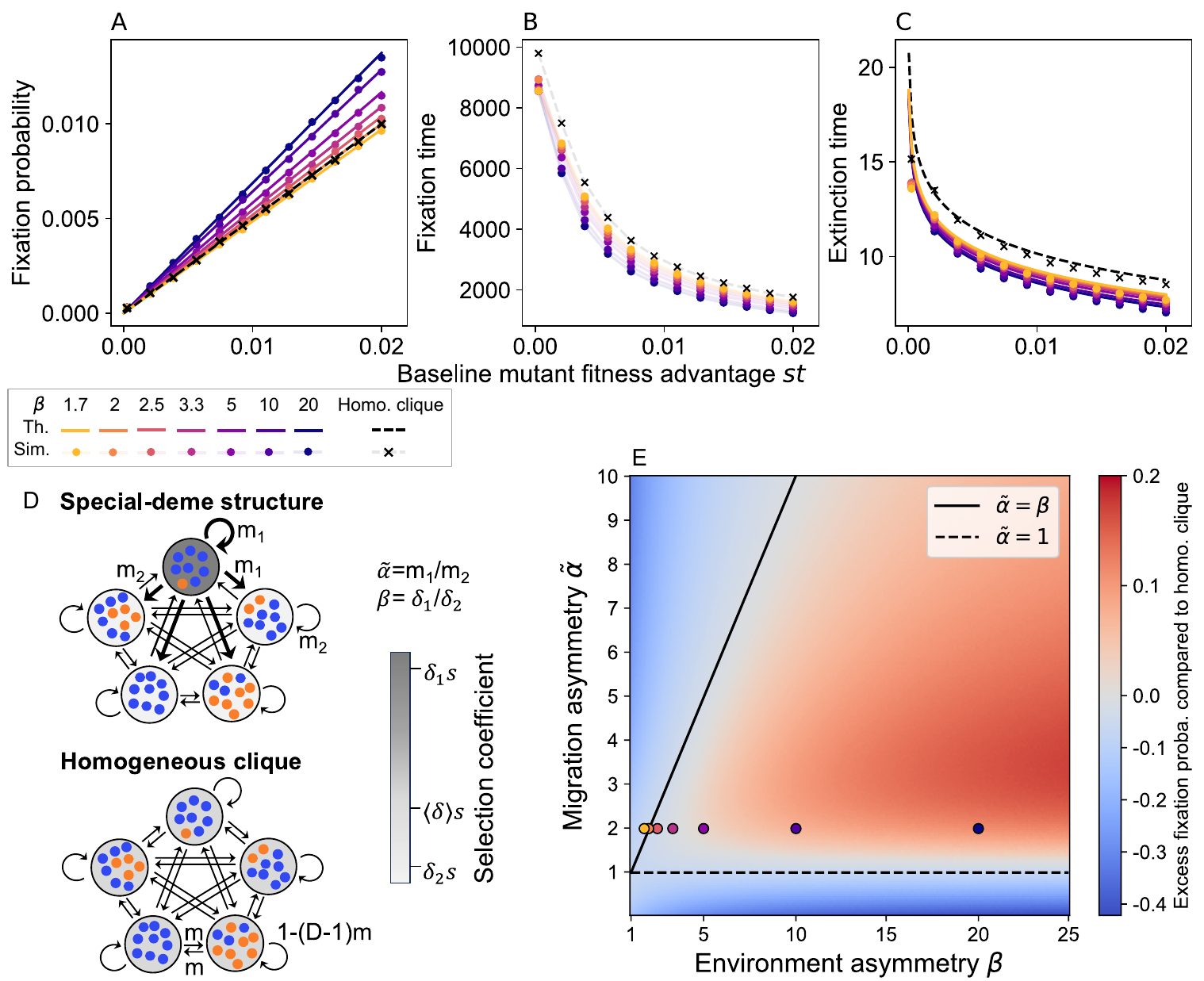}
    \caption{\textbf{Amplification of selection and accelerated dynamics in structures with a special deme in terms of mutant advantage and migration outflow.} Panel A: Mutant fixation probability versus baseline mutant fitness advantage $st$ for fully-connected population structures with one special deme, characterized by stronger migration outflow ($\tilde{\alpha}=m_1/m_2=2$) and greater mutant advantage than others ($\beta=\delta_1/\delta_2>1$), see schematics in D. Results for a homogeneous clique with the same $\langle\delta\rangle$ are shown for reference. Panel B: Time to mutant fixation, in bottlenecks, versus baseline mutant fitness advantage $st$, for the same structures as in A. Panel C: Same as in B, but for extinction time. ``Th.'' indicates branching process theory (Eq.~\ref{eq:pfix-dirdet} for A, Eq.~\ref{eq:text} for  C), ``Sim.'' stochastic simulations (A-C). Panel D: the structures considered in this figure: a fully-connected structure with a special deme and a homogeneous clique. Panel E: The excess fixation probability in a structure with one special deme, compared to a homogeneous clique with the same $\langle\delta\rangle$, is shown versus the migration asymmetry $\tilde{\alpha}$ and the environment asymmetry $\beta$, in the case where migration probabilities are drawn from Dirichlet distributions (no strong symmetry in migration probabilities, generalizing over the structure considered in A-D, see SI Section \ref{subs:alpha-eta}). The plotted values represent the difference between the first-order coefficient in $st$ of the fixation probability (denoted by $a$ in Eq.~\ref{expand}, computed numerically following SI Section \ref{sec:numerical_pfix}, and averaged over many sets of migration probabilities), and the value for the homogeneous clique with the same $\langle\delta\rangle$, which is $2\langle\delta\rangle$. Solid and dashed black lines mark theoretical boundaries for amplification in the strongly symmetric structure considered in A-D (see Eq.~\ref{eq:alphabelta} and main text): the solid line corresponds to $\tilde{\alpha} = \beta $ and the dashed line to $\tilde{\alpha} = 1$. Markers show the $(\tilde{\alpha},\beta)$ values considered in A-C. 
    Parameter values throughout: $D=5$, $\langle \delta \rangle = 0.25$; for A-C: $K=1000$, $\tilde{\alpha}=2$ and $m_1=0.33$ for the special-deme structure, $m=0.15$ for the homogeneous clique. Each value in the heatmap of E is obtained by averaging over $5\times10^3$ structures, whose migration probabilities are each time drawn independently in Dirichlet distributions.}
    \label{fig:Dirichlet-clique}
\end{figure}

We first address the highly symmetric case in which all outgoing migration probabilities from the special deme are equal to $m_1$, while all other migration probabilities are equal to $m_2$, and $\tilde{\alpha}=m_1/m_2$ denotes migration asymmetry (Figure~\ref{fig:Dirichlet-clique}A-D). When $\tilde{\alpha}>1$, meaning that the special deme has a stronger migration outflow than others, the spatial structure considered amplifies natural selection if and only if:
\begin{equation}
\frac{\delta_1}{\delta_2}\equiv \beta >\tilde{\alpha}\,,\label{eq:alphabelta}
\end{equation}
see SI Section \ref{subs:simplified-Dirichlet}. Hence, when the special deme sends out more individuals than others, selection is amplified if the mutant advantage in that deme is sufficiently stronger than elsewhere. More precisely, this environment contrast needs to be larger than the migration contrast. In the opposite case where $\tilde{\alpha}<1$, amplification occurs for $\beta<\tilde{\alpha}$, see SI Section \ref{subs:simplified-Dirichlet}. Hence, for amplification to exist in this structure, mutant advantage needs to be stronger upstream of the overall migration flow, and moreover, the environment needs to be more contrasted than the migration flows. Figure~\ref{fig:Dirichlet-clique}A shows that our prediction in Eq.~\ref{eq:alphabelta} is satisfied: with $\tilde{\alpha}>1$, amplification is obtained for $\beta>\tilde{\alpha}$, results coincide with those in the clique for $\beta=\tilde{\alpha}$, and suppression occurs for $\beta<\tilde{\alpha}$. Moreover, Figure~\ref{fig:Dirichlet-clique}B-C shows that amplification of selection is accompanied by acceleration of fixation and extinction. This is consistent with our findings for the star and the line, and suggests that locally increasing mutant fitness advantage and migration outflow constitutes a general mechanism for amplification of selection and acceleration.

Do these findings depend on the strong symmetry in migration probabilities assumed above? To address this question, we consider a structure with a special deme, as before, but where for each deme, the migration probabilities incoming to that deme from all other ones are drawn from the same Dirichlet distribution. The parameters of this distribution are chosen such that the special (first) deme has a stronger expected migration outflow than others, see SI Section \ref{subs:alpha-eta}, but with a substantial dispersion of migration probabilities, see Figure~\ref {fig:histo-migrates}. Figure~\ref{fig:Dirichlet-clique}E shows that, in this more general structure, substantial amplification of selection exists in the parameter region where $\beta>\tilde{\alpha}>1$, consistent with the conditions derived for the highly-symmetric case above. This indicates that the mechanism yielding amplification is robust to heterogeneity in migration intensities within a fully connected graph.

\paragraph{Amplification of selection extends to deleterious mutants and large graphs.} 

So far, we focused on beneficial mutants ($s>0$), allowing analytical insight using the branching process formalism. By definition, amplification of selection means that beneficial mutants have a higher fixation probability than in a well-mixed population or clique, while deleterious ones have a lower one~\cite{lieberman_evolutionary_2005}. To test whether amplification extends to deleterious mutants in our graph-structured populations with heterogeneous environments, we perform simulations in the star, the line and the highly-symmetric structure with one special deme, under conditions where overall migration flows from demes with the strongest mutant fitness advantage to those with weaker advantage, and where we previously observed amplification for $s>0$. We find that in all cases, amplification extends to deleterious mutants: their fixation probabilities are smaller than in a homogeneous clique with the same average environment $\langle\delta\rangle$, see Figure~\ref{fig:negs}.

Having confirmed that amplification extends to deleterious mutants, we next ask whether it persists in larger graphs. So far, we mainly considered relatively small graphs with five nodes. We now vary the number $D$ of demes while keeping $\langle\delta\rangle$ constant, in the same three highly-symmetric structures, using analytical expressions for the first-order coefficient in $st$ of the mutant fixation probability derived from the branching process approach (denoted by $a$ in Eq.~\ref{expand}). In the star where all leaves have the same environment, and in the structure with one special deme, we hold the environment contrast between demes constant in addition to $\langle\delta\rangle$. Under these conditions, we find that amplification can persist for all $D$, although its intensity decreases when $D$ increases, see Figure~\ref{fig:changeD-star}. We further derive a condition on migration asymmetry and on contrast for amplification in the large-$D$ limit in these structures, see SI Sections~\ref{subsec:varyD-star} and~\ref{subsec:highsym}. In contrast, in the line where all demes have the same environment, except the most upstream deme, amplification vanishes and suppression occurs beyond moderate $D$ when $\langle\delta\rangle$ and contrast are held constant, see Figure~\ref{fig:changeD-line}A. However, alternative conventions that hold $\langle\delta\rangle$ but not contrast constant can produce amplification that increases with $D$ in the line (Figure~\ref{fig:changeD-line}B; see SI Section~\ref{subsec:varyD-line} for analytical conditions). These results demonstrate that amplification of selection can exist in large graphs with heterogeneous environments, when overall migration flows from demes with the strongest mutant fitness advantage to those with weaker advantage. They also indicate that specific conditions on migration and environment are required to maintain significant amplification in large graphs.

\begin{figure}[htbp]
    \centering
    \includegraphics[width=\linewidth]{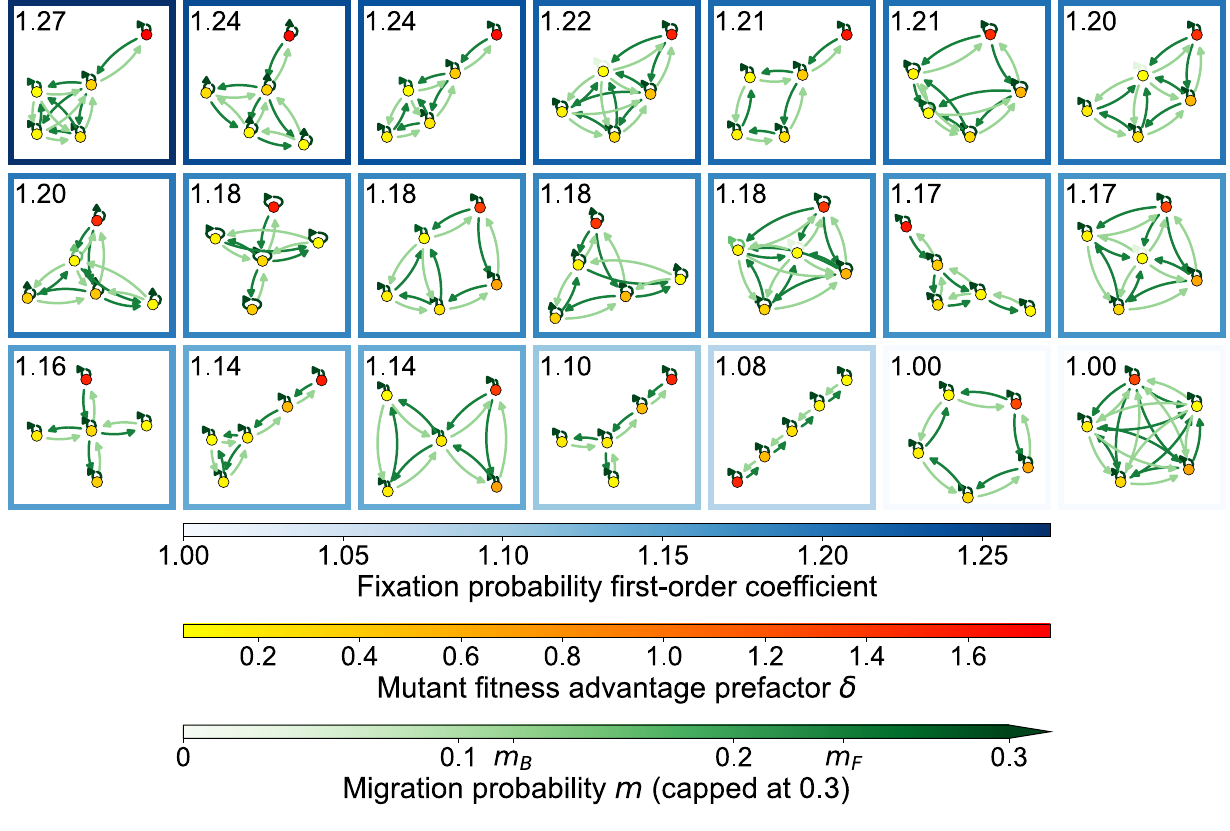}
    \caption{\textbf{Amplification of selection across five-node graphs with strong migration outflow from demes where mutants are advantaged.} Each panel shows a connected five-node graph, ordered by decreasing first-order coefficient in the expansion of the fixation probability in $st$ (denoted by $a$ in Eq.~\ref{expand}). These coefficients are indicated in the top left corner of each panel, and shown through the panel frame's color. They were calculated numerically as described in SI Section~\ref{sec:numerical_pfix}. Migration probabilities and environment heterogeneity follow convention 1 defined in SI Section~\ref{SI:gengraphs}, ensuring that overall migration flows from demes whith the strongest mutant fitness advantage to those with weaker advantage. For all graphs, $\langle \delta \rangle=0.5$. Hence, circulations have a first-order coefficient $2\langle \delta \rangle=1$, and coefficients larger than one indicate amplification of selection.}
    \label{fig:gengraphs}
\end{figure}

\paragraph{Amplification exists across graph structures with strong migration outflow from demes where mutants are advantaged.}
Does the amplification mechanism we identified persist in generic graph structures, beyond strongly symmetric or structurally simple graphs?
To address this question, we now consider all connected graphs with five demes~\cite{hindersin_counterintuitive_2014,Hindersin15,moller_exploring_2019}, shown in Figure~\ref{fig:gengraphs}. In all of them, we set migration probabilities and environment heterogeneity according to a convention that ensures that the overall migration flow is directed from the demes where mutants have the strongest fitness advantage to those where this advantage is smaller (see SI Section~\ref{SI:gengraphs}). Figure~\ref{fig:gengraphs} reports the value of the first-order coefficient in the expansion of the fixation probability in $st$ in each structure (denoted by $a$ in Eq.~\ref{expand}), sorted from the highest to the lowest first-order coefficient. Amplification of selection exists in all graphs considered, except the cycle and the clique, which are both circulations. We further observe that amplification of selection is stronger when the deme where mutants are most advantaged has a small degree, while other nodes are strongly connected together. These results show that the amplification mechanism identified above extends from specific analytically tractable structures to diverse structures, indicating that amplification of selection can generically exist when overall migration flows from demes with the strongest mutant fitness advantage to those with weaker advantage. Nevertheless, other conventions often lead to suppression of selection (see SI Section~\ref{SI:gengraphs} and Figure~\ref{fig:all-graphs-FO}), suggesting that suppression remains the most generic effect of spatial structure under frequent migrations in the branching process regime, as for homogeneous environments~\cite{Abbara__Frequent}. Specific situations, such as the overall migration flow going in the same direction as the environment gradient, are required to reverse this trend. 

\paragraph{Heterogeneous environments can induce amplification of selection in the rare migration regime.} So far, we considered frequent migrations, and focused on the branching process regime, assuming that $K\gg 1$ and $1/K\ll st\ll 1$. In this regime, we showed that environment heterogeneity can turn suppressors of selection into amplifiers. Can this also occur in the rare-migration regime, where the timescale of migrations is much slower than mutant fixation or extinction within a deme? To address this question, let us return to the star graph and to the line graph. Their mutant fixation probabilities were determined analytically for homogeneous environments in the rare migration regime, respectively in Refs.~\cite{marrec_toward_2021} and~\cite{servajean_impact_2025}. A key point under rare migrations is that a mutant that fixes always first takes over its deme of origin, before spreading to other demes, allowing a coarse-grained Markov chain description~\cite{Slatkin81,marrec_toward_2021}. Here, we generalize these calculations to heterogeneous environments. For the star, we consider the case where the center has a different environment from the leaves, associated to the mutant advantage prefactor $\delta_C$, but the leaves all have the same environment, associated to $\delta_L$, see SI Section \ref{subs:RM-star}. For the line, we perform a general calculation where all demes can have different environments, with a mutant advantage prefactor $\delta_i$ in deme $i$, see SI Section \ref{subs:RM-line}. 

\begin{figure}[htb!] 
\centering
\includegraphics[width=\textwidth]{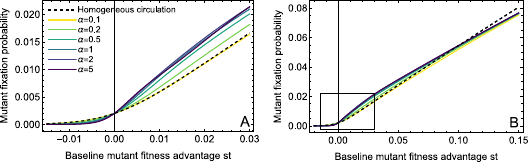}
    \caption{\textbf{Amplification in the heterogeneous star with rare migrations.} We show the mutant fixation probability, starting from one single mutant placed uniformly at random at a bottleneck (Eq.~\ref{PhiStar_2} using Eq.~\ref{Phi_star}), in a heterogeneous star with various migration asymmetries $\alpha$, and in a homogeneous circulation with the same mean mutant fitness advantage, i.e.\ with the same $\langle\delta\rangle$. All fixation probabilities are plotted versus the baseline mutant fitness advantage $st$. Panels A and B show different ranges of $st$. Parameter values: $D=5$, $K=100$. For the heterogeneous star, $\delta_C=1$ and $\delta_L=0.1$. }
    \label{fig:star_rare}
\end{figure}

Figure~\ref{fig:star_rare} shows that the heterogeneous star in the rare migration regime can amplify natural selection, compared to a homogeneous circulation with the same mean mutant fitness, in a broad range of migration asymmetries $\alpha=m_I/m_O$. Strikingly, this includes cases where $\alpha<1$, for which the star suppresses natural selection under rare migrations~\cite{marrec_toward_2021}. Hence, environment heterogeneity can turn suppressors of selection into amplifiers in the rare migration regime too. In Figure~\ref{fig:star_rare}, we observe amplification of selection for weakly deleterious and weakly beneficial mutants (recall that for deleterious mutants, amplification corresponds to having a smaller fixation probability than in the well-mixed or circulation case). The results of Figure~\ref{fig:star_rare} are obtained with $\delta_C$ substantially larger than $\delta_L$, i.e.\ when beneficial (resp.\ deleterious) mutants are substantially more advantaged (resp.\ disadvantaged) in the center than in leaves. In this case, consider $st$ such that $1/K\ll 2\delta_C st\ll 1$ but $2\delta_L st\lesssim 1/K$. Then, if a mutant starting in the center fixes there, it should ultimately take over the whole population. Indeed, wild-type individuals migrating to the center from other leaves cannot take over again in the center, due to their substantial fitness disadvantage (their fixation probability is exponentially suppressed). This existence of a special ``safe'' deme for mutants sets the star apart from the homogeneous circulation with the same $\langle\delta\rangle$, and causes amplification in this regime (see SI Section \ref{subs:RM-star} for further details, including an explanation of the suppression of selection observed for larger $st$). Recall that in the homogeneous star with $\alpha>1$, amplification of selection is observed under rare migrations~\cite{marrec_toward_2021}, but only when $st\lesssim 1/K$~\cite{Abbara__Frequent}. In fact, in the rare migration regime, amplification effects are associated to the finite size of demes and only exist in that regime. Our new findings are consistent with this, and shed light on how environment heterogeneity can induce amplification in the rare migration regime.

In the case of the line, which is a suppressor of selection in the homogeneous case~\cite{servajean_impact_2025}, we also find that environment heterogeneity can yield amplification of selection. This is shown in Figure~\ref{fig:p_mutant_line}, where we assume that only the leftmost deme has $\delta_L$ while the other ones have $\delta_R$ substantially smaller. A small amplification occurs in the regime where $1/K\ll 2\delta_L st\ll 1$ but $2\delta_R st\lesssim 1/K$, see SI Section \ref{subs:RM-line}. As in the star, the mutant benefits from the existence of a ``safe'' (leftmost) deme, where it cannot be re-invaded by wild-types. However, its location is less ideal than in the star, leading to a smaller effect. This illustrates the generality of the amplification mechanism by environment heterogeneity in the rare migration regime: it relies on refugia where mutants are safe from re-invasion, and is specific to finite-sized demes and to a range of small fitness differences between mutant and wild-type individuals.

\section*{Discussion}
In this work, we investigated how environment heterogeneity impacts mutant fixation in spatially structured populations with demes on nodes of a graph. We first focused on the frequent migration regime. We proved that the circulation theorem remains valid to first order in heterogeneous environments. However, to second order, environment heterogeneity increases the fixation probability of beneficial mutants in circulations. In non-circulation graphs, we demonstrated that environment heterogeneity can lead to amplification of selection, together with acceleration of mutant fixation and extinction. This result is striking for two reasons: first, without environment heterogeneity, no amplification is possible under frequent migrations~\cite{Abbara__Frequent}. Second, amplifiers of selection generally slow down fixation~\cite{frean_effect_2013, hindersin_counterintuitive_2014, hindersin_exact_2016, tkadlec_population_2019, tkadlec_fast_2021}. We found that this simultaneous amplification of selection and acceleration occurs in the star graph, and even more strongly in the line graph. We identified a key condition for amplification: mutants must be more favored in demes with stronger migration outflow. Qualitatively, individuals in a deme with strong outflow contribute more to next generations. Thus, a fitness increase in this deme helps mutants more than in other demes, contributing to amplification of selection. Under this condition, amplification can occur for deleterious as well as beneficial mutants and persists in large graphs. It also exists in more general graphs, including fully connected graphs with one special deme and all connected graphs with five nodes that are not circulations. Nevertheless, the most generic effect of spatial structure on mutant fixation remains suppression of selection, as in homogeneous environments~\cite{Abbara__Frequent}, since specific conditions on migration and environment are required to obtain amplification. 
Finally, in the rare migration regime, where mutant fixation or extinction in a deme is faster than migrations, environment heterogeneity can turn suppressors of selection into amplifiers.

While environment heterogeneity can foster amplification of selection under both frequent and rare migrations, the mechanisms differ between these two cases, as does the regime of mutant fitness advantage where amplification is observed. For frequent migrations, our branching process analysis focuses on the early and highly stochastic dynamics of a mutant lineage, long before it reaches a size comparable to that of a deme. In this regime, regions with both stronger mutant fitness advantage and larger migration outflow allow mutants to more successfully avoid extinction. 
Meanwhile, for rare migrations, fixation must occur within a deme before mutants can spread further. If mutant fitness advantage is much larger than $1/K$ in a deme, this deme essentially cannot be re-invaded by wild-type individuals if the mutant type fixes there, since wild-type fixation is extremely unlikely. Such demes with sufficient mutant advantage become refugia for mutants, compared to other demes where mutant fitness advantage is smaller and wild-types can take over again, thereby enabling amplification of selection. This effect is associated to the possibility that wild-types take over again in a deme, and thus requires small fitness advantages and finite deme sizes. This stands in contrast with the frequent-migration amplification effect. 

Our results were obtained by extending to heterogeneous environments a general model of spatially structured populations on graphs, which does not rely on update rules~\cite{Abbara__Frequent}, and is close to experimental protocols of serial transfer~\cite{Wahl01,LeClair18,Kryazhimskiy12,Nahum15,france_relationship_2019,Chakraborty23,abbara_mutant_2024}. Compared to previous studies of heterogeneous environments in evolutionary graph theory~\cite{maciejewski_environmental_2014, kamran_environmental_2019, kaveh_themoran_2020, brendborg_fixation_2022, svoboda_coexistence_2023, nemati_counterintuitive_2023}, which generally focus on two environments and their repartition across nodes, we addressed generic heterogeneity across demes. We obtained results both in the frequent migration branching process regime and in the rare migration regime. While the latter can in some cases be mapped to evolutionary graph theory models with one individual per node~\cite{marrec_toward_2021}, the frequent migration regime is fundamentally different, as states where several demes comprise both mutants and wild-type individuals become central to the mutant's fate. 
In population genetics, environment heterogeneity in spatially extended populations has also been studied in the context of species range limits. In these works, when a mutation is beneficial in one region but deleterious in another, gene flow from the region where the mutation is deleterious can hinder local adaptation in the one where it is beneficial, an effect known as swamping~\cite{haldane_relation_1956,slatkin_gene_1973,kirkpatrick_evolution_1997,lenormand_gene_2002,kawecki_conceptual_2004,polechova_limits_2015,kottler_draining_2021}. These studies typically consider continuous spatial structures and antagonistic effects of selection across space~\cite{slatkin_gene_1973,lenormand_gene_2002,kawecki_conceptual_2004} or quantitative traits whose optimum varies spatially~\cite{kirkpatrick_evolution_1997,polechova_limits_2015}, and focus on range expansion rather than mutant fixation or extinction. Despite these important differences, it is interesting to note that in swamping, migration from regions where a mutant is disadvantaged to regions where it is advantaged hampers adaptation, echoing our finding that the opposite migration flow (from more favorable to less favorable regions) can amplify natural selection. 

Our analysis is relevant to experimental microbial populations, particularly in the frequent migration regime~\cite{chakraborty_experimental_2023,abbara_mutant_2024}, whereas the rare migration regime leads to very long fixation times. Moreover, our results have direct implications for natural microbial populations in spatially structured habitats, such as the gut. In the line graph, we showed that a decreasing mutant fitness along the main flow direction can foster amplification of selection. 
The gut can be considered as a one-dimensional system to a first approximation, and features a directional hydrodynamic flow together with a fitness gradient due to nutrient availability~\cite{Cremer16, Cremer17, salari_diurnal_2025}. These ingredients have been shown to impact mutant fixation~\cite{labavic_hydrodynamic_2022, ghosh_emergent_2022}. Our results for a discrete line of demes suggest additional effects for mutants with spatially-dependent advantages in the gut, in particular promoting and accelerating their fixation.

Our results, obtained for spatially structured populations with (almost) fixed sizes, also resonate with known ones for expanding populations. 
In particular, the presence of antibiotic concentration gradients can foster resistance evolution in expanding populations~\cite{zhang_acceleration_2011, greulich_mutational_2012, hermsen_onthe_2012, baym_spatiotemporal_2016}. In these cases, expanding populations move toward zones where mutant fitness advantage is larger. It would be interesting to generalize our model to include initially empty demes where the population can grow. Other promising directions would be to consider larger and more complex graphs~\cite{kuo_theory_2024}, including scale-free~\cite{barabasi_emergence_1999} and small-world~\cite{watts_collective_1998} networks, and to include time-varying environments in demes~\cite{hernandez_coupled_2023, hernandez_ecoevolutionary_2024}. Moreover, while our focus has been on the fate of a single mutation, it would be very interesting to investigate the impact of heterogeneous spatially structured populations on fitness landscape exploration~\cite{visser2014,Das25,servajean_impact_2025}, longer-term evolution~\cite{Sharma22,sharma_graph-structured_2025}, as well as on population diversity under frequent mutations~\cite{Desai07,Good17,Blundell19}.

While here we focused on mutants whose fitness depends on the deme, the framework could also be extended to study social interactions, where fitness also depends on interactions between individuals within a deme. An intriguing perspective would be to investigate the impact of environment heterogeneity on social evolution in spatially structured populations, for instance in the context of evolutionary games and the evolution of cooperation~\cite{Turner99,doebeli2005,Nowak_2006,ohtsuki2006evolutionary,traulsen2009stochastic,gokhale_evolutionary_2014,Review_2019,Moawad_2023,RibierePreprint}. It would also be interesting to generalize such a study to group interactions~\cite{perc_coevolutionary_2010,perc_evolutionary_2013} and higher-order networks~\cite{alvarez-rodrigues_evolutionary_2021}, and to address cases where the environment or the graph coevolves with population composition~\cite{perc_coevolutionary_2010,perc_evolutionary_2013}.

\section*{Methods}

\paragraph{Serial dilution model on a graph with heterogeneous environment.}
We model a spatially structured population as a set of well-mixed demes, each sitting on one node of a connected graph with $D$ nodes. Migration probabilities $m_{ij}$ are defined between any pair of demes $(i,j) \in \{1,...,D\}^2$, including $i=j$, corresponding to individuals that stay in the same deme. They differ by their fitnesses, which are respectively $f_W=1$ for wild-types (taken as reference) and $f_M=1+s$ for mutants. Thus, $s$ is the relative fitness advantage of mutants compared to wild-types. The population is composed of wild-types of fitness $f_W=1$ (taken as reference) and mutants, whose fitness depends on the environment, and is denoted by $f_{M,i}=1+\delta_i s$ in deme $i$. Here, $s$ is the baseline mutant fitness advantage, and $\delta_i$ is an environment-dependent prefactor, which is nonnegative, and of order unity if it is nonzero. 

We generalize the model introduced in Ref.~\cite{Abbara__Frequent} to heterogeneous environments. The dynamics of the population, modeling serial passage with migrations, is composed of alternations of phases of growth and migration-dilution, see Figure~\ref{fig:methods}. 

Starting at a bottleneck, each deme separately undergoes deterministic exponential growth for a fixed time $t$. Fitnesses are understood as exponential growth rates. Therefore, the growth step is impacted by environment heterogeneity via the mutant fitness advantage $s \delta_i$, which depends on the deme $i$. Denoting by $x_i$ the initial fraction of mutants in deme $i$, the fraction of mutants after growth in deme $i$ is: 
\begin{equation}
x'_i = \frac{x_i e^{\delta_i st} }{1 + x_i (e^{\delta_i st} - 1)}\, . 
\label{eq:xpi}
\end{equation}
Note that the impact of the baseline selective advantage of mutants is encoded by the product $st$. We thus call $st$ the (effective) baseline fitness advantage. 

After a growth phase, a dilution (regulation) and migration phase takes place. The number of mutants going from deme $i$ to deme $j$ is sampled from a binomial distribution with $N_i'$ trials, with $N_i'$ the size of the deme after growth, and a probability of success $m_{ij} x_i' K/N_i'$. Similarly for wild types, but with a probability of success $m_{ij} (1-x_i') K/N_i'$.
Each deme $j$ thus receives on average $K \sum_{i=1}^D m_{ij}$ individuals, leading to a new bottleneck state. We assume that for all $j$, $\sum_{i=1}^D m_{ij}=1$, so that all demes have the same average bottleneck size $K$. 
For each pair of demes $(i,j)$, this phase corresponds to sampling approximately $K m_{ij}$ bacteria from the total $N_i'$ bacteria present in deme $i$, which contribute to the new bottleneck state of deme $j$. 

Growth and dilution-migration phases are iterated. 
Importantly, this serial dilution model with migrations is very close to a structured Wright-Fisher model~\cite{Lessard07,Burden18}. In fact, we showed in Ref.~\cite{Abbara__Frequent} that for $s\ll 1$, in the branching process approximation, results from structured Wright-Fisher models can be recovered from those of the serial dilution model by taking growth time $t=1$. As a consequence, all our results here hold for structured Wright-Fisher models.

\paragraph{Branching process description.} 
The state of the population is described with a multi-type branching process, where each type corresponds to mutants in each deme~\cite{Abbara__Frequent}. The branching process description assumes that all mutant lineages are independent~\cite{harris_thetheory_1963}. It holds when mutants are in small numbers and deme sizes are large, $K\gg 1$. For mutants with substantial selective advantage $st\gg 1/K$, extinction events happen when mutants are still rare~\cite{Desai07,Boenkost21}. Hence, the branching process approach holds when $K\gg 1$ and $st\gg 1/K$. 

In the branching process regime, assuming that all nonzero migration probabilities are of order unity, the binomial distributions used to model a dilution-migration phase (see above) can be approximated by Poisson ones. Under these conditions, starting from one mutant in deme $i$ at a bottleneck, the mutant extinction probability $p_i$ satisfies:
\begin{equation}
    p_i= \exp \left[ Kx'_i \sum_{j=1}^D m_{ij} (p_j -1) \right],
    \label{eq:fpe}
\end{equation}
for all $i$, where $x'_i$ is given by Eq.~\ref{eq:xpi} with $x_i=1/K$. These equations generalize those of the homogeneous case~\cite{Abbara__Frequent}, as here the mutant fraction $x'_i$ in deme $i$ after growth involves $\delta_i$. To solve Eq.~\ref{eq:fpe} for all $i$, we assume that the baseline mutant fitness advantage is positive but small, $0<st\ll 1$, and perform a Taylor expansion, see Eq.~\ref{expand} and SI Section \ref{sec:diff_env_all}. 
The fixation probability starting from one mutant in deme $i$ at a bottleneck is then $\rho_i=1-p_i$. 
We focus on the average fixation probability $ \rho = \sum_{i=1}^D \rho_i / D$ across the structure, assuming that mutants appear uniformly at random in the structure.

The average mutant extinction time can also be obtained in the branching process regime, see SI Section \ref{sec:ext_time}.

\paragraph{Rare migration regime.} When the timescale of migrations is much slower than mutant fixation or extinction within a deme, one can consider that each deme is always fully wild type or fully mutant as far as migration events are concerned. This allows a coarse-grained Markov chain description~\cite{Slatkin81,marrec_toward_2021}, which has allowed the determination of exact fixation probabilities in the homogeneous case~\cite{marrec_toward_2021, servajean_impact_2025}. Here, we extend this approach to the heterogeneous case, see SI Section \ref{subs:RM-star}. Contrary to the branching process results, these results hold for all values of $st$, including for deleterious, neutral and effectively neutral mutants. Besides, they connect to evolutionary graph theory models~\cite{marrec_toward_2021}. However, the rare migration regime is associated to slow dynamics. In contrast, the branching process calculations assume that migration probabilities are of order unity, so these two approaches cover entirely distinct regimes and provide complementary insights.

\paragraph*{Data availability statement.}
All the data relevant to this study is simulation data, and is included in the paper or in the supplementary material. The code used to generate this simulation data, allowing to reproduce our analyses, is available as specified below.

\paragraph*{Code availability statement.}
Our code is freely available at \url{https://github.com/Bitbol-Lab/StructPop_Heterogeneity} and archived on Zenodo  \cite{fruet2026environment_software}.

\section*{Acknowledgments}
This research was partly funded by the Swiss National Science Foundation (SNSF, grant No.~315230 208196, to A.-F.~B.), by the Chan-Zuckerberg Initiative (CZI, to A.-F.~B.), and by the European Research Council (ERC) under the European Union’s Horizon 2020 research and innovation programme (grant agreement No.~851173, to A.-F.~B.). C.~L. acknowledges funding by the Agence Nationale de la Recherche (grants ANR-20-CE30-0001, ANR-21-CE45-0015 and ANR-25-CE44-4822, to C.~L.).

\clearpage


\renewcommand{\thefigure}{S\arabic{figure}}
\setcounter{figure}{0}
\renewcommand{\thetable}{S\arabic{table}}
\setcounter{table}{0}
\renewcommand{\theequation}{S\arabic{equation}}
\setcounter{equation}{0}

\newpage

\begin{center}
\begin{Huge}
Supplementary Information
\end{Huge}
\end{center}

\vspace{1 cm}

\begingroup
\hypersetup{linkcolor=black}
\tableofcontents
\endgroup

\newpage

\section{Branching process approach for frequent migrations}  
\label{sec:diff_env_all}

\subsection{Poisson approximation}

For completeness, we briefly restate the stochastic process defined in the main text section ``Methods", in order to set notations, and to introduce the Poisson approximation for the migration-dilution process. We then calculate mutant fixation probabilities in the branching process regime.

We consider spatially structured populations with $D$ demes, each located on each node of a graph. We assume that there are two types of individuals, mutants and wild-types. Let us denote by $s$ the baseline relative fitness advantage of mutants with respect to wild-types. Let us further model environment heterogeneity by introducing a deme-dependent prefactor $\delta_i$  in the mutant relative fitness advantage, such that in deme $i$, we have a relative fitness difference of $\delta_i s$ between the two types. In other words, taking wild-type fitness as reference ($f_W=1$), we have $f_{M,i}=1+\delta_i s$, where $f_{M,i}$ denotes mutant fitness in deme $i$, while $f_W$ denotes wild-type fitness. We assume that $\delta_i$ is nonnegative, and of order unity if it is nonzero.

We start from the serial dilution model presented in Ref. \cite{Abbara__Frequent}, and we extend it to the heterogeneous case, where mutant fitness advantage is deme-dependent. In this model, demes undergo alternations of local deterministic exponential growth for time $t$, and of events of simultaneous dilution and migration leading to new bottlenecks, see Figure~\ref{fig:methods}. We call $m_{ij}$ the migration probability from deme $i$ to deme $j$ and assume $\sum_{i=1}^D m_{ij}=1$ for all $j$. 

We focus on the fate of a mutant that appears in deme $i$ at a bottleneck, giving an initial mutant fraction $x_0^{(i)}=1/K$ in that deme, where $K$ is bottleneck size, and 0 in other demes. For this, we consider the first growth and migration-dilution events happening. The mutant fraction after exponential growth reads in deme $i$~\cite{Abbara__Frequent}:
\begin{equation}
    x_0'^{(i)} =\frac{e^{\delta_i s t}}{K + e^{\delta_i s t} -1} \equiv \frac{\lambda^{(i)}}{K}\,,
    \label{eq:x0'}
\end{equation}
where we introduced $\lambda^{(i)}$, which is a deme-specific growth factor. Then, we model the subsequent migration-dilution event through binomial sampling from each deme to each other demes. In particular, mutants are sampled from the grown deme $i$, of size $N'_{i}$, to a destination deme $j$ through the binomial law $\mathcal{B}\left( N_i', Km_{ij}x_0'^{(i)}/N_i' \right)$. Assuming $N_i'\gg 1$, $Km_{ij}x_0'^{(i)} /N_i'\ll 1$ and $\lambda_{ij} = Km_{ij} x_0'^{(i)} \sim 1$, we approximate the binomial law by a Poisson law with mean
\begin{equation}
    \lambda_{ij} = K m_{ij} x_0'^{(i)}=\lambda^{(i)}m_{ij}\,.
\end{equation}
Compared to the homogeneous case considered in Ref. \cite{Abbara__Frequent}, the impact of heterogeneity is that the growth factor $\lambda^{(i)}$ depends on the deme $i$.

\subsection{Expansion of the mutant fixation probability}
\label{SI-expand}

Assuming $s>0$, $K\gg 1$ and $1/K\ll st\ll 1$, and following a multi-type branching process description, as in Ref.~\cite{Abbara__Frequent}, the probability $p_i$ that mutants gets extinct, starting from one single mutant in deme $i$ at the bottleneck, satisfies:
\begin{equation}
    p_i = \exp\left[\lambda^{(i)}\sum_{j=1}^D m_{ij} (p_j -1) \right]. \label{bpeq}
\end{equation}
Note that, while this equation regards the mutant extinction probability, the mutant fixation probability can then be obtained as $\rho_i=1-p_i$.

To solve the equations in Eq.~\ref{bpeq}, we perform a perturbative expansion in $st\ll 1$:
\begin{equation}
    p_j = 1 - a_j s t + \frac{b_j}{2} (s t)^2 -\frac{c_j}{6}(st)^3 +o((s t)^3) \hspace{1 cm} j = 1, \dots, D\,,
    \label{eq:pjsd-grad}
\end{equation}
where $a_j$, $b_j$ and $c_j$ are coefficients to be determined. Expanding Eq.~\ref{eq:x0'} yields
\begin{equation}
   \lambda^{(j)} = 1 + \delta_j s t + \frac{\delta_j ^2(st)^2}{2}  +o((s t)^2) \hspace{1 cm} j = 1, \dots, D. \label{lambdaj}
\end{equation}
Eq.~\ref{bpeq} then yields, for $j = 1, \dots, D$:
\begin{equation}
    \begin{split}
        p_j &=  \exp\left[\left( 1 + \delta_j s t + \delta_j ^2 \frac{(st)^2}{2}  +o((s t)^2)  \right) \sum_{k=1}^D m_{jk} \left( - a_k  s t + \frac{b_k}{2} ( s t)^2 -\frac{c_k}{6}( st)^3 +o(( st)^3)\right) \right]\\
        & = \exp\left[ - s t  \sum_{k=1}^D m_{jk} a_k   + (st )^2 \sum_{k=1}^D m_{jk}\left(\frac{b_j}{2} -a_j\delta_j\right)   + (st)^3 \sum_{k=1}^D m_{jk}\left( \frac{b_k}{2} \delta_j - a_k \frac{\delta_j^2}{2} -\frac{c_k}{6} \right)+o(( st)^3)\right].
    \end{split}
\end{equation}
This gives, to third order in $st\ll 1$:
\begin{equation}
\begin{split}
p_j & = 1 - s t \sum_k m_{jk}a_k  + \frac{(st)^2}{2} \left[ \left( \sum_{k=1}^D m_{jk}a_k \right)^2 +\sum_{k=1}^D m_{jk}\left(b_k -2\delta_j a_k\right)  \right]  \\
& + (st)^3 \left[ -\frac{1}{6} \left(\sum_{k=1}^D m_{jk} a_j \right)^3 - \sum_{k=1}^D m_{jk} a_k \, \sum_{l=1}^D m_{jl}\left( \frac{b_l}{2} -a_l\delta_j \right) +  \sum_{k=1}^D m_{jk} \left(\frac{b_k}{2} \delta_j -a_k \frac{\delta_j^2}{2} -\frac{c_k}{6}\right) \right].
\end{split}
\label{eq:pisd-grad-2}
\end{equation}
Eqs.~\ref{eq:pjsd-grad} and \ref{eq:pisd-grad-2} yield by identification, for $j=1, \dots, D$:
\begin{equation}
    \begin{cases}
        a_j =  \sum_{k=1}^D m_{jk}a_k\,, \\
        b_j =\sum_{k=1}^D m_{jk} b_k +a_j^2 -2 \delta_j a_j\,,\\
        c_j = \sum_{k=1}^D m_{jk} c_k-2 a_j^3 + 3a_jb_j   - 3\delta_j b_j +3\delta_j a_j^2 -3 \delta_j^2a_j.
    \end{cases}
    \label{eq:coeffs-abc}
\end{equation}
The first line in Eq.~\ref{eq:coeffs-abc} is the same in a homogeneous environment~\cite{Abbara__Frequent}, but the other ones differ, as they involve $\delta_j$.

In the following sections, we use Eq.~\ref{eq:coeffs-abc} to calculate the first-order coefficients $a_i$ first numerically for any strongly connected graph in Section~\ref{sec:numerical_pfix} and then analytically for different graphs with strong symmetries or simple structures in Section~\ref{sec:fo-highsym}. 

\subsection{General determination of the mutant fixation probability to first order in $st$} 
\label{sec:numerical_pfix}

Before performing explicit analytical calculations in symmetric structures (see Section~\ref{sec:fo-highsym}), we present a general method that allows one to numerically obtain the first-order coefficient in $st$ of the fixation probability in any strongly connected graph. A strongly connected graph is a graph where there is a path between each pair of
vertices of the graph, in both directions. 

Let us first consider the first line in Eq.~\ref{eq:coeffs-abc}, which can be rewritten in vector form as:
\begin{equation}
    M  \bm{a}=\bm{a}\,,
    \label{eq:gen-dire}
\end{equation}
with $\bm{a}=(a_1,\dots,a_D)$. 
Thus, we are looking for an eigenvector of $M$ for eigenvalue $1$. 
As the graph is strongly connected, its migration matrix $M$ is irreducible. 
Since $M$ is an irreducible, non-negative matrix, and satisfies $\sum_i m_{ij}=1$ for all $j$, the Perron–Frobenius theorem guarantees that 1 is an eigenvalue of $M$, and is simple. 

To determine the norm of the corresponding eigenvector $\bm{a}$, we use the second line in Eq.~\ref{eq:coeffs-abc}. 
We isolate $b_j$, use $\sum_{k=1}^D m_{kj}=1$, and sum over all possible $j$:
\begin{align}
        b_j = m_{jj} b_j + \sum_{k\ne j} m_{jk} b_k + a_j^2 -2\delta_j a_j\,,\,\,\textrm{i.e.}\nonumber\\
        b_j - \biggl(1 - \sum_{k\ne j} m_{kj} \biggr) b_j = \sum_{k\ne j} m_{jk} b_k + a_j^2 -2\delta_j a_j\,,\,\,\textrm{thus}\nonumber\\
        \sum_j \sum_{k\ne j} m_{kj} b_j = \sum_j \sum_{k\ne j} m_{jk} b_k + \sum_j a_j^2 -\sum_j2\delta_j a_j\,,\,\,\textrm{i.e.}\nonumber\\
        0 = \sum_ja_j^2 -2\sum_j\delta_j a_j\,.
        \label{eq:relafromb}
    \end{align}
For given $\delta_j$ values, this equation provides a constraint on the $a_i$, which allows to determine the norm of $\bm{a}$. Indeed, considering an eigenvector $\bm{a}'$ of $M$ with eigenvalue 1, the vector $\bm{a}$ we are looking for can be written as $\bm{a} = C \bm{a}'$, where $C$ is a constant, which should satisfy:
\begin{equation}
   C^2 \sum_j (a'_j)^2 = 2C \sum_j a'_j \delta_j\,. 
\end{equation}
With $C\ne 0$, this finally yields $\bm{a} = C \bm{a}'$ with
\begin{equation}
    C = \frac{2 \sum_j a'_j \delta_j}{\sum_j (a'_j)^2}\,.
    \label{eq:gen-norm}
\end{equation}

\subsection{Average mutant extinction time}
\label{sec:ext_time}
The average extinction time $t_i^{(\textrm{ex})}$ of the lineage of a single mutant introduced in deme $i$ at a bottleneck can be calculated in the branching process approximation from the probability generating function \cite{harris_thetheory_1963, Abbara__Frequent}. Specifically, in a multi-type branching process, the vector of extinction times, expressed in number of generations, $\bm{t^{(\textrm{ex})}} = \left( t_1^{(\textrm{ex})}, t_2^{(\textrm{ex})}, \dots, t_D^{(\textrm{ex})}  \right)$, is given by:
\begin{equation}
    \bm{t^{(\textrm{ex})}} = \bm{f}(0) + \sum_{n=1}^{\infty} n \left[ \bm{f}^{(n)}(0)-\bm{f}^{(n-1)}(0) \right],
    \label{eq:text}
\end{equation}
with $\bm{f}$ the generating function, whose component $i$ satisfies:
\begin{equation}
    f_i : \bm{x} \mapsto \exp\left[ \lambda^{(i)} \sum_j m_{ij} (x_j -1)\right],
\end{equation}
with $\lambda^{(i)}$ given by Eq.~\ref{lambdaj}.

\section{Mutant fixation probability in analytically tractable spatial structures with heterogeneous environment}
\label{sec:fo-highsym}

\begin{figure}[htbp]
    \centering
    \includegraphics[width=0.8\linewidth]{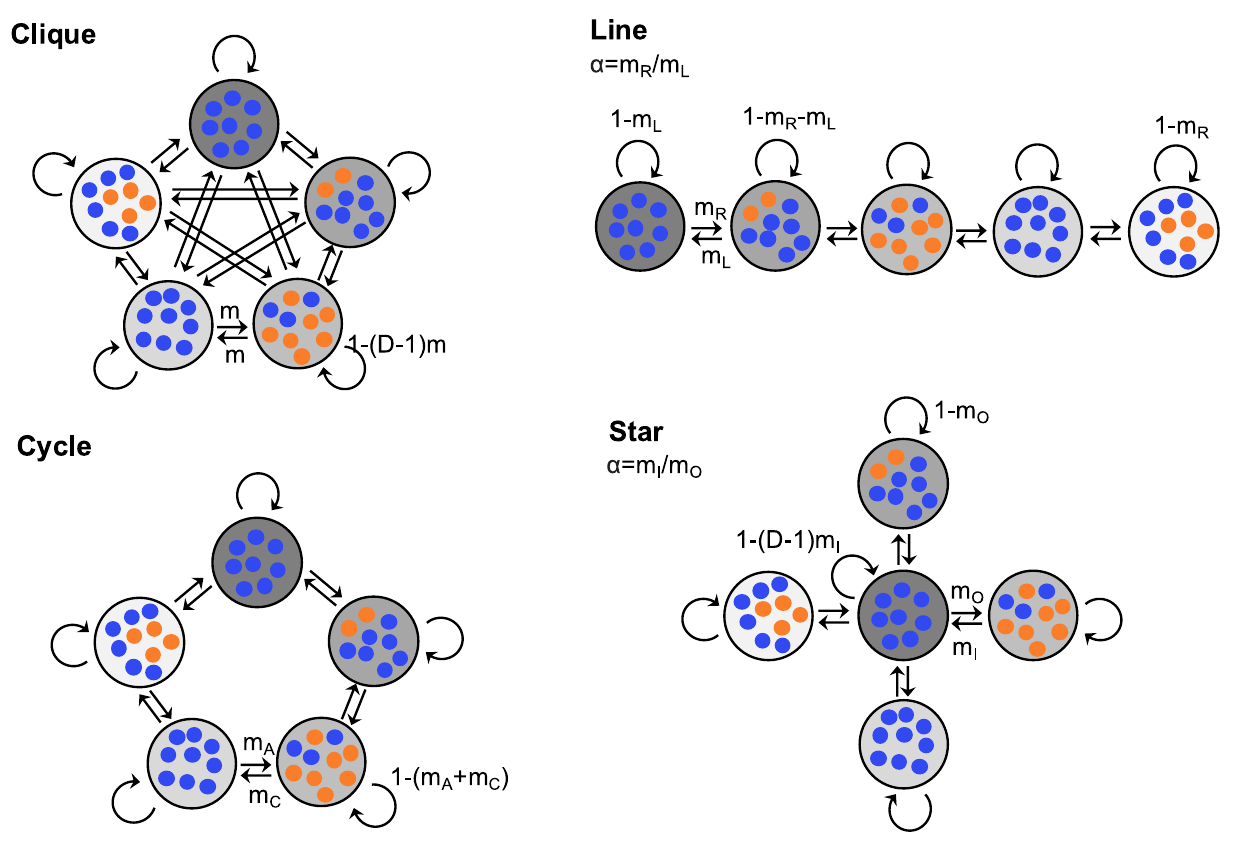}
    \caption{\textbf{Some spatial structures with environment heterogeneity considered in this study.} Schematics are shown for the clique with migration probability $m$, the line with migration probability to the right $m_R$ and to the left $m_L$, the cycle with migration probability $m_A$ in the anticlockwise direction and  $m_C$ in the clockwise direction, and the star, with incoming migration probability $m_I$ from each leaf to the center, and outgoing migration probability $m_O$ from the center to each leaf. Blue markers represent wild-type individuals and orange markers represent mutants. Gray backgrounds with different darkness represent different environments.}
    \label{fig:structures-sym}
\end{figure}

In this Section, we consider specific graphs with strong symmetries or simple structures, and we investigate the impact of environment heterogeneity across demes on the fixation probability of a mutant appearing uniformly at random in one of the demes of the population. Using the branching process approach introduced above, we provide analytical expressions of the fixation probability to first order in $st$ in the structures shown in Figure~\ref{fig:structures-sym}, and we push the expansion to second order in the clique and in a particular cycle. These two graphs are circulations, and we first analyze this category of graphs.

\subsection{Circulation graphs}
\label{sec:circulation_hetero}

Circulations are graphs such that for each node $i$, $\sum_jm_{ji} = \sum_j m_{ij}$. In other words, for each deme $i$, the total incoming migration flow to deme $i$ is equal to the total outgoing migration flow from deme~$i$. Concrete examples are the clique and the cycle in Figure~\ref{fig:structures-sym}. With homogeneous environments, these structures, which possess strongly symmetric migrations, are known to exhibit the same mutant fixation probability as a well-mixed population with the same size, whatever the initial location of the mutant. A version of this result is known in population genetics models with demes on the nodes of highly-symmetric graphs~\cite{maruyama70,maruyama74} as \textit{Maruyama's theorem}, and it was further formalized in evolutionary graph theory models with one individual per node of a graph~\cite{lieberman_evolutionary_2005}, as the \textit{circulation theorem}. In Ref.~\cite{marrec_toward_2021}, we showed that the circulation theorem extends to our model with demes in the rare migration regime, up to a finite-size effect correction. Then, in Ref.~\cite{Abbara__Frequent}, we showed that it holds exactly under the branching process approximation, in the case of frequent migrations between demes. 
Thus, circulations are a class of graphs where spatial structure has (essentially) no effect on mutant fixation probability for homogeneous environments. In this Section, we investigate whether the circulation theorem still holds with environment heterogeneity. 
As in Section~\ref{sec:diff_env_all}, we focus on the regime of frequent migrations and use the branching process approach.

\subsubsection{Circulation theorem to first order in $st$ with heterogeneous environment}
\label{subs:circ-fo}

For circulation graphs, 
using the Perron-Frobenius theorem and the first line of Eq.~\ref{eq:coeffs-abc} (which is not affected by environment heterogeneity) allows us to show that all the $a_i$ coefficients equal, using the approach presented in Ref.~\cite{Abbara__Frequent}. Let us briefly recall this proof for completeness. The the first line of Eq.~\ref{eq:coeffs-abc} shows that the vector $(a_1,\dots,a_D)$ is an eigenvector of the matrix $M$ of migration probabilities with eigenvalue 1. As the matrix $M$ is non-negative and irreducible~\cite{Abbara__Frequent}, the Perron-Frobenius theorem ensures that the largest real eigenvalue $r$ of $M$ is simple, positive, and satisfies:
\begin{equation}
     \min_i \sum_j m_{ij} \leq r \leq \max_i \sum_j m_{ij}\,.
\end{equation}
As circulations satisfy $\sum_i m_{ij} = \sum_j m_{ij}=1$, we have $r=1$. Since $(1, \dots, 1)$ is an eigenvector of $M$ associated to 1, it generates the eigenspace associated to the eigenvalue 1, and thus $(a_1,\dots,a_D)$ is proportional to $(1, \dots, 1)$, i.e.\ all the $a_i$ coefficients equal. Let us call them $a$. 

Moreover, the second line of Eq.~\ref{eq:coeffs-abc} then provides $a=2\langle \delta \rangle$ (see Eq.~\ref{eq:relafromb} in Section \ref{sec:numerical_pfix}, written in the specific case where all $a_i$ are equal). Hence, to first order in $st$, the probability of fixation of a mutant starting from any deme in a circulation is:
\begin{equation}
        \rho=2\langle \delta \rangle st\,,
        \label{eq:circ-fo}
\end{equation}
Hence, only the mean relative fitness advantage of the mutant across the structure matters for the mutant fixation probability in heterogeneous circulations. Furthermore, the result in Eq.~\ref{eq:circ-fo} is the same as in a well-mixed population with the same $\langle \delta\rangle$~\cite{haldane_amathematical_1927}.
Thus, to first order in the branching process regime, the circulation theorem extends to structures with heterogeneous fitness differences across demes.
Concretely, to first order, neither the spatial structure nor the environment heterogeneity within a circulation have any impact on fixation probability. All heterogeneous circulations have the same fixation probability as the corresponding homogeneous well-mixed population.

\begin{figure}[htb!]
    \centering
    \includegraphics[width=\linewidth]{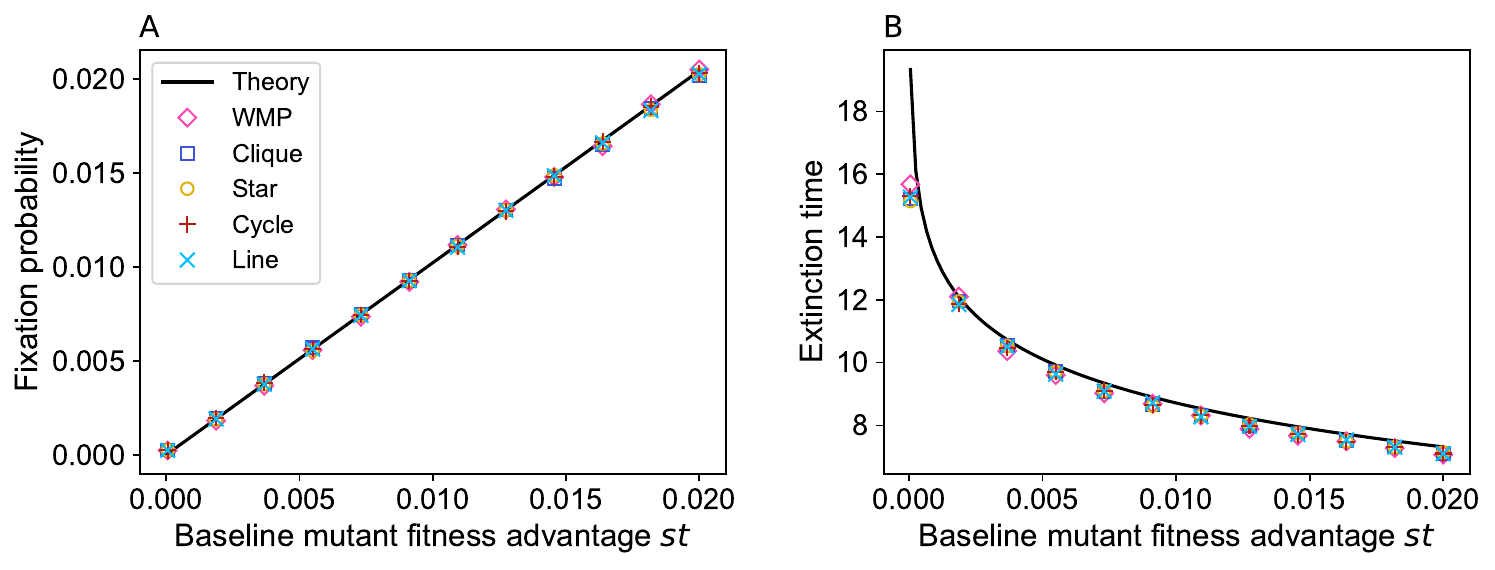}
    \caption{\textbf{Fixation probability and extinction time for structured populations on circulation graphs with environmental heterogeneity across demes.} Panel A: Fixation probability for four different graph structures: cycle, clique, line, and star (see Figure~\ref{fig:structures-sym}), under the condition $\sum_i m_{ij}= \sum_i m_{ji}$ for all nodes $j$, is shown as a function of the baseline mutant fitness advantage $st$. Panel B: Mutant extinction time, expressed in number of bottlenecks, as a function of the baseline mutant fitness advantage $st$, for the same structures. Solid lines represents results from the branching process theory (Eq.~\ref{eq:circ-fo} in panel A, Eq.~\ref{eq:text} in panel B), while markers represent results from stochastic simulations. The well-mixed population (``WMP") simulation result for the same $\langle \delta 
    \rangle$ is included in both panels for reference. Parameter values: $D=5$; $K=1000$; $\delta_i$ values: $0.49, 0.32, 0.70, 0.41, 0.62$ (independently sampled from a uniform distribution on the interval $[0,1]$), resulting in $\langle \delta 
    \rangle = 0.51$; $m=m_I=m_O=0.15$; $m_L=m_R=m_A=m_C=0.3$; each marker comes from \SI{5e5}{} simulation realizations.} 
    \label{fig:fi2-circulations}
\end{figure}

We show in Figure~\ref{fig:fi2-circulations}A our analytical prediction for the first-order fixation probabilities, given by Eq.~\ref{eq:circ-fo}, together with simulation results for four different circulations with heterogeneous environments: a clique, a cycle, a star with migration asymmetry $\alpha=m_I/m_O=1$, and a line with migration asymmetry $\alpha=m_R/m_L=1$ (see Figure~\ref{fig:structures-sym}). We also report results for the well-mixed population with mutant fitness advantage $\langle \delta \rangle s$. Our simulation results agree with our analytical prediction. We also show the mutant extinction time in Figure~\ref{fig:fi2-circulations}B, both from simulations and from a numerical calculation (see Section \ref{sec:ext_time}), for the four circulation graphs and the corresponding well-mixed population. We again observe good agreement between results from simulations and calculations. 

\subsubsection{\texorpdfstring{Specific circulation graphs: clique and cycle with heterogeneous environment}{Specific circulation graphs: clique and cycle with heterogeneous environment}}
\label{subs:pfix_clique}

In SI Sec.~\ref{subs:circ-fo}, we derived the fixation probability of a mutant in any circulation graph with heterogeneous environment to first order in $st$. The clique and the cycle graphs are circulations, for all values of $m$ in the clique and of $m_A$ and $m_C$ in the cycle (see Figure~\ref{fig:structures-sym}). Note that this stands in contrast with the star and the line, which are circulations only when $\alpha=1$ (see Figure~\ref{fig:structures-sym}). The clique and the cycle are thus fully covered by the general result in Eq.~\ref{eq:circ-fo}. Given their particularly high symmetry, more direct calculations of the first-order fixation probability can also be performed in the clique and the cycle. We provide them below for completeness.

\paragraph{Clique with heterogeneous environment.} 
In the clique, $m_{ij}=m$ for all $i\neq j$. Hence, the first line of Eq.~\ref{eq:coeffs-abc} becomes $a_j= m\sum_{k=1}^D a_k$ for all $j$. This implies that all $a_j$ coefficients are equal, say to $a$, consistently with our general result derived above for circulations. 

To obtain the value of $a$, we now use the second line of Eq.~\ref{eq:coeffs-abc}:
\begin{equation}
    b_j m (D-1) = m \sum_{k\ne j } b_k + a^2 -2a\delta_j  \hspace{1 cm} j = 1, \dots, D\,.
\label{eq:bj-deltas-clique}
\end{equation}
Summing Eq.~\ref{eq:bj-deltas-clique} over $j$ yields:
\begin{align}
    m (D-1) \sum_{j=1}^D b_j &= m \sum_{j=1}^D \sum_{k\ne j} b_k + Da^2 -2a \sum_{j=1}^D \delta_j\nonumber\\
    &= m (D-1) \sum_{j=1}^D b_j+ Da^2 -2a \sum_{j=1}^D \delta_j\,.
    \label{eq:bj-deltas-clique-2}
\end{align}
Thus, as $a\ne 0$, we obtain the first-order coefficient $a=2\langle \delta\rangle$ for the clique, with $\langle \delta\rangle= \sum_{j=1}^D \delta_j/D$. 

Therefore, the fixation probability $\rho_i$ of a single mutant starting in deme $i$ at a bottleneck in a clique reads to first order in $st$:
\begin{equation}
    \rho_i=1-p_i=a st =2\langle \delta\rangle st\,,\,\,\,\textrm{for all}\,\,i\,,
    \label{eq:a-deltaj-clique}
\end{equation}
which coincides with our general result for circulations in Eq.~\ref{eq:circ-fo}.

\paragraph{Cycle with heterogeneous environment.}
\label{subs:pfix_cycle}

Let us now consider the cycle, with migration probabilities to nearest neighbor demes that can differ in the clockwise and anticlockwise directions, and are denoted by $m_C$ and $m_A$, respectively (see Figure~\ref{fig:structures-sym}).

Here, we use the fact that all $a_i$ are equal to the same value $a$ in a circulation, see the beginning of Section~\ref{subs:circ-fo}. 
The second line of Eq.~\ref{eq:coeffs-abc} then reads:
\begin{equation}
        b_j = \sum_{k=1}^D m_{jk}b_k +a^2 -2\delta_j a = m_{jj} b_j + \sum_{k\ne j} m_{jk} b_k +a^2 - 2\delta_j a \hspace{1 cm}    j = 1, \dots, D.
        \label{eq:coeffs-abc-2}
\end{equation}
In the cycle (see Figure~\ref{fig:structures-sym}), we have $m_{ii} = 1- m_A - m_C$. Hence, Eq.~\ref{eq:coeffs-abc-2} becomes:
\begin{equation}
    \begin{cases}
        (m_A + m_C) b_1 = m_A b_2 + m_C b_D + a^2 -2\delta_1 a\,,\\
        (m_A + m_C) b_j =  m_A b_{j+1} + m_C b_{j-1} +a^2 - 2\delta_j a \hspace{0.5 cm} j = 2, \dots, D-1\,,\\
        (m_A + m_C) b_D = m_A b_{1} + m_C b_{D-1} +a^2 - 2\delta_D a\,.
    \end{cases}
    \label{eq:cycle-gradient}
\end{equation}
The sum of Eqs.~\ref{eq:cycle-gradient} reads:
\begin{equation}
    \sum_{j=1}^D (m_A + m_C) b_j = \sum_{j=1}^D (m_A + m_C) b_j + D a^2- 2 a \sum_{j=1}^D \delta_j.
\end{equation}
Thus, $a=2\langle \delta\rangle$, which entails that the mutant fixation probability $\rho_i$ to first order in $st$ is
\begin{equation}
    \rho_i=2\langle \delta\rangle st\,,\,\,\,\textrm{for all}\,\,i\,.
    \label{eq:cycle-a-deltaj}
\end{equation}
Again, this coincides with our general result for circulations in Eq.~\ref{eq:circ-fo}.

\subsubsection{\texorpdfstring{Beyond the circulation theorem: second order in $st$}{Beyond the circulation theorem: second order in st}}
\label{sec:circ-SO}

Does the circulation theorem hold for heterogeneous environments beyond the first order in $st$? To investigate this, we calculate the second-order coefficient of the fixation probability. We start from the second line of Eq.~\ref{eq:coeffs-abc}, written in matrix form as:
\begin{equation}
    \bm{b} = M \bm{b} + a^2 \mathbf{1} -2a\bm{\delta}\,.
    \label{eq:B-so}
\end{equation}
Here, we introduced the vector of second-order coefficients $\bm{b} = (b_1, \dots, b_D)$, as well as the vectors $\bm{\delta} = (\delta_1, \dots, \delta_D)$, and $\mathbf{1}=(1,\dots,1)$. Meanwhile, $M$ is the migration matrix with coefficients $m_{ij}$, and $a=2\langle \delta\rangle$ is the first-order coefficient (see Section \ref{subs:circ-fo}). Recall that the first line of Eq.~\ref{eq:coeffs-abc} implies that $\mathbf{1}$ is an eigenvector of $M$ with eigenvalue 1.

To determine $\bm{b}$, we generalize the approach developed for homogeneous environments in Ref.~\cite{Abbara__Frequent}, and use a Jordan decomposition of the migration matrix $M$. It allows us to complete $\mathbf{1}$ into a complex basis such that the complementary space $\mathcal{E}_{\perp}$ of the space $\mathcal{E}$ generated by $\mathbf{1}$ is stable under the action of $M$. We decompose $\bm{b}$ as follows: $\bm{b} = \bm{b}_{\perp} + \bm{b}_{\parallel}$, with $\bm{b}_{\parallel}=b_{\parallel}\mathbf{1}\in \mathcal{E}$ and $\bm{b}_{\perp}\in \mathcal{E}_{\perp}$. Similarly, we write $\bm{\delta} = \bm{\delta}_{\perp} + \bm{\delta}_{\parallel}$, with $\bm{\delta}_{\parallel}=\delta_{\parallel}\mathbf{1}$. We can then rewrite Eq.~\ref{eq:B-so} as:
\begin{equation}
    \bm{b}_{\parallel} + \bm{b}_{\perp} = M \bm{b}_{\parallel} + M \bm{b}_{\perp} + a^2 \mathbf{1} -2a\bm{\delta}_{\parallel} - 2a\bm{\delta}_{\perp}.
\end{equation}
Thus, in $\mathcal{E}$, since $M \bm{b}_{\parallel}=\bm{b}_{\parallel}$:
\begin{align}
     \bm{b}_{\parallel} &=  \bm{b}_{\parallel}+ a^2 \mathbf{1}  -2a\bm{\delta}_{\parallel}\,,\,\,\,\textrm{i.e.}\nonumber\\
     a^2 &= 2a \delta_{\parallel},
     \label{eq:parallel-space}
\end{align}
and in $\mathcal{E}_{\perp}$:
\begin{equation}
    \bm{b}_{\perp} =  M \bm{b}_{\perp} - 2a\bm{\delta}_{\perp}.
    \label{eq:perp-space}
\end{equation}
Since $a=2\langle \delta \rangle$ (see Section \ref{subs:circ-fo}), Eq.~\ref{eq:parallel-space} ensures that $\delta_{\parallel}=\langle\delta\rangle$. 
Thus, we have:
\begin{equation}
\bm{\delta}_{\perp} =\bm{\delta}- \bm{\delta}_{\parallel}= (\delta_1 -\langle \delta \rangle, \delta_2 - \langle \delta \rangle, \dots, \delta_D - \langle \delta \rangle)\,.\label{eq:deltaperp}
\end{equation}

As shown in Section \ref{subs:circ-fo}, the largest eigenvalue of $M$ is 1 and is associated to the eigenspace $\mathcal{E}_{\parallel}$, and all other eigenvalues have a strictly smaller modulus.
Thus, $(\mathbb{I} - M)$ restricted to $\mathcal{E}_{\perp}$, where $\mathbb{I}$ denotes the identity matrix, has no zero eigenvalue, and is invertible. 
Hence, Eq.~\ref{eq:perp-space} yields:
\begin{equation}
\bm{b}_{\perp} =  -4\langle\delta\rangle(\mathbb{I} - M )^{-1}\bm{\delta}_{\perp}\,,
    \label{eq:bperp}
\end{equation}
with $\bm{\delta}_{\perp}$ given in Eq.~\ref{eq:deltaperp}.

Next, to find $b_{\parallel}$, we use the third line of Eq.~\ref{eq:coeffs-abc}. Using $\sum_{k}m_{kj}=1$, and $a_j=a$ for all $j$, the third line of Eq.~\ref{eq:coeffs-abc} becomes, for all $j\in[1,D]$,
\begin{equation}
 \sum_{k=1}^D m_{kj} c_j =  \sum_{k=1}^D m_{jk} c_k-2 a^3 + 3ab_j   - 3\delta_j b_j +3\delta_j a^2 -3 \delta_j^2a\,.
\end{equation}
Summing this equation over all $j\in[1,D]$ yields:
\begin{equation}
    \sum_{j=1}^D\sum_{k=1}^D m_{kj} c_j =  \sum_{j=1}^D\sum_{k=1}^D m_{jk} c_k-2 D a^3 + 3a \sum_{j=1}^D b_j   - 3 \sum_{j=1}^D \delta_j b_j +3a^2 \sum_{j=1}^D \delta_j  -3a \sum_{j=1}^D \delta_j^2\,,
\end{equation}
which reduces to
\begin{equation}
    3\sum_{j=1}^Db_j (a-\delta_j)  = 2Da^3 -3a^2 D \langle \delta \rangle +3a \sum_{j=1}^D \delta_j^2\,.
    \label{eq:third-order}
\end{equation}
Writing $b_j=b_\parallel+b'_j$ for all $j$, where $b'_j$ is the $j$-th component of $\bm{b}_{\perp}$, and recalling that for circulations $a= 2\langle \delta \rangle$ (see Section \ref{subs:circ-fo}), we obtain:
\begin{align}
    b_\parallel &=\frac{4}{3} \langle \delta \rangle^2+2 \langle \delta^2 \rangle-\frac{1}{D}\sum_j b'_j\left(2-\frac{\delta_j}{\langle\delta\rangle}\right)= \frac{4}{3}\langle\delta\rangle^2+2\langle\delta^2\rangle-2\langle b'\rangle+\frac{\langle b'\delta\rangle}{\langle\delta\rangle}\,,
    \label{eq:bparallel}
\end{align}
where $b'_j$ can be obtained from Eq.~\ref{eq:bperp}.

\paragraph{Clique with heterogeneous environment.}
\label{subs:so-clique}
Recall that in the clique (see Figure~\ref{fig:methods}B), $m_{ij}=m$ for all $i\neq j$. 
For a heterogeneous clique, $\bm{b}_\perp$ with components 
\begin{equation}
b_i' = 4 \langle \delta \rangle \frac{\langle \delta \rangle -\delta_i }{Dm}
\label{eq:bperp-cl}
\end{equation}
satisfies Eq.~\ref{eq:bperp}.

Let us now calculate $b_\parallel$ for a heterogeneous clique, using Eq.~\ref{eq:bparallel}. Eq.~\ref{eq:bperp-cl} entails that $\langle b'\rangle=0$, and allows us to write:
\begin{equation}
    \langle b'\delta\rangle=\frac{1}{D}\sum_j  b'_j \delta_j = \frac{4 \langle \delta \rangle}{D^2 m} \sum_j \left(\langle \delta \rangle - \delta_j \right)\delta_j = \frac{4 \langle \delta \rangle}{Dm} \left( \langle \delta \rangle^2 - \langle \delta^2\rangle    \right),
\end{equation}
so Eq.~\ref{eq:bparallel} yields:
\begin{equation}
    b_\parallel=\frac{4}{3}\langle\delta\rangle^2+2\langle\delta^2\rangle+\frac{4}{Dm} \left( \langle \delta \rangle^2 - \langle \delta^2\rangle    \right)\,.
    \label{eq:bpara-cl}
\end{equation}

Eq.~\ref{eq:bperp-cl} and Eq.~\ref{eq:bpara-cl} show that, to second order in $st$, the mutant fixation probability in a heterogeneous clique generically depends on the deme $i$ where the mutant starts, and on migration intensity $m$. Therefore, the circulation theorem does not hold to second order in $st$ for heterogeneous structures.

In the particular case of the homogeneous clique with $\delta_i=\Delta$ for all $i$, Eq.~\ref{eq:bperp-cl} reduces to $b'_i=0$ for all $i$, and Eq.~\ref{eq:bpara-cl} reduces to $b_\parallel=10\Delta^2/3$, yielding a mutant fixation probability $\rho_i=2\Delta st -5(\Delta s t)^2/3 + o(( st)^2)$ for all $i$. This is in agreement with the result of Ref.~\cite{Abbara__Frequent} for homogeneous circulations (see Eq.~S16 in the Supplementary Material of that paper, written for a mutant with effective fitness advantage $\Delta st$). Recall that the circulation theorem holds for homogeneous environments in the branching process approximation~\cite{Abbara__Frequent}.

Note that the mutant fixation probability averaged over starting demes reads:
\begin{equation}
    \rho = 2\langle\delta\rangle st - \frac{b_\parallel+\langle b'\rangle}{2} (st)^2 + o(( st)^2) = 2\langle\delta\rangle st - \frac{b_\parallel}{2}  (st)^2+ o(( st)^2)\,,
    \label{eq:pfix-2}
\end{equation}
where $b_\parallel$ is given by Eq.~\ref{eq:bpara-cl}. Thus, to second order, this fixation probability does not involve $\bm{b}_\perp$, and only depends on environment heterogeneity through $\langle\delta\rangle$ and $\langle\delta^2\rangle$.

Let us compare the mutant fixation probability, averaged over starting demes, in a heterogeneous clique and in a homogeneous clique with the same $\langle\delta\rangle$. Calling $\rho$ and $\rho^\textrm{homo}$ these two fixation probabilities, Eq.~\ref{eq:pfix-2} yields:
\begin{align}
    \rho-\rho^\textrm{homo}&=\frac{(st)^2}{2}\left(b_\parallel^\textrm{homo}-b_\parallel\right) = \frac{(st)^2}{2}\left[\frac{10}{3}\langle\delta\rangle^2-\frac{4}{3}\langle\delta\rangle^2-2\langle\delta^2\rangle-\frac{4}{Dm} \left( \langle \delta \rangle^2 - \langle \delta^2\rangle    \right)\right]\nonumber\\
    &=\frac{2-Dm}{Dm} (st)^2 \left( \langle \delta^2\rangle-\langle \delta \rangle^2    \right)\,.
    \label{eq:so-cmp-cl}
\end{align}
In the clique, $\sum_i m_{ij}=1$ entails that $m_{ii}=1-(D-1)m$, and as all migration probabilities are between 0 and 1, we have $(D-1)m\leq 1$ and, and thus $Dm\leq 1+m\leq 2$. This entails that $\rho\geq\rho^\textrm{homo}$.

\begin{figure}[htb!]
    \centering
    \includegraphics[width=\linewidth]{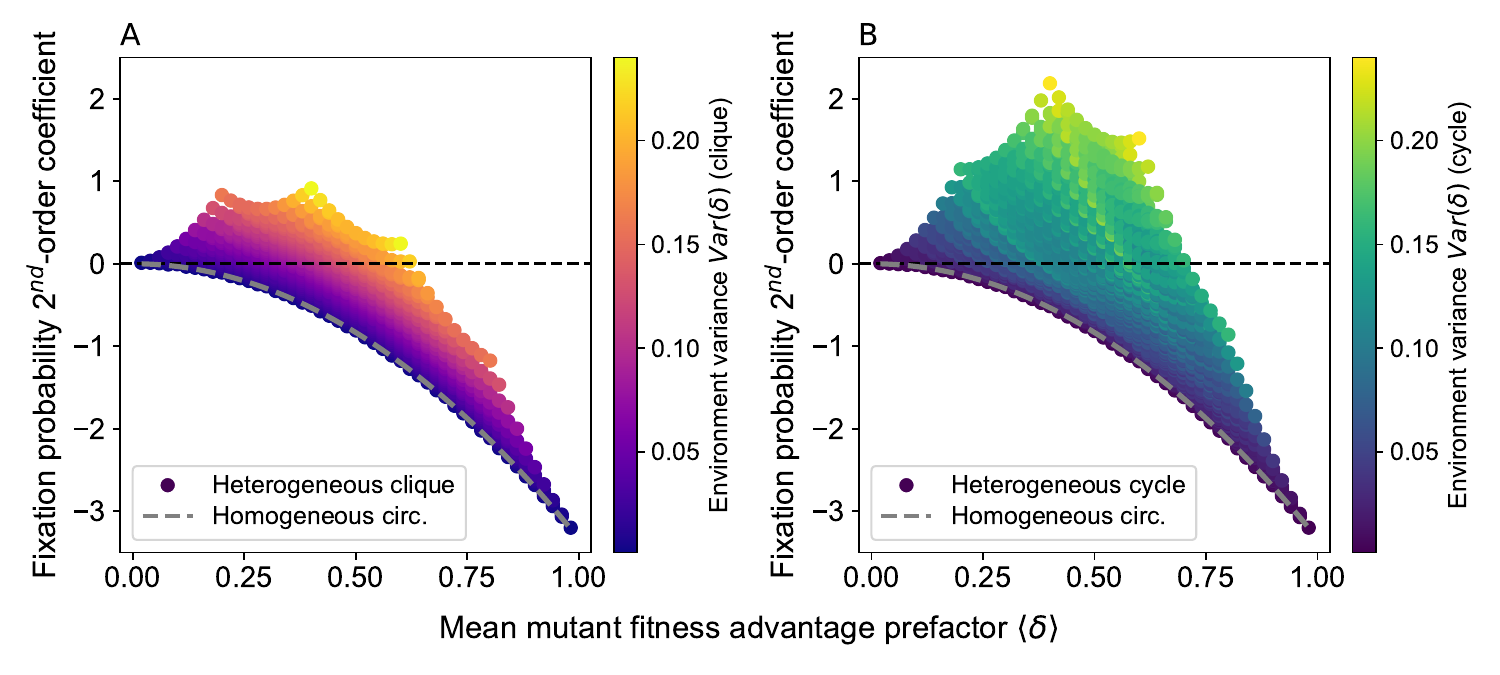}
    \caption{\textbf{Impact of environment heterogeneity on fixation probability in circulations.} As the fixation probability is the same in all circulations to first order in $st$, we focus on the impact of the second-order term. Specifically, we show the second-order coefficient $-b_\parallel$ in the expansion of fixation probability  (Eqs.~\ref{eq:pfix-2} and~\ref{eq:pfix-2b}; see also Eq.~\ref{expand} in the main text). Panel A: This second-order coefficient (expressed using Eq.~\ref{eq:bpara-cl}) is shown versus the mean mutant fitness advantage prefactor $\langle \delta \rangle$ across the structure, for various heterogeneous cliques. Markers are colored according to the variance of the deme-specific mutant fitness advantage prefactor $\delta_i$ coefficient across demes $i$, a proxy for environment heterogeneity. Horizontal dashed black line: no deviation from the first-order approximation. Dashed gray line: homogeneous environment case. Panel B: Same as A, but for heterogeneous cycles with equal clockwise and anticlockwise migration probabilities (using Eq.~\ref{eq:SO-cycle}). Parameters: $D=5$, $m=0.1$ in the clique, $m=0.2$ in the cycle (same total of exchanges in the two structures). Numbers in $\{0,0.1,\dots,0.9,1\}$ are employed for each $\delta_i$, and all combinations of them are considered.}
    \label{fig:fig3-SO}
\end{figure}

Thus, in the clique, the fixation probability averaged over all initial mutant locations is increased by environment heterogeneity, compared to that of a homogeneous clique with the same $\langle\delta\rangle$. In other words, environment heterogeneity generically amplifies natural selection to second order in $st$. Furthermore, this effect increases with the environment variance $\textrm{Var}(\delta)$ of the $\delta_i$ coefficients across demes is. It also increases when the migration probability $m$ decreases. 
Figure~\ref{fig:fig3-SO}A shows the second-order coefficient in $st$ of the fixation probability. Besides being increased by environment heterogeneity, we observe that it can be positive, in which case the second order term increases the fixation probability, while for homogeneous environments, it always decreases it.

\newpage

\paragraph{A particular cycle with heterogeneous environment.}
\label{subs:so-cycle}

Recall that in the cycle, migration probabilities are nonzero only to nearest neighbor demes (and to the original deme itself). Migrations to nearest neighbor demes can differ in the clockwise and anticlockwise directions, and are denoted by $m_C$ and $m_A$, respectively (see Figure~\ref{fig:methods}B). Here, we present explicit calculations in the specific case of a heterogeneous cycle with $D=5$ and $m_A = m_C=m$. Solving Eq.~\ref{eq:bperp} in this case leads to:
\begin{equation}
    \bm{b}_{\perp} =\begin{pmatrix}
         b'_1\\
        b'_2\\
         b'_3\\
        b'_4\\
        b'_5
    \end{pmatrix} = \frac{4\langle\delta\rangle}{5m}\begin{pmatrix}
        \delta_3 +\delta_4 - 2\delta_1 \\
        \delta_4 +\delta_5 -2\delta_2\\
        \delta_5+\delta_1 -2\delta_3\\ 
        \delta_1+\delta_2 -2\delta_4\\ 
        \delta_2 +\delta_3 -2\delta_5
    \end{pmatrix}.
    \label{eq:bperp-cy}
\end{equation}

Let us now calculate $b_\parallel$ in this case, using Eq.~\ref{eq:bparallel}. Eq.~\ref{eq:bperp-cy} entails that $\langle b'\rangle=0$, and allows us to write:
\begin{equation}
    \langle b'\delta\rangle=\frac{1}{5}\sum_{j=1}^5 b'_j \delta_j=-\frac{8\langle\delta\rangle}{25m} \left(5\langle \delta^2 \rangle-\sum_{j=1}^5 \delta_j \delta_{(j+2)\textrm{mod}\,5}\right),
    \label{eq:bpara-cy}
\end{equation}
so Eq.~\ref{eq:bparallel} yields:
\begin{equation}
    b_{\parallel}=\frac{4}{3}\langle\delta\rangle^2+2\langle\delta^2\rangle-\frac{8}{25m} \left(5\langle \delta^2 \rangle-\sum_{j=1}^5 \delta_j \delta_{(j+2)\textrm{mod}\,5}\right).
     \label{eq:SO-cycle}
\end{equation}

Note that in a homogeneous cycle with $D=5$ and $m_A=m_C=m$, where $\delta_i=\Delta$ for all $i$, Eq.~\ref{eq:bperp-cy} reduces to $\bm{b}_{\perp}=\bm{0}$, and Eq.~\ref{eq:SO-cycle} reduces to $b_\parallel=10\Delta^2/3$, which is the same as above for the homogeneous clique. This is consistent with the fact that the circulation theorem holds for homogeneous environments in the branching process approximation~\cite{Abbara__Frequent}.

The mutant fixation probability averaged over starting demes reads:
\begin{equation}
    \rho = 2\langle\delta\rangle st - \frac{b_\parallel+\langle b'\rangle}{2} (st)^2 + o(( st)^2) = 2\langle\delta\rangle st - \frac{b_\parallel}{2}  (st)^2+ o(( st)^2)\,,
    \label{eq:pfix-2b}
\end{equation}
where $b_\parallel$ is given by Eq.~\ref{eq:bpara-cy}. As for the clique, to second order, this fixation probability does not involve $\bm{b}_\perp$.

Let us compare the mutant fixation probability, averaged over starting demes, in a heterogeneous cycle and in a homogeneous cycle with the same $\langle\delta\rangle$, still in the particular case where $D=5$ and $m_A=m_C=m$. Calling $\rho$ and $\rho^\textrm{homo}$ these two fixation probabilities, Eq.~\ref{eq:pfix-2b} yields:
\begin{align}
    \rho-\rho^\textrm{homo}&=\frac{(st)^2}{2}\left(b_\parallel^\textrm{homo}-b_\parallel\right) = \frac{(st)^2}{2}\left[\frac{10}{3}\langle\delta\rangle^2-\frac{4}{3}\langle\delta\rangle^2-2\langle\delta^2\rangle+\frac{8}{25m} \left(5\langle \delta^2 \rangle-\sum_{j=1}^5 \delta_j \delta_{(j+2)\textrm{mod}\,5}\right)\right]\nonumber\\
    &=(st)^2\left[\langle\delta\rangle^2-\langle\delta^2\rangle+\frac{4}{25m} \left(5\langle \delta^2 \rangle-\sum_{j=1}^5 \delta_j \delta_{(j+2)\textrm{mod}\,5}\right)\right].
    \label{eq:cycle-so-cmp}
\end{align}

Let us now investigate the sign of $\rho-\rho^\textrm{homo}$. Noticing that
\begin{equation}
\left(\sum_{i=1}^5\delta_i\right)^2=\sum_{i=1}^5\delta_i^2+2\sum_{i=1}^5 \delta_i \delta_{(i+1)\textrm{mod}\,5}+2\sum_{i=1}^5 \delta_i \delta_{(i+2)\textrm{mod}\,5}\,,
\end{equation}
we can write, in terms of rough orders of magnitude, 
\begin{equation}
    \sum_{i=1}^5 \delta_i \delta_{(i+2)\textrm{mod}\,5}\simeq\sum_{i=1}^5 \delta_i \delta_{(i+1)\textrm{mod}\,5}\simeq\frac{1}{4}\left[\left(\sum_{i=1}^5\delta_i\right)^2-\sum_{i=1}^5\delta_i^2\right]=\frac{25}{4}\langle\delta\rangle^2-\frac{5}{4}\langle\delta^2\rangle\,.
    \label{eq:roughappx}
\end{equation}
Eq.~\ref{eq:cycle-so-cmp} then yields:
\begin{equation}
\rho-\rho^\textrm{homo}\simeq(st)^2\left[\langle\delta\rangle^2-\langle\delta^2\rangle+\frac{4}{25m} \left(5\langle \delta^2 \rangle-\frac{25}{4}\langle\delta\rangle^2+\frac{5}{4}\langle\delta^2\rangle\right)\right]\simeq \frac{1-m}{m}(st)^2\left( \langle \delta^2\rangle-\langle \delta \rangle^2    \right).
\label{eq:cycle-so-cmp-2}
\end{equation}
In our particular cycle with $m_A = m_C=m$, the condition $\sum_i m_{ij}=1$ for all $j$ gives $2m+m_{ii}=1$, i.e.\ $m_{ii}=1-2m$. Since all migration probabilities are between 0 and 1, we have $m\leq 1/2$. 
Hence, under the rough assumption of Eq.~\ref{eq:roughappx}, Eq.~\ref{eq:cycle-so-cmp-2} indicates that $\rho \gtrsim\rho^\textrm{homo}$ and that the smaller $m$ is, and the larger the variance of the $\delta_i$ is, the larger $\rho-\rho^\textrm{homo}$ is expected to become. These approximate results are reminiscent of the exact results obtained for the clique in Eq.~\ref{eq:so-cmp-cl}. Accordingly, we observe in Figure~\ref{fig:fig3-SO}B a similar behavior of the second-order coefficient in $st$ of the fixation probability in this particular cycle as in the clique (considered in Figure~\ref{fig:fig3-SO}A).

\begin{figure}[htb!]
    \centering
    \includegraphics[width=\linewidth]{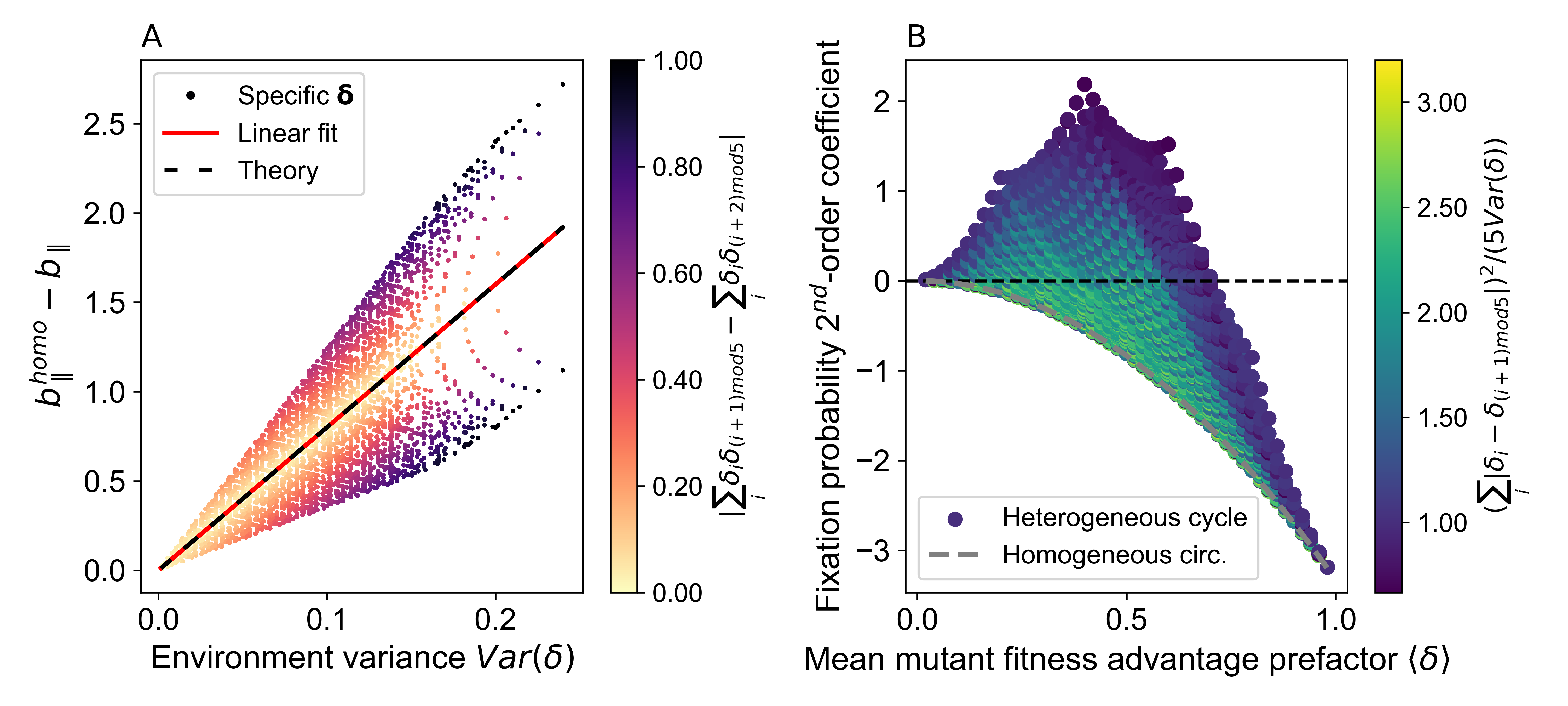}
    \caption{\textbf{Impact of environment heterogeneity on the second-order coefficient of the fixation probability in the cycle.} Panel A: Difference $b_\parallel^\textrm{homo}-b_\parallel$ between the second-order coefficients for a cycle with heterogeneous and homogeneous environment (see Eq.~\ref{eq:cycle-so-cmp}), plotted against the environment variance $\textrm{Var}(\delta)$. Markers are colored according to $\left|\sum_{i=1}^5 \delta_i \delta_{(i+2)\textrm{mod}\,5}-\sum_{i=1}^5 \delta_i \delta_{(i+1)\textrm{mod}\,5}\right|$. Solid red line: linear fit of the data. Dashed black line: approximate theoretical prediction from Eq.~\ref{eq:cycle-so-cmp-2}. Panel B: The second-order coefficient $-b_\parallel$ in the expansion of fixation probability (Eq.~\ref{eq:pfix-2b}; see also Eq.~\ref{expand} in the main text) is shown versus the mean mutant fitness advantage prefactor $\langle \delta \rangle$ across the structure, for various heterogeneous cycles, as in Figure~\ref{fig:fig3-SO}B, but colored according to the ratio of local to global environmental variation, defined in Eq.~\ref{eq:ratio-hetero-cycle}. Dashed gray line: homogeneous environment case (Eq.~\ref{eq:bpara-cl}). Parameter values for both panels: $D=5$, $m_A=m_C=m=0.2$ for the cycle. Numbers in $\{0,0.1,\dots,0.9,1\}$ are employed for each $\delta_i$, and all combinations of them are considered. All cycles considered here have equal clockwise and anticlockwise migration probabilities.}
    \label{fig:SO-cycle-SI}
\end{figure}

Figure~\ref{fig:SO-cycle-SI}A shows the exact value of the deviation with respect to the homogeneous case of $b_\parallel$, from Eq.~\ref{eq:cycle-so-cmp}, for multiple heterogeneous cycles, differing by their environmental heterogeneity, via $\bm{\delta}$. We observe that the approximate second-order coefficient, from Eq.~\ref{eq:cycle-so-cmp-2}, yields the correct trend of its dependence in $\mathrm{Var}(\delta)=\langle\delta^2\rangle-\langle\delta\rangle^2$ (in agreement with a linear fit), but that there is substantial spread around that trend. We also observe that the discrepancy between the exact and the approximate formulas is small (resp.\ large) when $\left|\sum_{i=1}^5 \delta_i \delta_{(i+2)\textrm{mod}\,5}-\sum_{i=1}^5 \delta_i \delta_{(i+1)\textrm{mod}\,5}\right|$ is small (resp.\ large), as expected given the approximation made.

Since for the cycle the variance does not fully capture the impact of environment heterogeneity, we further explore the role of the spatial placement of heterogeneity in the cycle. A metric to quantify local variation between neighboring demes compared to the global variation is:
\begin{equation}
   V = \frac{\left(\sum_{i=1}^5 |\delta_i - \delta_{(i+1) \textrm{mod 5}} |\right)^2}{5\textrm{Var}(\delta)},
    \label{eq:ratio-hetero-cycle}
\end{equation}
where $\delta_i$ denotes the deme-specific mutant fitness advantage prefactor, and $\textrm{Var}(\delta)$ is the variance of these prefactors across demes. The numerator of $V$ quantifies local environment variation, and the denominator overall environment heterogeneity. At a fixed value of the variance, a gradual local variation results in a small numerator, resulting in a small value of $V$. Meanwhile, repeated abrupt alternations in $\delta_i$ for nearby demes result in a large numerator and hence a large value of $V$. In Figure~\ref{fig:SO-cycle-SI}B, we color the excess fixation probability in the cycle due to the second order term (i.e.\ $-b_\parallel$) based on the value of $V$ in Eq.~\ref{eq:ratio-hetero-cycle}. We observe that when similar values of $\delta_i$ are clustered together (i.e.\ for small $V$), the excess fixation probability tends to be higher.

\subsection{Star with heterogeneous environment} 

Let us now consider the star, with incoming migration probabilities from a leaf to the center that can differ from outgoing ones from the center to a leaf, and are denoted by $m_I$ and $m_O$, respectively (see Figure~\ref{fig:methods}B).

As noted earlier, the first line in Eq.~\ref{eq:coeffs-abc} is not impacted by environment heterogeneity. Thus, as in Ref.~\cite{Abbara__Frequent}, the first-order coefficients for the fixation probability for a mutant starting in the center and in a leaf, denoted respectively by $a_C$ and $a_L$, satisfy $a_L = \alpha a_C$, where $\alpha=m_I/m_O$ represents migration asymmetry. Note that environment heterogeneity across leaves does not make these first-order coefficients dependent on the leaf where the mutant start.

Let $\mathcal{L}$ be the set of the indexes labeling the leaves of the star. The second line in Eq.~\ref{eq:coeffs-abc}, written for the center (with second-order coefficient $b_C$) and for one of the leaves $j\in \mathcal{L}$ (with second-order coefficient $b_j$), yields:
\begin{equation}
    \begin{cases}
       (D-1) m_I b_C = m_O \sum_{k\in \mathcal{L}} b_k + a_C^2 - 2 \delta_C a_C\,, \\
        m_O b_j =m_I b_C + a_L^2 -2\delta_j a_L\,, \hspace{0.5 cm } j \in \mathcal{L}.
    \end{cases}
    \label{eq:star-deltaj-b}
\end{equation}
Summing Eqs.~\ref{eq:star-deltaj-b} over all demes gives:
\begin{equation}
    (D-1) m_I b_C + m_O \sum_{j\in \mathcal{L}} b_j =  (D-1) m_I b_C + m_O \sum_{k\in \mathcal{L}} b_k  + a_C^2 + (D-1) a_L^2 -2 \delta_C a_C -2\sum_{j \in \mathcal{L}} \delta_j a_L\,,
    \label{eq:star-deltaj-b-2}
\end{equation}
i.e., after simplification, and using $a_L = \alpha a_C$:
\begin{equation}
    a_C^2[ 1 + \alpha^2 (D-1) ] = 2a_C \delta_C +2\alpha a_C \sum_{j \in \mathcal{L}} \delta_j.
\end{equation}
Hence, as $a_C \ne0$, we obtain:
\begin{equation}
a_C = \frac{2 (\delta_C + \alpha \sum_{i\in \mathcal{L}} \delta_i)}{1+\alpha^2 (D-1)}\,\,\,\,\textrm{and}\,\,\,\,a_L = \alpha  a_C\,.
\label{eq:a-star-deltajs}
\end{equation}

Let us now consider a mutant appearing at a bottleneck in a deme chosen uniformly at random. Its fixation probability $\rho$ reads, to first order in $st$:
\begin{equation}
\begin{split}
    \rho &= st\,\frac{a_C + (D-1) a_L }{D} = 2st\,\frac{[1 + \alpha (D-1)](\delta_C +\alpha \sum_{i\in \mathcal{L}} \delta_i) }{D[1+\alpha^2 (D-1)] }\\&=2\langle \delta\rangle st\,\frac{[1 + \alpha (D-1) ][\sigma_C(1-\alpha) +1+\alpha (D-1)] }{D[1+\alpha^2 (D-1)] }\,,
    \label{eq:fo-star}
    \end{split}
\end{equation}
where we introduced $\langle\delta\rangle=\sum_{i=1}^D \delta_i/D$, and $\sigma_C = (\delta_C - \langle \delta \rangle)/\langle \delta \rangle$. 
Importantly, the result in Eq.~\ref{eq:fo-star} only depends on selection through the mean mutant fitness effect, via $\langle\delta\rangle$, and through the relative fitness excess in the center, via $\sigma_C$. In other words, the heterogeneity of selection across leaves does not impact this fixation probability to first order. We show the first-order coefficient of the fixation probability as a function of $\alpha$ and $\sigma_C$ in Figure~\ref{fig:fig4-star}A.

Note that for a homogeneous star, where $\sigma_C=0$ and $\delta$ takes a unique value, say $\Delta$, across the whole population, Eq.~\ref{eq:fo-star} reduces to:
\begin{equation}
    \rho = 2\Delta st\,\frac{[1+\alpha (D-1)]^2 }{D[1+\alpha^2 (D-1)] }\,,
    \label{eq:fo-star2}
\end{equation}
i.e.\ the result from Ref.~\cite{Abbara__Frequent} for the homogeneous star. Note also that for $\alpha=1$, when the star is a circulation, Eq.~\ref{eq:fo-star} yields the same result as for other circulations considered so far, namely $\rho=2\langle \delta\rangle st$ (see Eqs.~\ref{eq:a-deltaj-clique} and \ref{eq:cycle-a-deltaj}).

\subsection{Line with heterogeneous environment}

Let us now consider the line, with migration probabilities to the left that can differ from those to the right, and are denoted by $m_L$ and $m_R$, respectively (see Figure~\ref{fig:methods}B). Contrary to other structures discussed so far, the line was not considered in Ref.~\cite{Abbara__Frequent} in the homogeneous case. Hence, the results below are new even in the homogeneous case.

The first line in Eq.~\ref{eq:coeffs-abc}, written for each node of the line, yields:
\begin{equation}
    \begin{cases}
        a_1 =  (1-m_L) a_1 + m_R a_2\,,\\
        a_j = (1- m_L - m_R) a_j +m_R a_{j+1} + m_L a_{j-1} \hspace{0.5 cm} j=2, \dots, D-1\,,\\
        a_D = (1-m_R) a_D + m_L a_{D-1}\,.
    \end{cases}
    \label{eqs:a-line}
\end{equation}
Introducing migration asymmetry $\alpha = m_R/m_L$, we get to:
\begin{equation}
        \begin{cases}
        a_1=  \alpha a_2\,,\\
        a_j (1+\alpha) = \alpha a_{j+1} + a_{j-1} \hspace{0.5 cm} j=2, \dots, D-1\,,\\
        a_D = a_{D-1}/\alpha\,.
    \end{cases}
    \label{eq:aj+1-D}
\end{equation}
It can be shown by induction that
\begin{equation}
    a_j = \frac{B}{\alpha^{j-1}}\,,
    \label{eq:gensol-D}
\end{equation}
where $B$ is a constant, is a solution of Eq.~\ref{eq:aj+1-D}.

Let us now use the equations corresponding to the second line of Eq.~\ref{eq:coeffs-abc}, written for each node of the line, to determine the values of the constant $B$:
\begin{equation}
\begin{cases}
    b_1 m_L =  m_R b_2 + a_1^2 - 2\delta_1 a_1\,,\\
    b_j (m_L + m_R) =  m_R b_{j+1} + m_L b_{j-1} + a_j^2 - 2\delta_j a_j\hspace{0.5 cm} j = 2, \dots, D-1\,,\\
    b_D m_R =  m_L b_{D-1} + a_D^2 - 2\delta_D a_D\,.
\end{cases}
\label{eq:b-line-deltajD}
\end{equation}
Summing Eqs.~\ref{eq:b-line-deltajD} over all nodes of the line yields:
\begin{equation}
    m_L b_1 + \sum_{j=2}^{D-1} b_j (m_L + m_R) + b_D m_R = m_R b_2 +m_L b_{D-1} + \sum_{j=2}^{D-1} b_{j+1}m_R + \sum_{j=2}^{D-1} b_{j-1}m_L + \sum_{j=1}^{D} a_j^2 -2 \sum_{j=1}^D \delta_j a_j\,, 
    \label{eq:mlr_sum}
\end{equation}
which simplifies into
\begin{equation}
    \sum_{j=1}^{D} a_j^2 = 2 \sum_{j=1}^D \delta_j a_j\,. 
    \label{eq:mlr_sum2}
\end{equation}
Using Eqs.~\ref{eq:gensol-D} and \ref{eq:mlr_sum2} then gives:
\begin{equation}
    B = 2 \frac{\sum_{j=1}^D \delta_j \alpha^{-j+1}}{\sum_{j=1}^D \alpha^{-2j+2}} =2 \frac{\alpha^2 - 1}{\alpha^2-\alpha^{2(1-D)}} \sum_{j=1}^D \delta_j \alpha^{-j+1},
    \label{eq:B-dcase}
\end{equation}
The first-order coefficient $a_k$ of the fixation probability for a mutant starting in deme $k$, thus reads:
\begin{equation}
    a_k = \frac{B}{\alpha^{k-1}} = 2\alpha^{-k} \frac{\alpha^2 -1}{ 1 - \alpha^{-2D} } \sum_{j=1}^D \delta_j \alpha^{-j}.
\end{equation}

Let us now consider a mutant appearing at a bottleneck in a deme chosen uniformly at random. Its fixation probability $\rho$ reads, to first order in $st$:
\begin{equation}
        \rho = st\frac{1}{D} \sum_{k=1}^D a_k = \frac{2st}{D} \frac{1+\alpha}{1+\alpha^{-D}}\sum_{j=1}^D \delta_j \alpha^{-j}.
        \label{eq:aline}
\end{equation}

In particular, for a homogeneous line where all $\delta_i$ are equal, say to $\Delta$, this reduces to:
\begin{equation}
        \rho = \frac{2\Delta st}{D} \frac{1+\alpha}{1+\alpha^{-D}}\sum_{j=1}^D \alpha^{-j}=\frac{2\Delta st}{D} \frac{(1+\alpha)(1-\alpha^{D})}{(1-\alpha)(1+\alpha^{D})}\,.
        \label{eq:aline-homo}
\end{equation}

Note that if $\alpha=1$, when the line is a circulation, Eq.~\ref{eq:gensol-D} reduces to $a_j = C$, with $C$ a constant. Then, Eq.~\ref{eq:mlr_sum2} yields $C=2\langle \delta\rangle$, and hence $\rho=2\langle \delta\rangle st$, which is the same result as for the other circulation graphs we considered so far.

\section{Comparison of fixation probabilities between structures} 

\subsection{Comparison of fixation probabilities in the star and in the clique} 
\label{sec:SIstar}
Let us focus on the heterogeneous star structure, and on its impact on mutant fixation probability. First, for reference, in Section \ref{subs:homostar-cl}, we briefly compare a homogeneous star to a clique. Next, in Section \ref{subs:homo-hetero-star}, we determine under what conditions heterogeneity can increase the fixation probability in the star, compared to a homogeneous star. Finally, in Section \ref{sub:sup-star}, we compare a heterogeneous star to a clique, to determine if a heterogeneous star can suppress or amplify selection. 

In all cases, when we compare a structure with heterogeneous environment to another one with homogeneous environment, we match their average mutant fitness advantage, i.e.\ in our notations, we match their $\langle\delta\rangle$. Our comparisons all rely on the first-order fixation probabilities derived in Section \ref{sec:diff_env_all} in the branching process approximation.

\subsubsection{Homogeneous star versus clique}
\label{subs:homostar-cl}
We first consider a homogeneous star with relative mutant fitness advantage $s\Delta$ in all demes. We compare it to a heterogeneous clique with relative mutant fitness advantage $s\delta_i$ in deme $i$, satisfying $\langle \delta \rangle=\Delta$. Using the fixation probabilities in Eqs.~\ref{eq:fo-star2} and \ref{eq:a-deltaj-clique}, we find that the homogeneous star has a first-order fixation probability lower than the heterogeneous clique if and only if:
\begin{equation}
    \frac{1}{D}\frac{[1+(D-1)\alpha]^2}{[1+\alpha^2(D-1)]}<1\,.
    \label{eq:cond-star-homo}
\end{equation}
Note that the heterogeneity of the clique does not impact its fixation probability to first order (see Eq.~\ref{eq:a-deltaj-clique}), and thus, this condition is the same as that found in Ref.~\cite{Abbara__Frequent}, when comparing the homogeneous star and the homogeneous clique. As shown in Ref.~\cite{Abbara__Frequent}, this condition is always satisfied, except for $\alpha=1$, when the two fixation probabilities are equal. Thus, the homogeneous star suppresses natural selection under frequent asymmetric migrations.

\subsubsection{Homogeneous star versus heterogeneous star}
\label{subs:homo-hetero-star}

Let us compare the fixation probabilities in a heterogeneous star and in a homogeneous star, given respectively in Eqs.~\ref{eq:fo-star} and~\ref{eq:fo-star2} to first order in $st$. Our homogeneous star has relative mutant fitness advantage $s\Delta$ in all demes, while our heterogeneous star has relative mutant fitness advantage $s\delta_i$ in deme $i$, satisfying $\langle \delta \rangle=\Delta$. The first-order fixation probability is larger in the heterogeneous star than in the homogeneous star if and only if:
\begin{equation}
\begin{split}
\frac{[1 + \alpha (D-1) ][\sigma_C(1-\alpha) +1+\alpha (D-1)] }{D[1+\alpha^2 (D-1)] } >\frac{[1+\alpha (D-1)]^2 }{D[1+\alpha^2 (D-1)] }\,,
    \end{split}
\end{equation}
As the denominator and the term $1+\alpha(D-1)$ are positive, this is equivalent to
\begin{equation}
    \sigma_C(1-\alpha)  >0\,.
\end{equation}
Thus, the relative fitness excess in the center should have the same sign as $1-\alpha$ for heterogeneity to enhance mutant fixation probability. This means that if $\alpha<1$ (i.e.\ $m_I<m_O$, more outflow from the center than inflow), heterogeneity benefits mutant fixation if the center features a larger mutant fitness advantage than the leaves. Conversely, if $\alpha>1$ (i.e.\ $m_I>m_O$, more inflow to the center than outflow), heterogeneity benefits mutant fixation if the center features a smaller mutant fitness advantage than the leaves.

\subsubsection{Heterogeneous star versus clique}
\label{sub:sup-star}

We have shown that a homogeneous star has a smaller mutant fixation probability than a clique, but that a heterogeneous star can have a larger mutant fixation probability than a homogeneous one. What is the overall effect of a heterogeneous star structure compared to a clique? In particular, can environment heterogeneity allow the star to overcome the suppression of selection effect it has under homogeneous environments? To address this question, we compare a heterogeneous star to a clique. Recall that only average fitness matters to first order for the mutant fixation probability in the clique. Hence, we consider a clique which can be homogeneous or heterogeneous, but which has the same average fitness as our heterogeneous star. 

The fixation probabilities in a heterogeneous star and in a clique are given respectively in Eqs.~\ref{eq:fo-star} and~\ref{eq:a-deltaj-clique} to first order in $st$. The first-order fixation probability is larger in the heterogeneous star than in a clique if and only if:
\begin{equation}
\begin{split}
\frac{[1 + \alpha (D-1) ][\sigma_C(1-\alpha) +1+\alpha (D-1)] }{D[1+\alpha^2 (D-1)] } >1\,.
    \end{split}
\end{equation}
This condition is equivalent to:
\begin{align}
    \sigma_C(1-\alpha)[1+\alpha(D-1)]&>D[1+\alpha^2(D-1)]-[1+\alpha(D-1)]^2\,,\,\,\,\mathrm{i.e.}\nonumber\\
    \sigma_C(1-\alpha)[1+\alpha(D-1)]&>(D-1)(1-\alpha)^2\,.
    \label{eq:CS-pfixa}
\end{align}
The condition in Eq.~\ref{eq:CS-pfixa} can be expressed as: 
\begin{itemize}
    \item $\sigma_C > f_{\textrm{star}}(\alpha,D)$ if $\alpha <1$\,,
    \item $\sigma_C < f_{\textrm{star}}(\alpha,D)$ if $\alpha >1$\,,
\end{itemize}
where we introduced
\begin{equation}
    f_{\textrm{star}}(\alpha,D) = \frac{(D-1)(1-\alpha)}{1+\alpha(D-1)}\,.
    \label{eq:fstar}
\end{equation}

\begin{figure}[htb!]
    \centering
    \includegraphics[width=\linewidth]{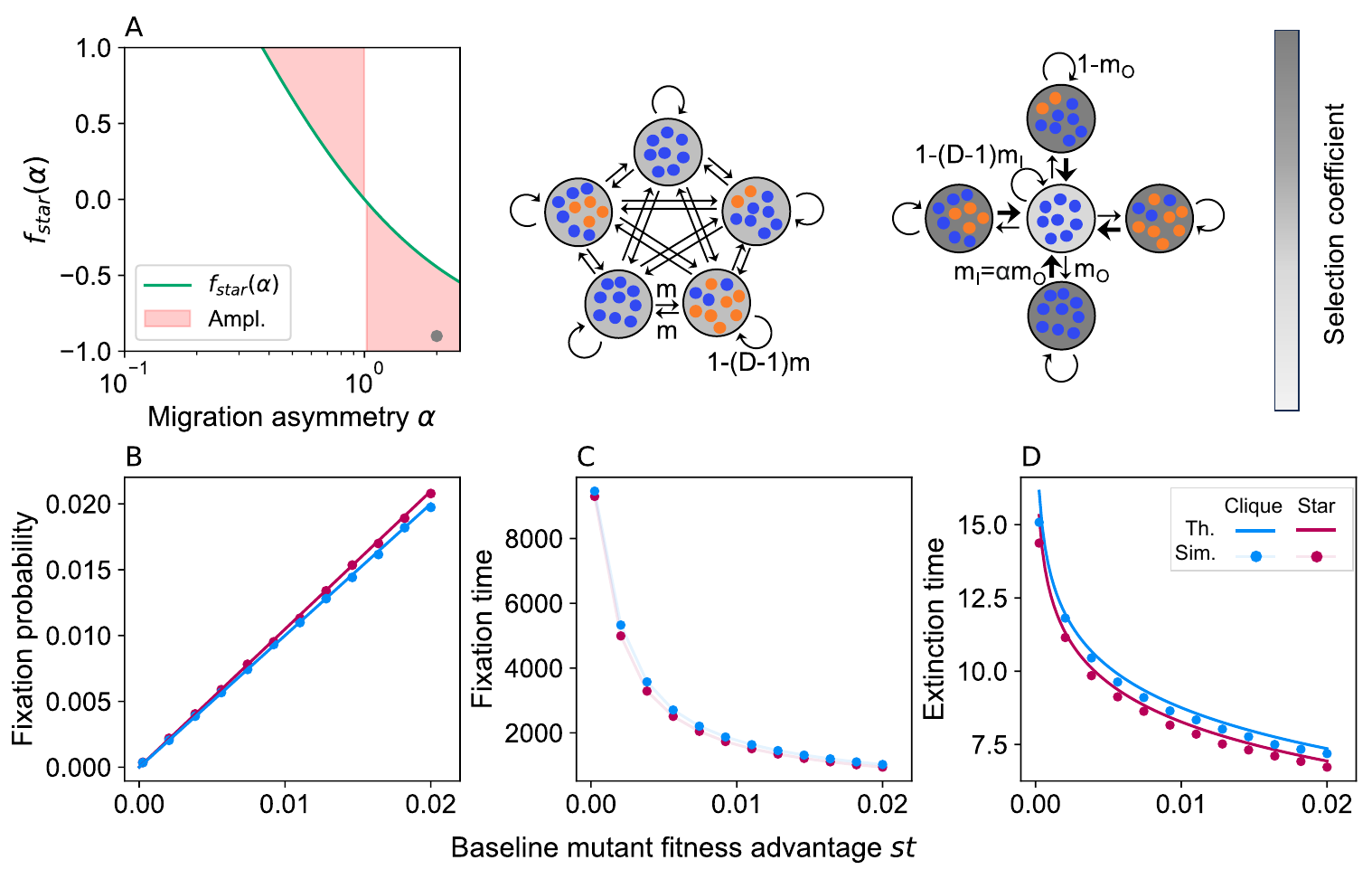}
    \caption{\textbf{Amplification of natural selection by the heterogeneous star with frequent migrations.} Panel A: $f_\textrm{star}(\alpha, D=5)$ defined in Eq.~\ref{eq:fstar} is plotted as a function of $\alpha$.  The pink region highlights the parameter range where Eq.~\ref{eq:CS-pfixa} predicts that a heterogeneous star amplifies selection relative to a clique. The gray marker indicates the parameter set used in the other panels. Panel B: Fixation probability for a heterogeneous star and a clique with the same $\langle \delta\rangle$ as a function of the baseline mutant fitness advantage $st$. Panels C-D: Mutant fixation and extinction time, respectively, expressed in numbers of bottlenecks, as a function of the baseline mutant fitness $st$. Results from the branching process theory (``Th.'') are shown in panels B (Eq.~\ref{eq:fo-star}) and D (Eq.~\ref{eq:text}). Stochastic simulation results (``Sim.'') are shown in panels B-D. Parameters: $D=5$, $K=1000$; for the star: $\alpha = 2$, $m_O=0.1$; $\delta_C=0.05$ in the center; $\delta_L=0.6125$ in all leaves, yielding $\langle \delta\rangle=0.5$ and $\sigma_C = -0.9$; for the clique: $\delta = 0.5$ in all demes and $m=0.15$. Each marker comes from $5\times 10^5$ stochastic simulation realizations.}
    \label{fig:faD}
\end{figure}

We show $ f_{\textrm{star}}(\alpha,D)$ versus $\alpha$ in Figure~\ref{fig:faD}A, and highlight in pink the regions of the plane $(\alpha, \sigma_C)$ where amplification of selection exists (see also Figure~\ref{fig:fig4-star}). We further present an example of amplification of selection in Figure~\ref{fig:faD}B. The associated fixation and extinction times are shown in Figure~\ref{fig:faD}C-D. As observed with the line in the main text (see Figure~\ref{fig:line-proba-time}), we simultaneously obtain amplification of selection and acceleration of fixation and extinction, albeit to a smaller extent.

\subsection{Comparison of fixation probabilities in the line and in the clique} 
In this section, we focus on the line structure, and on its impact on mutant fixation probability. First, in Section \ref{sec:SI-homline-homclique}, we compare a homogeneous line to a clique, to assess the baseline impact of the line spatial structure. Then, in Section \ref{subs:homo-hetero-line} we compare a homogeneous line to a heterogeneous line, to assess the impact of environment heterogeneity in a line. Finally, in Section \ref{sec:heteroC-L}, we compare a heterogeneous line to a clique, to determine if a heterogeneous line can suppress or amplify selection.

As done above for the star, in all cases, when we compare a structure with heterogeneous environment to another one with homogeneous environment, we match their average mutant fitness advantage, i.e.\ in our notations, we match their $\langle\delta\rangle$. Besides, our comparisons again rely on the first-order fixation probabilities derived in Section \ref{sec:fo-highsym} in the branching process approximation.

\subsubsection{Homogeneous line versus clique}
\label{sec:SI-homline-homclique}

\begin{figure}[htb!]
    \centering
    \includegraphics[width=0.55\textwidth]{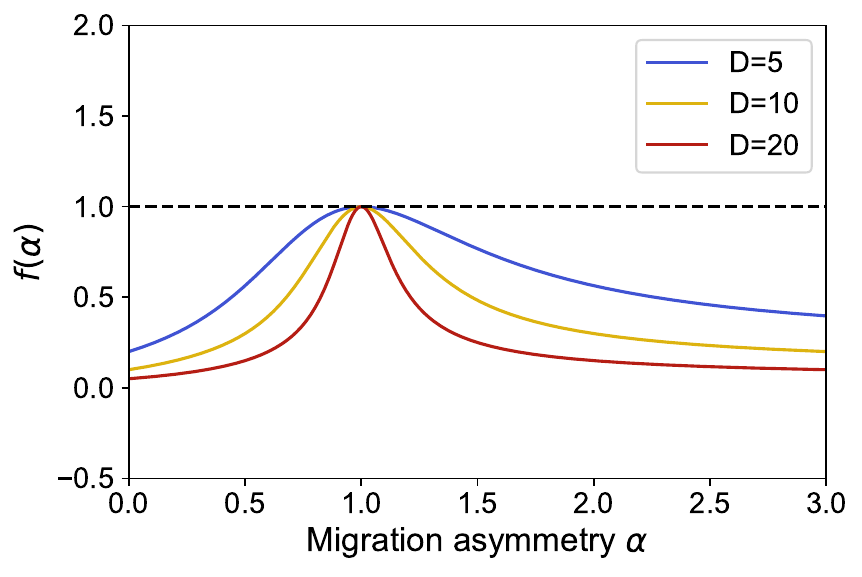}
    \caption{\textbf{Suppression of selection in the homogeneous line with migration asymmetry.} The function $f(\alpha)=(1+\alpha)(1-\alpha^{D})/[D(1-\alpha)(1+\alpha^{D})]$ involved in Eq.~\ref{eq:cond-aD0} is plotted versus the migration asymmetry $\alpha$ in the line for $D=5,10,20$.}
    \label{fig:cond-aD}
\end{figure}

Let us consider a homogeneous line with relative mutant fitness advantage $s\Delta$ in all demes. Let us compare it to a heterogeneous clique with relative mutant fitness advantage $s\delta_i$ in deme $i$, satisfying $\langle \delta \rangle=\Delta$. We derived their fixation probabilities in Eqs.~\ref{eq:aline-homo} and \ref{eq:a-deltaj-clique}, respectively. Recall that the result for the clique was unaffected by environment heterogeneity -- thus, our comparison will hold for homogeneous and heterogeneous cliques. A homogeneous line has a first-order mutant fixation probability lower than a heterogeneous clique if and only if:
\begin{equation}
\frac{1}{D} \frac{(1+\alpha)(1-\alpha^{D})}{(1-\alpha)(1+\alpha^{D})} < 1\,. 
\label{eq:cond-aD0}
\end{equation}
This condition is always satisfied, except if $\alpha=1$, when the two fixation probabilities are equal. It is illustrated in Figure~\ref{fig:cond-aD}. Hence, the homogeneous line suppresses selection, as long as it features migration asymmetry. This is consistent with our general result on homogeneous spatially structured populations from Ref.~\cite{Abbara__Frequent}.

\subsubsection{Homogeneous line versus heterogeneous line}
\label{subs:homo-hetero-line}

Let us compare the fixation probabilities in a heterogeneous line and in a homogeneous line, given respectively in Eqs.~\ref{eq:aline} and~\ref{eq:aline-homo} to first order in $st$. Our homogeneous line has relative mutant fitness advantage $s\Delta$ in all demes, while our heterogeneous line has relative mutant fitness advantage $s\delta_i$ in deme $i$, satisfying $\langle \delta \rangle=\Delta$. The first-order fixation probability is larger in the heterogeneous star than in the homogeneous star if and only if:
\begin{align}
        \frac{1}{1+\alpha^{-D}}\sum_{j=1}^D \delta_j \alpha^{-j}&>\Delta  \frac{(1-\alpha^{D})}{(1-\alpha)(1+\alpha^{D})}\,,\,\,\,\mathrm{i.e.}\nonumber\\
        \sum_{j=1}^D \frac{\delta_j}{\Delta} \alpha^{-j}&>\frac{1-\alpha^{-D}}{\alpha-1}\,.
        \label{eq:aline-cmp}
\end{align}
Let us now introduce the relative mutant fitness excess $\sigma_j$ in deme $j$ such that for all $j$: $\delta_j = \langle \delta \rangle (1 +\sigma_j) =\Delta (1 +\sigma_j)$. Note that the $\sigma_j$ satisfy $\sum_{j=1}^D \sigma_j=0$. 
Eq.~\ref{eq:aline-cmp} then becomes:
\begin{align}
        \sum_{j=1}^D \alpha^{-j}+ \sum_{j=1}^D \sigma_j \alpha^{-j} &>\frac{1-\alpha^{-D}}{\alpha-1}\,,\,\,\,\mathrm{i.e.}\nonumber\\
        S(\alpha)\equiv\sum_{j=1}^D \sigma_j \alpha^{-j}&>0\,.
        \label{eq:cond-sigmasj}
\end{align}

\paragraph{Monotonic mutant fitness advantage variation along the line.} 
Let us consider the case where the mutant fitness advantage varies monotonically along the line. Without loss of generality, let us assume that it monotonically decreases left to right, from deme 1 to deme $D$. Since the $\sigma_j$ satisfy $\sum_{j=1}^D \sigma_j=0$, there exists $k\in[1,D]$ such that 
\begin{equation}
\sigma_1 >\dots \sigma_k \geq0 >\sigma_{k+1} > \dots > \sigma_D.
\label{eq:grd}
\end{equation}

If $\alpha>1$, then for all $j>k\geq 1$, we have $\alpha^{-j}<\alpha^{-k}$, while for all $j$ satisfying $1\leq j\leq k$, we have $\alpha^{-j}\geq\alpha^{-k}$. Combining this with Eq.~\ref{eq:grd} yields:
\begin{equation}
S(\alpha) = \sum_{j=1}^k \sigma_j\alpha^{-j} + \sum_{j=k+1}^D \sigma_j \alpha^{-j}> \alpha^{-k}\sum_{j=1}^k \sigma_j + \alpha^{-k}\sum_{j=k+1}^D \sigma_j = \alpha^{-k}\sum_{j=1}^D \sigma_j=0\,,
\end{equation}
which gives $S(\alpha)>0$. Hence, Eq.~\ref{eq:cond-sigmasj} holds for all $\alpha>1$ in our line with monotonically decreasing fitness advantage from left to right.

If $\alpha<1$, a similar reasoning leads to $S(\alpha)<0$. Hence, Eq.~\ref{eq:cond-sigmasj} never holds for $\alpha<1$ in our line with monotonically decreasing fitness advantage from left to right.

Thus, the heterogeneous line with monotonically decreasing mutant fitness advantage from left to right features a larger mutant fixation probability than the homogeneous line for $\alpha>1$, but a smaller one for $\alpha<1$, and the same one for $\alpha=1$. When $\alpha>1$, most migrations are from left to right, i.e.\ demes where mutants are most advantaged are upstream of the overall migration flow. This favors mutant spread and fixation.

\subsubsection{Heterogeneous line versus clique}
\label{sec:heteroC-L}

Let us now compare a heterogeneous line to a clique.  Recall that only average fitness matters to first order for the mutant fixation probability in the clique. Hence, we consider a clique which can be homogeneous or heterogeneous, but which has the same average fitness as our heterogeneous line. The fixation probabilities in a heterogeneous line and in a clique are given respectively in Eqs.~\ref{eq:aline} and~\ref{eq:a-deltaj-clique} to first order in $st$. The first-order fixation probability is larger in the heterogeneous line than in a clique if and only if:
\begin{equation}
        \frac{1}{D} \frac{1+\alpha}{1+\alpha^{-D}}\sum_{j=1}^D \delta_j \alpha^{-j}>\langle\delta\rangle \,.
        \label{eq:aline-cmpcliq}
\end{equation}
Let us now introduce, as above, $\sigma_j$ such that $\delta_j = \langle \delta \rangle (1 + \sigma_j)$. Recall that $\sum_{j=1}^D \sigma_j =0$. Eq.~\ref{eq:aline-cmpcliq} becomes:
\begin{align}
    \frac{1}{D} \frac{1+\alpha}{1+\alpha^{-D}}\left(\sum_{j=1}^D\alpha^{-j}+S(\alpha)\right)&>1\,,\,\,\,\mathrm{i.e.}\nonumber\\
    S(\alpha)&> D\frac{1+\alpha^{-D}}{1+\alpha}- \frac{1-\alpha^{-D}}{\alpha-1}\,,\,\,\,\mathrm{i.e.}\nonumber\\
    S(\alpha)&>\frac{(D-1)\alpha -(D+1) + (D+1)\alpha^{1-D} -(D-1)\alpha^{-D}}{\alpha^2-1}\equiv g(\alpha)\,.
    \label{eq:galpha}
\end{align}

If a heterogeneous line satisfies Eq.~\ref{eq:galpha}, it features a higher mutant fixation probability than a homogeneous clique, to first order in $st$. This means that environment heterogeneity then allows the line to overcome the suppression of selection that the homogeneous line features when compared to the clique.

\paragraph{No amplification is possible if the mutant fitness advantage monotonically decreases left to right and $\alpha<1$. } We showed in the previous section that if the mutant fitness advantage monotonically decreases left to right and $\alpha<1$, the mutant fixation probability in the heterogeneous line is smaller than in the homogeneous line. Because the homogeneous line itself suppresses selection, we conclude that no amplification is possible if the mutant fitness advantage monotonically decreases left to right and $\alpha<1$. 

Another way to see this, which connects more directly to Figure~\ref{fig:line-proba-time}A, is that if $\alpha<1$, $S(\alpha)<0$ (see previous section), while we then have $g(\alpha)>0$ (see below), which entails $S(\alpha)<g(\alpha)$. To show that $g(\alpha)>0$ if $\alpha<1$, we notice that the denominator of $g(\alpha)$ is then negative, and thus we aim to show that the numerator is negative too. Denoting the numerator as
\begin{equation}
    g_N(\alpha) = (D-1) \alpha -(D+1) + (D+1)\alpha^{1-D} - (D-1) \alpha^{-D},
\end{equation}
its derivative with respect to $\alpha$ reads:
\begin{equation}
    g_N'(\alpha)= (D-1) \left[ 1+\alpha^{-D}( -D-1 + D\alpha^{-1} ) \right],
\end{equation}
and its second derivative with respect to $\alpha$ is:
\begin{equation}
    g_N''(\alpha) = D (D-1)(D+1) \left( \alpha-1 \right)\alpha^{-D-2}.
\end{equation}
When $\alpha<1$, $g_N''(\alpha)<0$. Thus, $g_N'$ is a decreasing function on $(0,1)$. As $\lim_{\alpha \to 1^-}g'_N(\alpha)=0^+$, we have $g_N'(\alpha)>0$ on $(0,1)$, which entails that $g_N$ is an increasing function on $(0,1)$. As $g_N(1)=0$, we conclude that $g_N(\alpha)<0$ on $(0,1)$. Hence, $g(\alpha)>0$ if $\alpha<1$ as announced, and there cannot be amplification of selection for $\alpha<1$.

\paragraph{Step environmental profile with one special upstream deme.} Let us consider the case where the mutant is advantaged only in the leftmost deme ($\delta_1=1$, $\delta_i=0$ for $i>1$). We plot $S(\alpha)$ together with $g(\alpha)$ in such a case in Figure~\ref{fig:line-proba-time}A in the main text. This Figure~suggests that Eq.~\ref{eq:galpha} is valid for all $\alpha>1$, and that the difference between $S(\alpha)$ and $g(\alpha)$ becomes smaller and smaller when $\alpha$ increases. To better understand this, let us investigate asymptotics for $\alpha\to\infty$. In the simple case considered here, we have:
\begin{equation}
    S(\alpha) = (D-1)\alpha^{-1}-\sum_{i=2}^D\alpha^{-i}\sim (D-1) \alpha^{-1}-\alpha^{-2},
\end{equation}
where we have retained the two leading terms in powers of $\alpha$ when $\alpha\to\infty$. Doing the same for $g(\alpha)$, Eq.~\ref{eq:galpha} gives:
\begin{equation}
    g(\alpha) \sim (D-1) \alpha^{-1}-(D+1)\alpha^{-2}.
\end{equation}
Thus, in this case, $S(\alpha)>g(\alpha)$ in terms of asymptotic behavior for $\alpha\to \infty$.  Note however that for other choices of $\sigma_i$ that monotonically decrease from left to right, the curves representing $S(\alpha)$ and $g(\alpha)$ often cross again for some $\alpha>1$, meaning that amplification then exists in a finite range of $\alpha$.

\paragraph{Examples of other environmental profiles.}

\begin{figure}[htb]
    \centering
    \includegraphics[width=\linewidth]{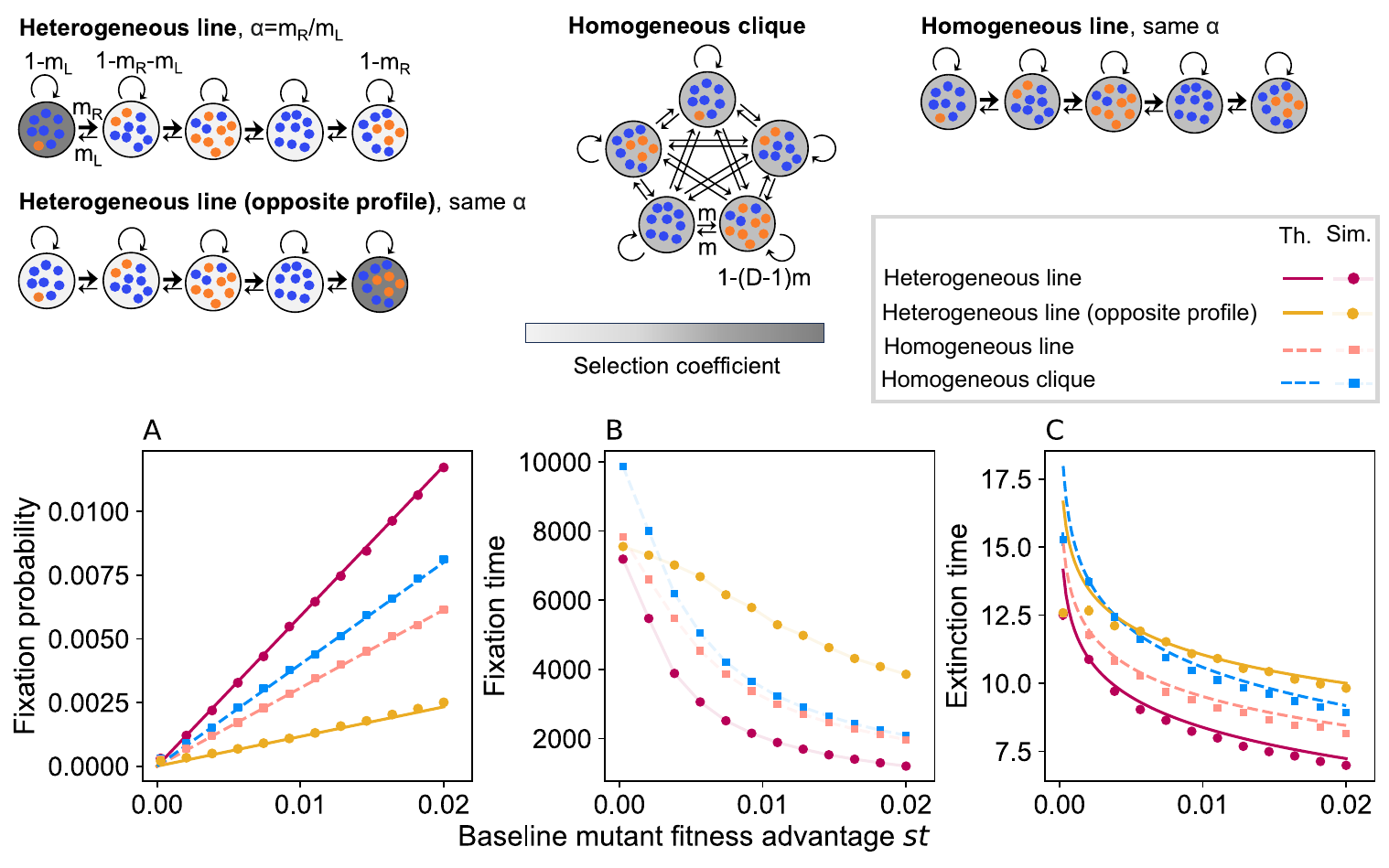}
    \caption{\textbf{Impact of the direction of the environmental profile in the line.} Top: Schematic of all spatial structures considered in this Figure: heterogeneous lines with overall migration flow and environmental profile that are either similarly oriented or opposite (left); homogeneous clique (middle), and homogeneous line with the same migration probabilities (right). All spatial structures have the same average mutant fitness advantage (same $\langle\delta\rangle$). Panel A: Mutant fixation probability in the different spatial structures considered, versus baseline mutant fitness advantage $st$. Panel B: Mutant fixation time, in number of bottlenecks, in the different spatial structures considered, versus baseline mutant fitness advantage $st$. Panel C: Same as in B, but for mutant extinction time. Results from the branching process theory (``Th.'') are shown in panels A (Eq.~\ref{eq:aline}) and C (Eq.~\ref{eq:text}). Stochastic simulation results (``Sim.'') are shown in panels A-C. Parameter values for all structures: $D=5$, $K=1000$; for all lines: $\alpha = 1.5$, $m_L = 0.3$; for the clique: $m=0.15$; for heterogeneous lines: $\delta=1$ in the special deme, $\delta=0$ in other ones. Each marker comes from $5\times 10^5$ stochastic simulation realizations.}
    \label{fig:grads-1}
\end{figure}

Let us first continue our analysis of the simple step environmental gradient. 
In Figure~\ref{fig:line-proba-time}B in the main text, we considered the case where mutants had an advantage only in the deme upstream of the main migration flow (special deme on the left and $\alpha>1$). Figure~\ref{fig:line-proba-time}A also shows that Eq.~\ref{eq:galpha} does not hold with the same environment profile when $\alpha<1$. In Figure~\ref{fig:grads-1}A, we further show the case where the mutant is advantaged in the rightmost deme with $\alpha>1$ (by symmetry, this is equivalent to having a special leftmost deme with asymmetry $1/\alpha$, giving $\alpha<1$), and as predicted above, we obtain suppression of selection. This suppression is even stronger than in a homogeneous line.
These results show that in the line, when the environmental profile has the same orientation as the overall migration flow, selection can be amplified. Meanwhile, when the environmental profile is homogeneous or opposes the overall migration flow, selection is suppressed.

Figure~\ref{fig:grads-1}B-C (see also Figure~\ref{fig:line-proba-time}C-D) also shows that while the line with similarly-oriented migration flow and environmental profile can accelerate fixation or extinction compared to the homogeneous clique and to the homogeneous line with the same migration asymmetry. Conversely, Figure~\ref{fig:grads-1}B-C shows that the line with opposite migration flow and environmental profile tends to feature slower fixation or extinction than the homogeneous clique and to the homogeneous line with the same migration asymmetry.

\newpage

\begin{figure}[htb]
    \centering
    \includegraphics[width=\linewidth]{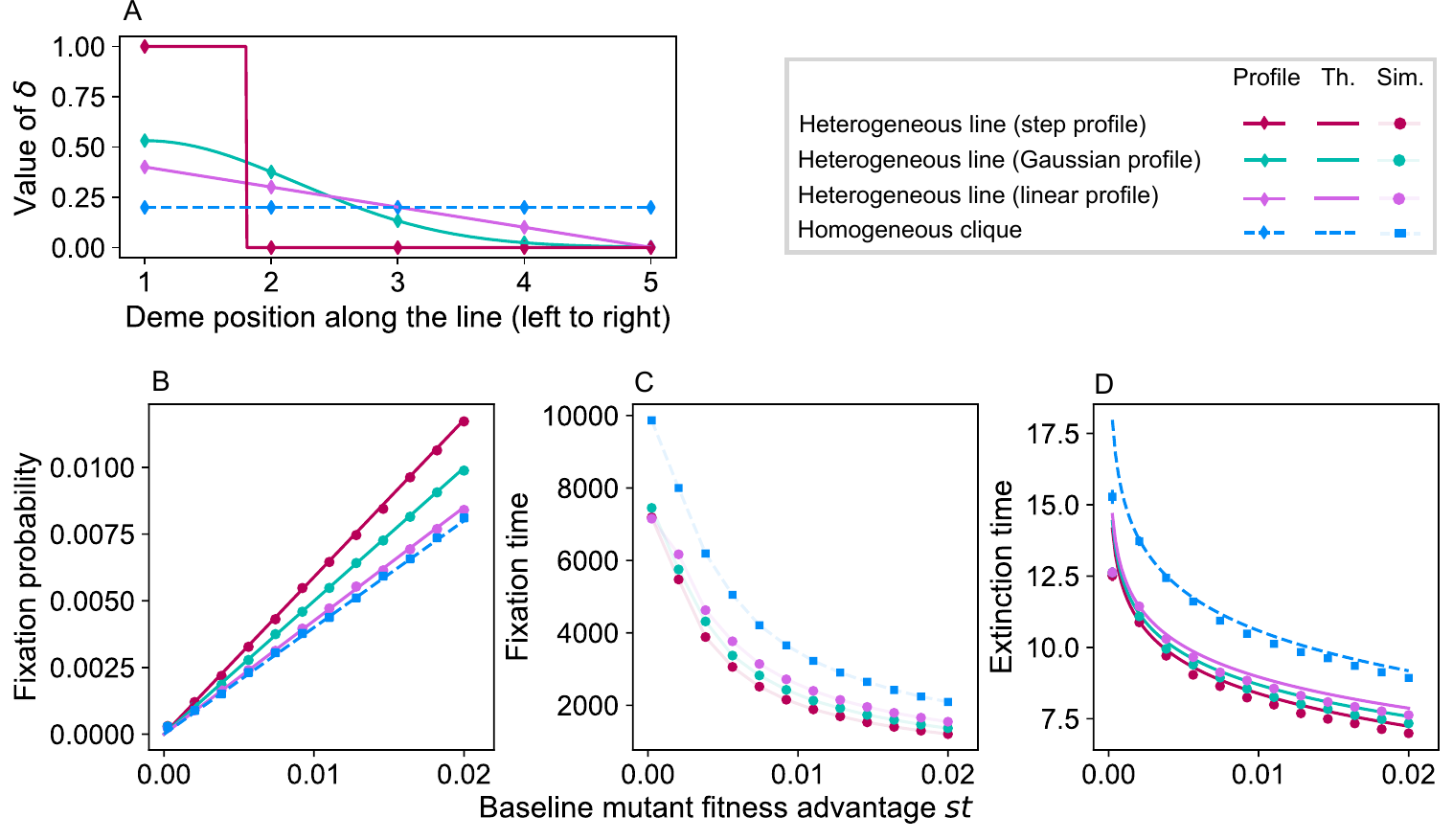}
    \caption{\textbf{Impact of different environmental profiles in the line.} Panel A: Different environmental profiles considered: step, Gaussian, linear, and homogeneous. For each of them, the mutant fitness advantage prefactor $\delta$ is shown versus position along the line. In what follows, we consider lines with $\alpha=1.5$ and values $\delta_i$ taken in each of these profiles at regular intervals for $i\in\{1,\dots,5\}$ (markers) for each of the 5 demes of the graph. The profiles are chosen so that all lines have the same $\langle\delta\rangle$.  Panel B: Mutant fixation probability in the different structures considered, versus baseline mutant fitness advantage $st$. Panel C: Mutant fixation time, in number of bottlenecks, in the different structures considered, versus baseline mutant fitness advantage $st$. Panel D: Same as in C, but for mutant extinction time. Results from the branching process theory (``Th.'') are shown in panels B (Eq.~\ref{eq:a-deltaj-clique}) and D (Eq.~\ref{eq:text}). Stochastic simulation results (``Sim.'') are shown in panels B-D. Parameter values for all structures: $D=5$, $K=1000$; for all lines: $\alpha = 1.5$, $m_L = 0.3$; for the clique: $m=0.15$. Each marker comes from $5\times 10^5$ stochastic simulation realizations.}
    \label{fig:grads-2}
\end{figure}

In Figure~\ref{fig:grads-2}, we further investigate the effect of different environmental profiles in a line with $\alpha>1$. We consider a linear and a Gaussian profile, both with the same orientation as the overall migration flow (i.e., the demes where the mutants are most advantaged are upstream). These profiles are intermediate cases between the step environmental profile and the homogeneous environment. We chose them so that the average mutant fitness advantage is the line is the same in all cases (same $\langle \delta \rangle$), see Figure~\ref{fig:grads-2}A for a depiction of the spatial profiles considered. 
We observe in Figure~\ref{fig:grads-2}B that these various cases yield some amplification of selection, and that larger environmental contrast between demes is associated to more amplification of selection. Figure~\ref{fig:grads-2}C-D further suggest that larger environmental contrast between demes leads to more acceleration of mutant fixation and extinction.

\newpage

\section{Spatial structures with one special deme}
\label{sec:dirichlet-cliques}

Our results for the heterogeneous star and of the heterogeneous line show that environment heterogeneity can allow spatially structured populations with frequent asymmetric migrations to amplify natural selection. They also suggest that this happens when the overall migration flow goes from a region where mutants are most beneficial to a region where they are less beneficial. To further explore this, we now consider a more general spatial structure, where all pairs of demes are connected by migrations, but where one deme has a migration outflow that differs from others, and can thus send out more individuals. We assess the impact of having a different environment in that special deme. 

First, in Section \ref{subs:simplified-Dirichlet}, we consider the simplest version of this structure with a special deme where the environment and the outflow both differ from those of other demes, assuming a highly symmetric migration matrix. We determine mutant fixation probability to first order in $st$, and obtain a condition on the asymmetry in migrations and the environment heterogeneity to obtain amplification of selection. Next, in Section \ref{subs:alpha-eta}, we consider more general and less symmetric migration matrices, with coefficients drawn in Dirichlet distributions (see Ref.~\cite{Abbara__Frequent} for the homogeneous environment case), with one special deme where both the environment and the migration outflow differ from those of other demes. 

\subsection{Highly symmetric structure with one special deme}
\label{subs:simplified-Dirichlet}

\subsubsection{Fixation probability}

Let us consider a simple structure where the first deme has a different migration outflow, i.e.\ it sends out more (or fewer) individuals than others. Specifically, we assume that all outgoing migrations from deme 1 happen with probability $m_1$, while all other migrations across the structure occur with probability $m_2$. Figure~\ref{fig:Dirichlet-clique} illustrates the case $m_2<m_1$. In addition to having a different outflow, the first deme is such that the relative mutant fitness advantage is $\delta_1 s$, while in all other demes it is $\delta_2 s$. Under these assumptions, only the first deme is different from others, and all other ones are equivalent to each other: this structure is highly symmetric.

Since $\sum_{k=1}^D m_{kj} = 1$ for all $j$, we have the following constraint on migration probabilities:
\begin{equation}
    m_1 +(D-1)m_2 = 1 \,.   \label{eq:m2-DC}
\end{equation}
The first line of Eq.~\ref{eq:coeffs-abc}, written for the first node and for another one, yields:
\begin{align}
        a_1 &=  m_1 a_1 + (D-1) m_1 a_2\,,    \label{eqs:a-detdir1}\\
        a_2 &= m_2 a_1 + (D-1) m_2 a_2\,, \label{eqs:a-detdir2}
\end{align}
Note that we have used the symmetry of the structure to write that $a_i=a_2$ for all $i\geq 2$. Using Eqs.~\ref{eq:m2-DC}, both Eq.~\ref{eqs:a-detdir1} and Eq.~\ref{eqs:a-detdir2} reduce to:
\begin{equation}
    a_2=\frac{a_1}{\tilde{\alpha}}\,,\,\,\,\textrm{with}\,\,\,\tilde{\alpha}=\frac{m_1}{m_2}\,. \label{eq:a2eq}
\end{equation}
Note that here we introduced $\tilde{\alpha}=m_1/m_2$ while the most analogous quantity to our work with the star (here and in Refs.~\cite{marrec_toward_2021,Abbara__Frequent}) would be $\alpha=m_2/m_1=\tilde{\alpha}^{-1}$. This is motivated by the fact that our results in the next Sections take a simpler form with $\tilde{\alpha}$. However, to facilitate the comparison with the star, we express fixation probabilities as a function of $\tilde{\alpha}^{-1}$. 

The second line of Eq.~\ref{eq:coeffs-abc}, written for the first node and for another one, yields:
\begin{align}
        b_1 &=  m_1 b_1 + (D-1) m_1 b_2+ a_1^2 - 2a_1\delta_1\,,    \label{eqs:b-detdir1}\\
        b_2 &= m_2 b_1 + (D-1) m_2 b_2+a_2^2 - 2a_2\delta_2\,, \label{eqs:b-detdir2}
\end{align}
Summing Eq.~\ref{eqs:b-detdir1} with $(D-1)$ times Eq.~\ref{eqs:b-detdir2}, and using Eq.~\ref{eq:m2-DC}, we obtain:
\begin{equation}
  a_1^2 +(D-1) a_2^2 -2a_1\delta_1+2(D-1)a_2\delta_2=0\,,
\end{equation}
which yields
\begin{equation}
    a_1 = 2\frac{\delta_1  +(D-1) \delta_2 \tilde{\alpha}^{-1}}{1 +(D-1)\tilde{\alpha}^{-2} }\,. \label{eq:a1eq}
\end{equation}

Eqs.~\ref{eq:a2eq} and~\ref{eq:a1eq} then provide the fixation probability $\rho$, to first order in $st$, of a mutant appearing at a bottleneck in a deme chosen uniformly at random in the structure:
\begin{equation}
    \rho  = st\,\frac{a_1 + (D-1) a_2}{D} = 2st\,\frac{[1+\tilde{\alpha}^{-1}(D-1)] [\delta_1+ (D-1) \delta_2\tilde{\alpha}^{-1} ] }{D\left[ 1 +(D-1)\tilde{\alpha}^{-2}\right] }.\label{eq:pfix-dirdet}
\end{equation}
Note that, for $\tilde{\alpha}=1$, this reduces to $\rho=2\langle\delta\rangle st$, as in all other structures considered so far.

\subsubsection{Condition for amplification of selection}
\label{subs:ampli-Diri}
The fixation probabilities in our structure and in a clique are given respectively in Eqs.~\ref{eq:pfix-dirdet} and~\ref{eq:a-deltaj-clique} to first order in $st$. The first-order fixation probability is larger in our structure than in a clique with the same average relative mutant fitness advantage if and only if
\begin{equation}
\begin{split}
   \frac{[1+\tilde{\alpha}^{-1}(D-1)] [\delta_1+ (D-1) \delta_2 \tilde{\alpha}^{-1} ] }{1 +(D-1)\tilde{\alpha}^{-2} } &> \delta_1 + (D-1) \delta_2\,,\,\,\,\,\textrm{i.e.}\\
   \delta_1\tilde{\alpha}^{-1}\left(1-\tilde{\alpha}^{-1}\right) &> \delta_2 \left(1-\tilde{\alpha}^{-1}\right) \,.\label{cdn:pfix-dirdet}
\end{split}
\end{equation}
Let us introduce
\begin{equation}
    \beta=\frac{\delta_1}{\delta_2}\,,
\end{equation}
which characterizes environment heterogeneity in this structure. 
If $\tilde{\alpha}>1$, i.e.\ if the special deme has a stronger outflow than others, Eq.~\ref{cdn:pfix-dirdet} reduces to:
\begin{equation}
    \beta>\tilde{\alpha}\,.
\end{equation}
Since here $\tilde{\alpha}>1$, we find that the mutant should have a sufficiently stronger advantage in the special deme than in the other demes for the structure to amplify selection ($\delta_1$ should be sufficiently larger than $\delta_2$).

If $\tilde{\alpha}<1$, i.e.\ if the special deme has a weaker outflow than others, Eq.~\ref{cdn:pfix-dirdet} reduces to: 
\begin{equation}
    \beta<\tilde{\alpha}\,.
\end{equation}
Since here $\tilde{\alpha}<1$, we find that the mutant should have a sufficiently stronger advantage in the other demes than in the special deme for the structure to amplify selection ($\delta_2$ should be sufficiently larger than $\delta_1$).

These results generalize our previous observations in the star and line that environment heterogeneity can allow spatially structured populations with frequent asymmetric migrations to amplify natural selection when the overall migration flow goes from a region where mutants are most beneficial to a region where they are less beneficial. 

\subsection{More general structures with one special deme}
\label{subs:alpha-eta}

Let us now consider more general structures with one special deme. Specifically, let us consider Dirichlet cliques with one special deme. Dirichlet cliques are spatial structures with tunable migration asymmetries on each edge of a clique, which were introduced, and studied with a homogeneous environment, in Ref.~\cite{Abbara__Frequent}. They have migration probabilities $(m_{1j},m_{2j},\dots,m_{Dj})$ sampled from a Dirichlet distribution with parameters $\eta_1, \dots, \eta_D$ for each $j$. The Dirichlet distribution has probability distribution function:
\begin{equation}
    \Phi_{(\eta_1, \dots, \eta_D)}(x_1, \dots, x_D) = \frac{1}{B(\eta_1, \dots, \eta_D)} \prod_{i=1}^D x_i^{\eta_i -1},
\end{equation}
with $B(\eta_1, \dots, \eta_D) = \prod_{i=1}^D \Gamma(\eta_k)/ \Gamma(\sum_{i=1}^D \eta_k)$ and where $\Gamma$ denotes the gamma function. Random variables $(X_1, \dots, X_D)$ sampled from this distribution have respective expectation values $\mathbb{E}(X_i) = \eta_i / \sum_j \eta_j$.
Considering $(\eta_1, \eta_2, \dots, \eta_D) = (\eta, 1, \dots, 1)$ allows us to make deme 1 special in terms of migration outflow. In fact, the expectation values are such that $\mathbb{E}(m_{1i})/\mathbb{E}(m_{i1})=\eta$ for all $i$, which entails that, ignoring the variability of these migration probabilities, $\eta$ maps to $\tilde{\alpha}=m_1/m_2$ in the highly-symmetric structure considered in the previous sections. 

\paragraph{Construction of Figure~\ref{fig:Dirichlet-clique}.} 
 
\begin{figure}[htb!]
    \centering
    \includegraphics[width=0.9\linewidth]{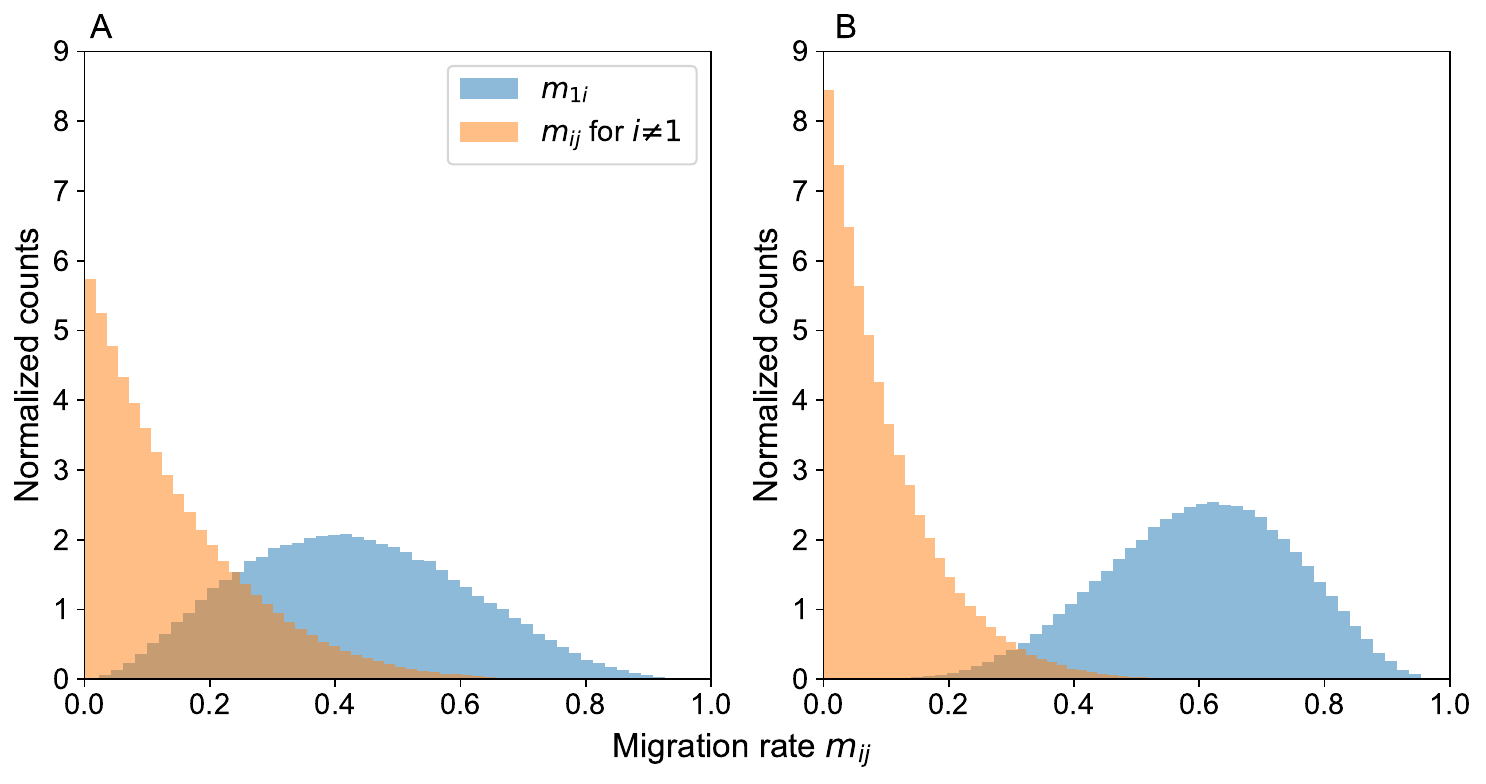}
    \caption{\textbf{Distribution of migration probabilities in Dirichlet cliques with one special deme.} Histograms of $m_{1i}$ for all $i$, whose expectation value is $m_1$, are shown in blue, and histograms of $m_{ij}$ for all $i\ne1$ and all $j$, whose expectation value is $m_2$, are shown in orange. Parameter values: Panel A: $\eta = 3$, yielding $ m_1  =0.43$, $ m_2  =0.14$ and $\tilde{\alpha} = 3$; Panel B: $\eta =6$, yielding $m_1= 0.60$, $m_2 = 0.10$ and $\tilde{\alpha} = 6$. For each panel, $5\times 10^4$ Dirichlet clique migration matrices were generated.}
    \label{fig:histo-migrates}
\end{figure}

In Figure~\ref{fig:Dirichlet-clique}, we consider values of $\eta$ both greater than and less than 1, in order to produce cases with either strong outflow or weak outflow from deme 1. For each value of $\eta$, we generate several (in practice, $5\times10^3$) Dirichlet cliques, and we associate all of them to $\tilde{\alpha}=\eta$. 
Note that despite this mapping to the highly-symmetric structure considered above, the migration probabilities are variable across various edges in a given Dirichlet clique, and across different Dirichlet cliques. As an illustration, Figure~\ref{fig:histo-migrates} shows example histograms for two different values of $\eta$ of $m_{1i}$ for all $i$, with expectation value $m_1$, and of $m_{ij}$ for all $i\ne1$ and all $j$, with expectation value $m_2$.

In Figure~\ref{fig:Dirichlet-clique}, we also vary environment heterogeneity, via the ratio $\beta=\delta_1/\delta_2$, while keeping fixed the average mutant fitness advantage $\langle \delta \rangle$ in each of the structures we consider. This leads to:
\begin{equation}
\delta_2 = \frac{D \langle \delta \rangle}{\beta + D - 1}\,, \hspace{0.5cm}\mathrm{and}\hspace{0.5cm} \delta_1 = \beta \delta_2\,.
\end{equation}

We finally use the general method presented in Section \ref{sec:numerical_pfix} to obtain mutant fixation probability to first order in $st$ for the Dirichlet cliques associated to a given $\tilde{\alpha}$. Comparing to the first-order mutant fixation probability for a standard clique then gives the regions of amplification of selection shown in Figure~\ref{fig:Dirichlet-clique}.

\section{Deleterious mutants}
\label{SI-del}

We demonstrated that environment heterogeneity can induce amplification of selection in deme-structured populations on graphs, under frequent migrations. The branching process calculations that we performed so far hold under the conditions $s>0$, $K\gg 1$ and $1/K\ll st\ll 1$~\cite{Abbara__Frequent}. In particular, under the branching process approximation, the extinction of deleterious mutants ($s<0$) is certain~\cite{Abbara__Frequent}. However, deleterious mutants have a nonzero fixation probability in finite-size populations due to genetic drift, which can be calculated under the diffusion approximation in well-mixed populations~\cite{CrowKimura}. Does the amplification of selection we evidenced extend to deleterious mutations? The complete definition of amplification of selection by spatial structure is that it increases the fixation probability of beneficial mutants but decreases that of deleterious mutants, compared to a well-mixed population~\cite{lieberman_evolutionary_2005}. Here, we explore this point by using numerical simulations.

\begin{figure}[htbp]
    \centering
    \includegraphics[width=\linewidth]{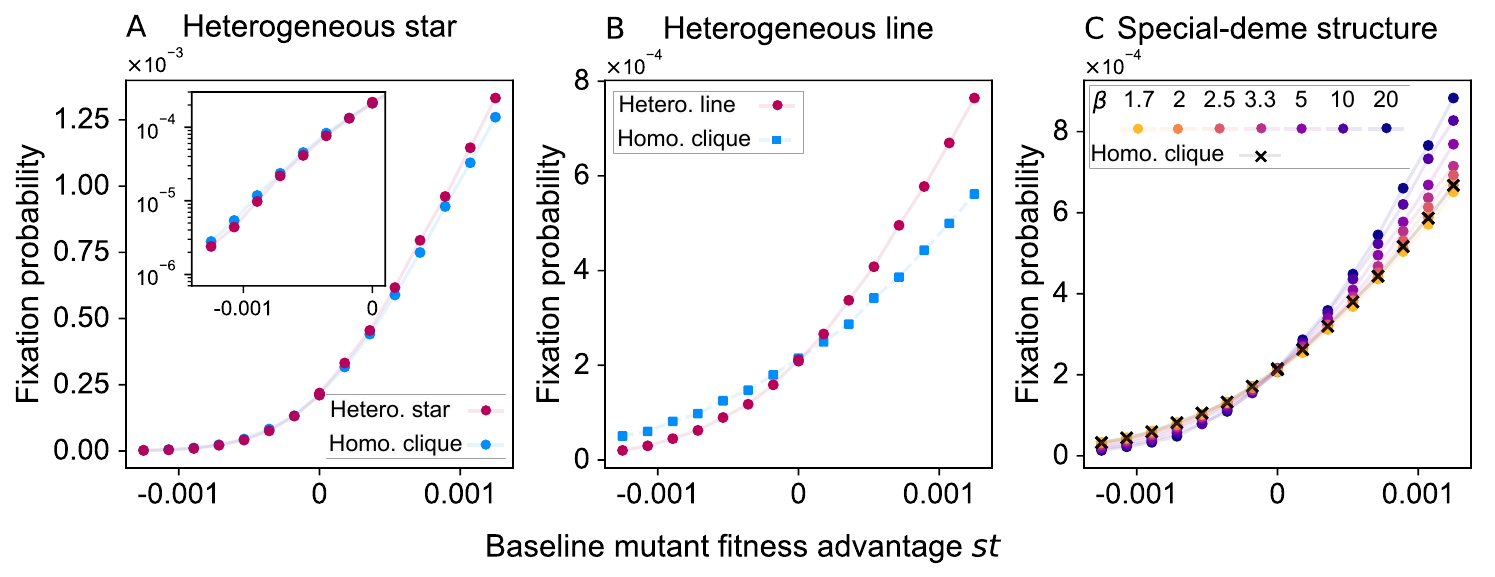}
    \caption{\textbf{Amplification of selection also holds for deleterious mutants.} Mutant fixation probability versus baseline mutant fitness advantage $st$ over a range of $st$ centered around zero, in different spatial structures where we observed amplification for $s>0$. In all cases, the homogeneous clique with the same $\langle\delta\rangle$ is shown for reference. Panel A: heterogeneous star. The inset shows a magnification of the negative $st$ region, with logarithmic vertical axis. Parameter values: as in Figure~\ref{fig:faD}, $\alpha=2$, $m_O=0.1$, $\delta_C=0.05$ in the center and $\delta_L=0.625$ in all leaves, giving $\langle \delta \rangle=0.5$. Panel B: heterogeneous line. Parameter values: as in Figure~\ref{fig:line-proba-time}, $\alpha =1.5$, $m_L=0.3$, $\delta=1$ in the leftmost deme and $\delta=0$ in all others, resulting in $\langle \delta \rangle=0.2$. Panel C: structure with one special deme for several values of environment heterogeneity $\beta=\delta_1/\delta_2$. Parameter values: as in Figure~\ref{fig:Dirichlet-clique}, $\tilde{\alpha}=2$ and $\langle \delta \rangle =0.25$. Recall that for $s>0$, in the branching process regime, we expect amplification for $\beta>\tilde{\alpha}=2$ (see Section~\ref{subs:simplified-Dirichlet}). Parameter values common to all panels: $K=1000$, $D=5$, and $m=0.15$ for the clique. Each marker comes from $5\times 10^6$ stochastic simulation realizations (except for the first three values of $st$ in panel A, for which $5\times 10^7$ realizations were used).}
    \label{fig:negs}
\end{figure}

In Figure~\ref{fig:negs}, we consider three different spatial structures where we demonstrated amplification of selection for $s>0$, and we report the fixation probability for a range of selection coefficients that includes $s<0$, obtained from stochastic simulations. In all cases where beneficial mutants have an increased fixation probability compared to the homogeneous clique with the same $\langle\delta\rangle$, we also observe that deleterious mutants have a decreased fixation probability compared to the homogeneous clique with the same $\langle\delta\rangle$. This suggests that the amplification of selection we evidenced extends to $s<0$.

\newpage

\section{Impact of the number of demes on the fixation probability}
\label{SI-varyD}

In this Section, we analyze the impact of varying the number $D$ of demes on the fixation probability of a mutant in a spatially structured population on graphs with heterogeneous environments, in the branching process regime. We consider the spatial structures discussed above, namely the star, the line and highly-symmetric structures with one special deme. In all cases, we compare results to those obtained for a clique with a homogeneous environment that has the same number of demes, the same deme size and the same average mutant advantage (i.e., in practice, the same $\langle\delta\rangle$). 

Note that, in the branching process approximation, we assume $K\gg 1$, and $K$ does not enter in our expressions of the mutant fixation probability. Hence, we do not discuss further the impact of $K$ here, and we vary $D$ while assuming $K\gg 1$.

\subsection{Star}
\label{subsec:varyD-star}

The fixation probability of a mutant in a heterogeneous star graph is given by Eq.~\ref{eq:fo-star} to first order in $st$, considering a single mutant introduced at a bottleneck in a deme chosen uniformly at random. Let us now study the impact of $D$ on this fixation probability. For this, we focus on the case where all leaves have the same environment, but the center may have a different one, i.e., $\delta_i=\delta_L$ for all $i\in\mathcal{L}$, but we may have $\delta_C\neq \delta_L$. We vary $D$ while keeping $\langle\delta\rangle$ constant, as well as the contrast $C=(\delta_C-\delta_L)/\langle\delta\rangle$ between the center and the leaf environment. The fixation probability of a mutant in a heterogeneous star graph then reads
\begin{align}
    \rho &= 2st\,\frac{[1 + \alpha (D-1)]\left[\delta_C +(D-1)\alpha \delta_L\right] }{D[1+\alpha^2 (D-1)] }\nonumber\\&=2\langle \delta\rangle st\,\frac{1 + \alpha (D-1)  }{D^2[1+\alpha^2 (D-1)] }\left[CD+D-C+\alpha(D-1)(D-C)\right]\,,
    \label{eq:star-varyD}
\end{align}
where we used $\delta_C=(CD+D-C)\langle\delta\rangle/D$ and $\delta_L=(D-C)\langle\delta\rangle/D$. 

\paragraph{Large-$D$ limit.} For large $D$, Eq.~\ref{eq:star-varyD} yields the following expansion:
\begin{align}
    \rho &=2\langle \delta\rangle st\,\left[1+\frac{1}{D}\left(\frac{1}{\alpha}-1\right)\left(1-\frac{1}{\alpha}+C\right)+O\left(\frac{1}{D^2}\right)\right]\,.
    \label{eq:star-varyD-large}
\end{align}
This fixation probability tends to that in the clique, $\rho=2\langle \delta\rangle st$, see Eq.~\ref{eq:a-deltaj-clique}, when $D\to\infty$. Furthermore, in the large-$D$ limit, the fixation probability of a mutant is larger in the heterogeneous star than in the clique if and only if 
\begin{align}
    \left(\frac{1}{\alpha}-1\right)\left(1-\frac{1}{\alpha}+C\right)>0\,.
    \label{cdn-star}
\end{align}
This condition is satisfied for the values of $\alpha$ and $C$ considered in Figure~\ref{fig:faD}.

\begin{figure}[htbp]
\centering
\includegraphics[width=\textwidth]{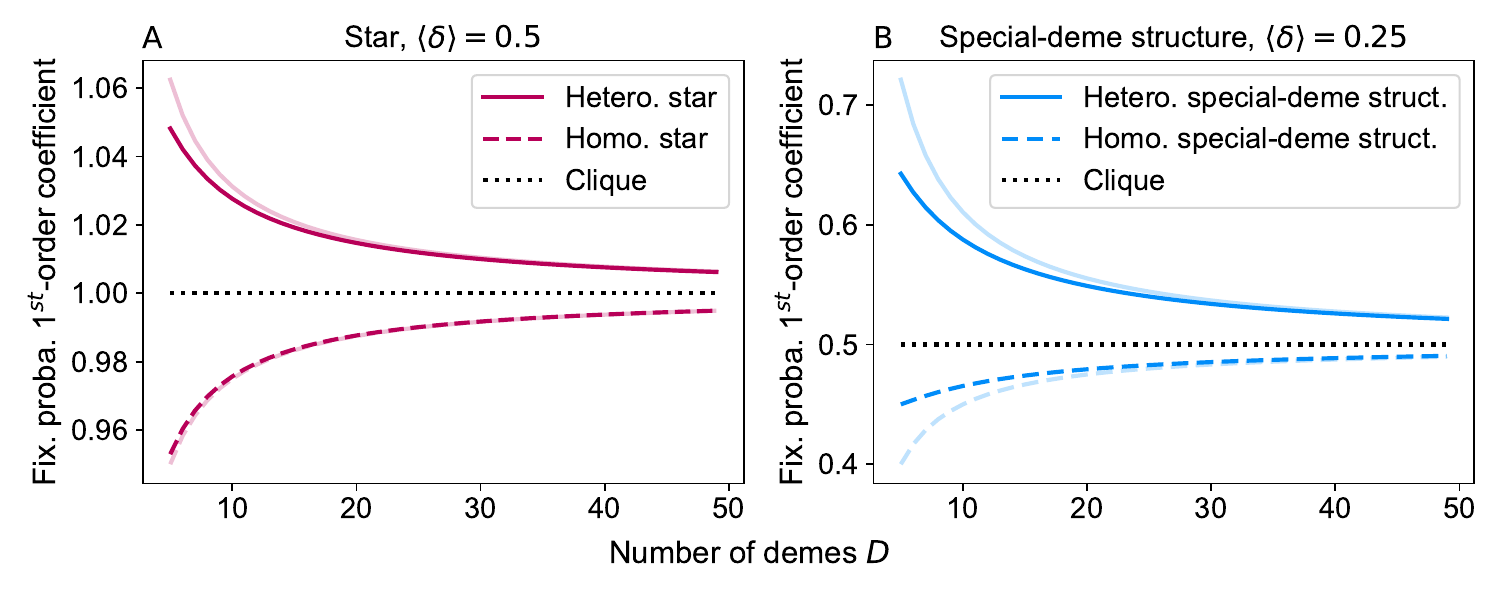}
\caption{\textbf{Impact of the number of demes on the fixation probability in the star and in the structure with one special deme.} Panel A: First-order coefficient in the expansion of the fixation probability in $st$, given in Eq.~\ref{eq:star-varyD}, versus the number of demes $D$, for a star structure. Solid lines: heterogeneous environment; dashed lines: homogeneous environment; pale lines: corresponding large-$D$ asymptotic predictions from Eq.~\ref{eq:star-varyD-large}. The case of the clique with the same $\langle \delta \rangle$ is shown for reference. Parameter values: as in Figure~\ref{fig:faD}, $\langle \delta \rangle=0.5$, $\alpha=2$, and $C=-1.125$. Panel B: Same as in A, but for a highly symmetric structure with one special deme. First-order coefficient in $st$ in Eq.~\ref{eq:star-varyD} with the substitution $\alpha\to\tilde{\alpha}^{-1}$ and large-$D$  asymptotic predictions from Eq.~\ref{eq:star-varyD-large} are shown for heterogeneous and homogeneous environments. In both panels, the average mutant fitness advantage prefactor $\langle \delta \rangle$ and the contrast $C$ are held fixed while varying $D$. Parameter values: as in Figure~\ref{fig:Dirichlet-clique}, $\langle \delta \rangle=0.25$, $\tilde{\alpha}=2$, and $C=3.214$ (corresponding to $\beta=10$).}
\label{fig:changeD-star}
\end{figure}

Figure~\ref{fig:changeD-star}A shows the first-order coefficient in the expansion of the fixation probability in $st$, given by Eq.~\ref{eq:star-varyD}, versus $D$, in a star with other parameter values matching those of Figure~\ref{fig:faD}. We observe that amplification persists but weakens when $D$ is increased for this heterogeneous star, while suppression persists and weakens for the corresponding homogeneous star. These behaviors are well-predicted by our asymptotic expansion for large $D$ from Eq.~\ref{eq:star-varyD-large}.

\subsection{Highly-symmetric structure with one special deme}
\label{subsec:highsym}

The fixation probability of a mutant in a highly-symmetric structure with one special deme is given by Eq.~\ref{eq:pfix-dirdet} to first order in $st$, considering a single mutant introduced at a bottleneck in a deme chosen uniformly at random. With the substitution $\alpha\to\tilde{\alpha}^{-1}$ (see Section~\ref{subs:simplified-Dirichlet}), this is the same as the fixation probability of a mutant in a heterogeneous star graph in the particular case where all leaves have the same environment, which we considered in Section~\ref{subsec:varyD-star}, see Eq.~\ref{eq:star-varyD}. Hence, the conclusions we obtained there hold in this case too. In particular, Eq.~\ref{cdn-star} gives the condition for the probability of fixation to be larger than in the clique in the large-$D$ limit, namely:
\begin{align}
\left(\tilde{\alpha}-1\right)\left(1-\tilde{\alpha}+C\right)>0\,.
    \label{cdn-spec}
\end{align}

Figure~\ref{fig:changeD-star}B shows the first-order coefficient in the expansion of the fixation probability in $st$, given by Eq.~\ref{eq:star-varyD} with $\alpha\to\tilde{\alpha}^{-1}$, versus $D$. Other parameter values match those of Figure~\ref{fig:Dirichlet-clique}, with $\beta=10>2=\tilde{\alpha}$. As for the star (see Figure~\ref{fig:changeD-star}A), we observe that amplification persists but weakens when $D$ is increased for this heterogeneous structure, while suppression persists and weakens for the corresponding homogeneous structure. 

\subsection{Line}
\label{subsec:varyD-line}

The fixation probability of a mutant in a heterogeneous line graph is given by Eq.~\ref{eq:aline} to first order in $st$, considering a single mutant introduced at a bottleneck in a deme chosen uniformly at random. As done for the star in the previous paragraph, we will now study the impact of $D$ on this fixation probability. We focus on the case where all demes have the same environment, except the most upstream deme (with respect to the migration flow direction), labeled deme 1 under the assumption that $\alpha>1$, which may have a different environment, i.e., $\delta_i=\delta_2$ for all $i>1$, but we may have $\delta_1\neq \delta_2$. As for the star, we vary $D$ while keeping $\langle\delta\rangle$ constant. However, at first, we do not make the hypothesis that the environment contrast is constant. The fixation probability of a mutant in a heterogeneous line graph then reads
\begin{equation}
        \rho = \frac{2st}{D} \frac{1+\alpha}{1+\alpha^{-D}}\left(\delta_1 \alpha^{-1}+\delta_2 \sum_{j=2}^D \alpha^{-j}\right)=\frac{2st}{\alpha D} \frac{1+\alpha}{1+\alpha^{-D}}\left(\delta_1 +\delta_2 \frac{1-\alpha^{1-D}}{\alpha-1}\right).
        \label{line-varyD}
\end{equation}

\paragraph{Large-$D$ limit.} Since our focus is on $\alpha>1$, we obtain the following leading order term for $D\gg 1$: 
\begin{equation}
        \rho \sim 2st\frac{1+\alpha}{\alpha D}\left(\delta_1 + \frac{\delta_2}{\alpha-1}\right).
\end{equation}
Meanwhile, in the clique with the same $\langle \delta\rangle=[\delta_1+(D-1)\delta_2]/D$, the fixation probability reads $\rho=2 st [\delta_1+(D-1)\delta_2]/D$, see Eq.~\ref{eq:a-deltaj-clique}. Hence, in the large-$D$ limit, the fixation probability of a mutant is larger in the heterogeneous line than in the clique if and only if 
\begin{align}
    \delta_1>\alpha D \delta_2\,.
    \label{cdn}
\end{align}

\paragraph{Convention 1: fixed average environment and contrast.}

Let us now keep the contrast $C=(\delta_1-\delta_2)/\langle\delta\rangle$ between the deme 1 and others constant, in addition to $\langle\delta\rangle$, similarly to what we did for the star and the structure with one special deme. Eq.~\ref{line-varyD} then gives
\begin{equation}
        \rho =2\langle\delta\rangle st \frac{1+\alpha}{D\left(1+\alpha^{-D}\right)}\left(\frac{C}{\alpha} +\frac{D-C}{D} \frac{1-\alpha^{-D}}{\alpha-1}\right),
        \label{eq:line-1}
\end{equation}
where we used $\delta_1=(CD+D-C)\langle\delta\rangle/D$ and $\delta_2=(D-C)\langle\delta\rangle/D$. 

To leading order for $D\gg 1$, this yields
\begin{equation}
        \rho \sim2\langle\delta\rangle st \frac{1+\alpha}{D}\frac{C\alpha-C+\alpha}{\alpha(\alpha-1)},
        \label{eq:line-largeD1}
\end{equation}
which tends to 0 as $D\to\infty$, and hence becomes smaller than the fixation probability in the clique. Consistently, we note that in this case, Eq.~\ref{cdn} is not satisfied in the large-$D$ limit. Hence, in this convention, the mutant fixation probability in the heterogeneous line becomes smaller than in the clique for large $D$, even though it can be larger for small $D$, as in Figure~\ref{fig:line-proba-time}.

\begin{figure}[htbp]
\centering
\includegraphics[width=\textwidth]{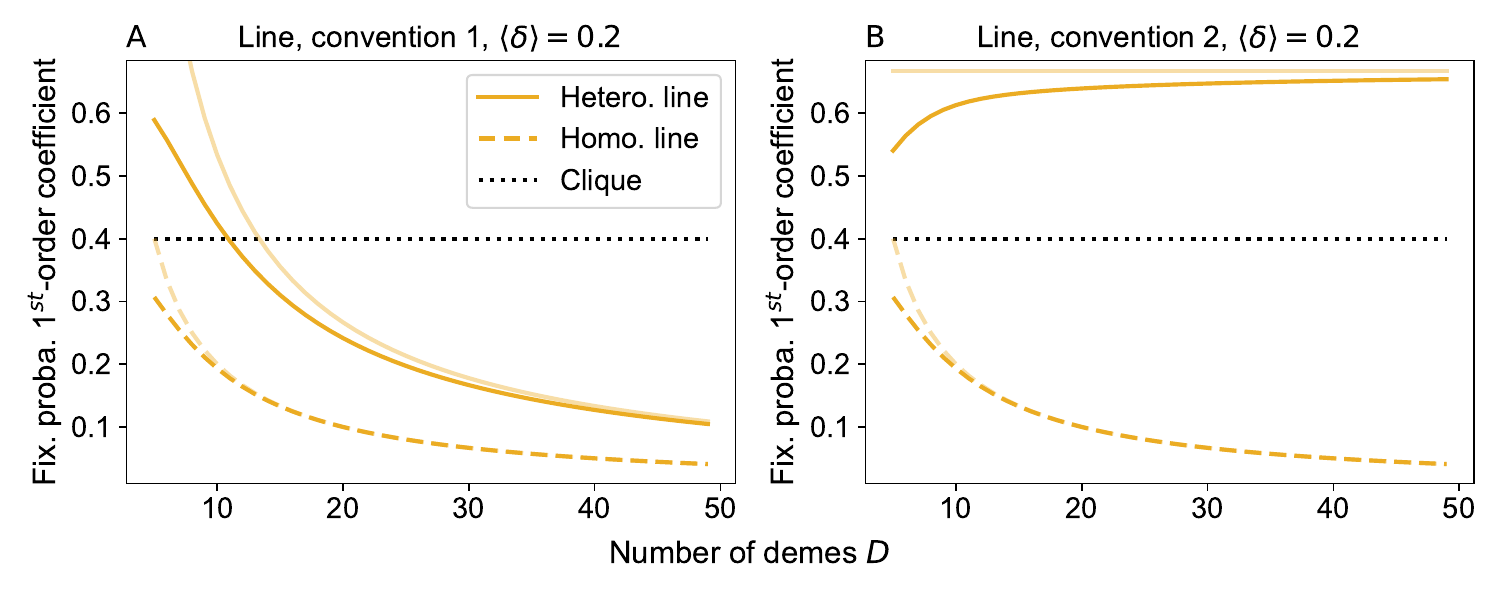}
\caption{\textbf{Impact of the number of demes on the fixation probability in the line.} First-order coefficient in the expansion of the fixation probability in $st$, versus the number of demes $D$, for a line structure, shown under two conventions that specify how the environment changes with $D$. Panel A uses Eq.~\ref{eq:line-1}, derived under ``convention 1'', where the average mutant fitness advantage prefactor $\langle \delta \rangle$ and the contrast $C$ are fixed while varying $D$. Solid lines: heterogeneous environment; dashed lines: homogeneous environment; pale lines: corresponding large-$D$ asymptotic predictions from Eq.~\ref{eq:line-largeD1}. The case of the clique with the same $\langle \delta \rangle$ is shown for reference. Panel B uses Eq.~\ref{eq:line-2} for the heterogeneous line. This result is derived under ``convention 2'', where the average mutant fitness advantage prefactor $\langle \delta \rangle$ is fixed and the ratio $\delta_1/\delta_2$ is set to $1/D^2$ while varying $D$. The large-$D$  asymptotic prediction for the heterogeneous line are from Eq.~\ref{eq:line-largeD2}. Line styles are as in panel A, and the homogeneous line case is the same as in panel A. Parameter values: as in Figure~\ref{fig:line-proba-time}, $\langle \delta \rangle = 0.2$, $\alpha=1.5$ for both panels, $C=5$ for panel A.}
\label{fig:changeD-line}
\end{figure}

Figure~\ref{fig:changeD-line}A shows the first-order coefficient in the expansion of the fixation probability in $st$, given by Eq.~\ref{eq:line-1}, when varying $D$ under convention 1, in a line with other parameter values matching those of Figure~\ref{fig:line-proba-time}. We observe that amplification exists for $D\leq 10$, but weakens when $D$ is increased for this heterogeneous line, and turns into suppression for $D>10$. Meanwhile, suppression increases with $D$ for the corresponding homogeneous line. Besides, the observed trends are well-predicted by our asymptotic expansion for large $D$ from Eq.~\ref{eq:line-largeD1}.

\paragraph{Convention 2: fixed average environment and $\delta_2/\delta_1=1/D^2$.} 

Let us now set $\delta_2/\delta_1=1/D^2$, in addition to keeping $\langle\delta\rangle$ constant. Eq.~\ref{line-varyD} then gives
\begin{equation}
        \rho =2\langle\delta\rangle st \frac{1+\alpha}{\alpha\left(1+\alpha^{-D}\right)\left(D^2+D-1\right)}\left(D^2 +\frac{1-\alpha^{1-D}}{\alpha-1}\right),
        \label{eq:line-2}
\end{equation}
where we used $\delta_1=D^3\langle\delta\rangle/(D^2+D-1)$ and $\delta_2=D\langle\delta\rangle/(D^2+D-1)$. 

To leading order for $D\gg 1$, this yields
\begin{equation}
        \rho \sim2\langle\delta\rangle st \frac{1+\alpha}{\alpha},
        \label{eq:line-largeD2}
\end{equation}
which is larger than the fixation probability in the clique $\rho =2\langle\delta\rangle st$. Consistently, we note that in this case, Eq.~\ref{cdn} is satisfied in the large-$D$ limit. Hence, in this convention, the mutant fixation probability in the heterogeneous line is larger than in the clique even for large $D$.

Figure~\ref{fig:changeD-line}B shows the first-order coefficient in the expansion of the fixation probability in $st$, given by Eq.~\ref{eq:line-2}, when varying $D$ under convention 2, in a line with other parameter values matching those of Figure~\ref{fig:line-proba-time}. We observe that amplification increases with $D$ for this heterogeneous line, in contrast with the results from convention 1 (see Figure~\ref{fig:changeD-line}A). The large-$D$ limit is well-predicted by Eq.~\ref{eq:line-largeD2}. The case of the corresponding homogeneous line is unchanged, and suppression increases with $D$. Thus, the discrepancy between the homogeneous and the heterogeneous lines in terms of fixation probability increases with $D$.

\newpage

\section{More general graph structures}
\label{SI:gengraphs}

\begin{figure}[htbp]
\centering
\includegraphics[width=\textwidth]{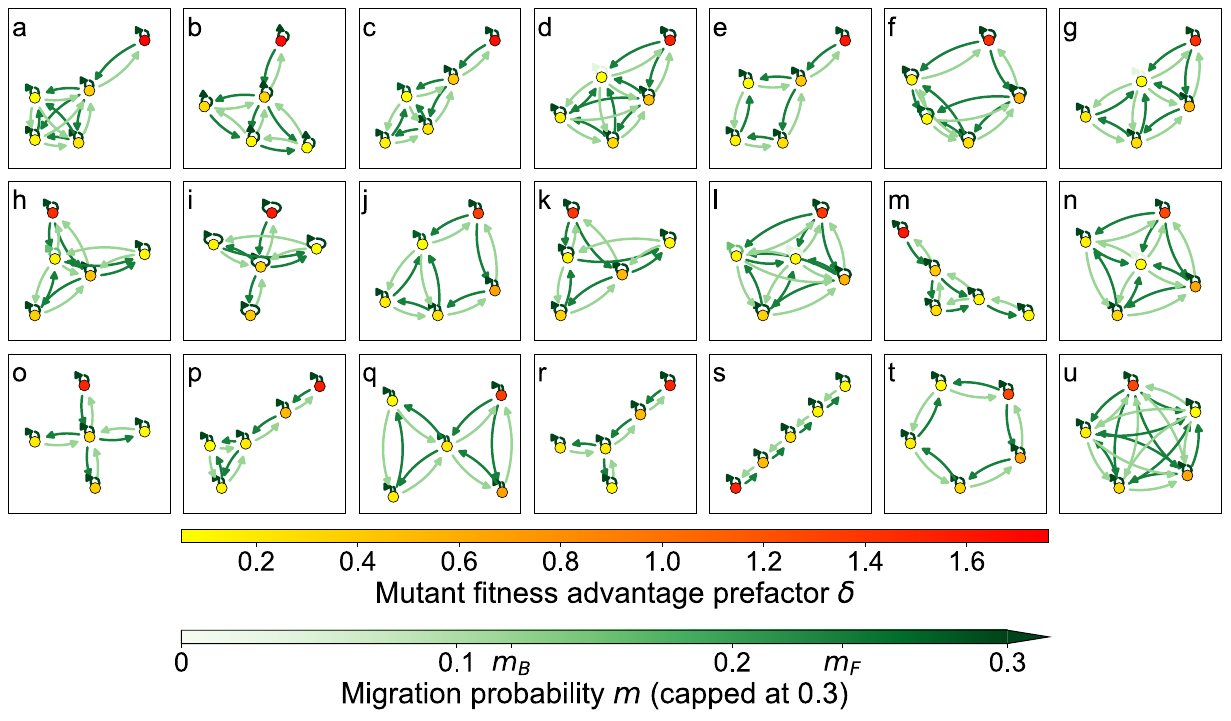}
\caption{\textbf{Visualization of all  distinct connected graphs with five nodes under convention~1.} Each lowercase letter (a–u) labels a distinct graph. Mutant fitness advantage prefactors and migration probabilities are those from convention 1 (see text), where the overall migration flow is directed from the demes where the mutant has the strongest fitness advantage to those where this advantage is smaller. Here, $\langle \delta \rangle=0.5$, $m_B=0.12$, $\alpha=m_F/m_B=2$.
}
\label{fig:all-graphs}
\end{figure}

So far, we mainly focused on graphs with strong symmetries or simple structures, namely circulations, the star, the line and the fully-connected structure with one special deme. For frequent migrations, in the branching process regime, we showed that the star, the line and the structure with one special deme can all amplify selection under some environment heterogeneities, while they are suppressors of selection when the environment is homogeneous. We found that this happened when there is an overall migration flow directed from a deme where mutants are strongly advantaged toward other demes where they are less advantaged. How general is this finding? To address this question, we consider all connected graphs with five demes, in the spirit of evolutionary graph theory works~\cite{hindersin_counterintuitive_2014,Hindersin15,moller_exploring_2019}. There are 21 distinct such graphs, shown in Figure~\ref{fig:all-graphs}. We compute the fixation probability in these graphs to first order in $st$ under the branching process approximation, using our general method from Section~\ref{sec:numerical_pfix}. Importantly, there is an infinity of ways to set the migration probabilities $m_{ij}$ and the mutant fitness advantage prefactors $\delta_i$ in these structures. We consider three different conventions setting migration probabilities and mutant fitness advantage prefactors. They are defined below.

\paragraph{Convention 1.} We set migration probabilities $m_{ij}$ in order to introduce a directional flow across the structure. We first number nodes as follows. Node number 1 is the node with smallest degree. If several nodes have that same degree, we select the node whose neighbors' degrees have the smallest sum, and if there is still ambiguity, node number 1 is selected uniformly at random among the possible ones. Next, we choose node number 2 as the neighbor of node number 1 that has the smallest degree (and if there is ambiguity, we use the same strategy as for node 1). We repeat this procedure until all nodes are labeled. 
We then introduce a forward migration probability $m_F$, and a backward migration probability $m_B$, such that $\alpha = m_F/m_B >1$. For each pair of nodes $(i,\,j\neq i)$ that is connected in the graph of interest, we set all non-zero elements of the adjacency matrix in the upper-right triangle ($i<j$) to $m_{ij}=m_F$ if $i<j$ and $m_{ij}=m_B$ if $i>j$. Finally, we set $m_{ii}=1-\sum_{j\neq i} m_{ji}$.

We define the mutant fitness advantage prefactor $\delta_i$ in each deme $i$ to be proportional to the inverse degree of the corresponding node, multiplied by a decreasing sequence of the node index, chosen as $\{2^{-i}\}_{i=1}^{5}$. We further normalize the values so that the average mutant fitness advantage prefactor $\langle \delta \rangle$ is the same across all structures. 

This procedure ensures that the overall migration flow is directed from the demes where the mutant has the strongest fitness advantage to those where this advantage is smaller.

\paragraph{Convention 2.} We set $m_{ii}=1/5$ for all $i$ and $m_{ij}=4/(5d_j)$ for each pair of nodes $(i,\,j\neq i)$ that is connected in the graph of interest, with $d_j$ the degree of node $j$. This ensures that $\sum_i m_{ij}=1$.

We define the mutant fitness advantage prefactor in each deme to be proportional to the degree of the corresponding node. Values are normalized so that the average mutant fitness advantage prefactor $\langle \delta \rangle$ is the same across all structures.

\paragraph{Convention 3.} Migration probabilities are the same as in convention 2.

We define the mutant fitness advantage prefactor in each deme to be proportional to the inverse degree of the corresponding node. Again, values are normalized so that the average mutant fitness advantage prefactor $\langle \delta \rangle$ is the same across all structures.

\newpage

\begin{figure}[htbp]
\centering
\includegraphics[width=\textwidth]{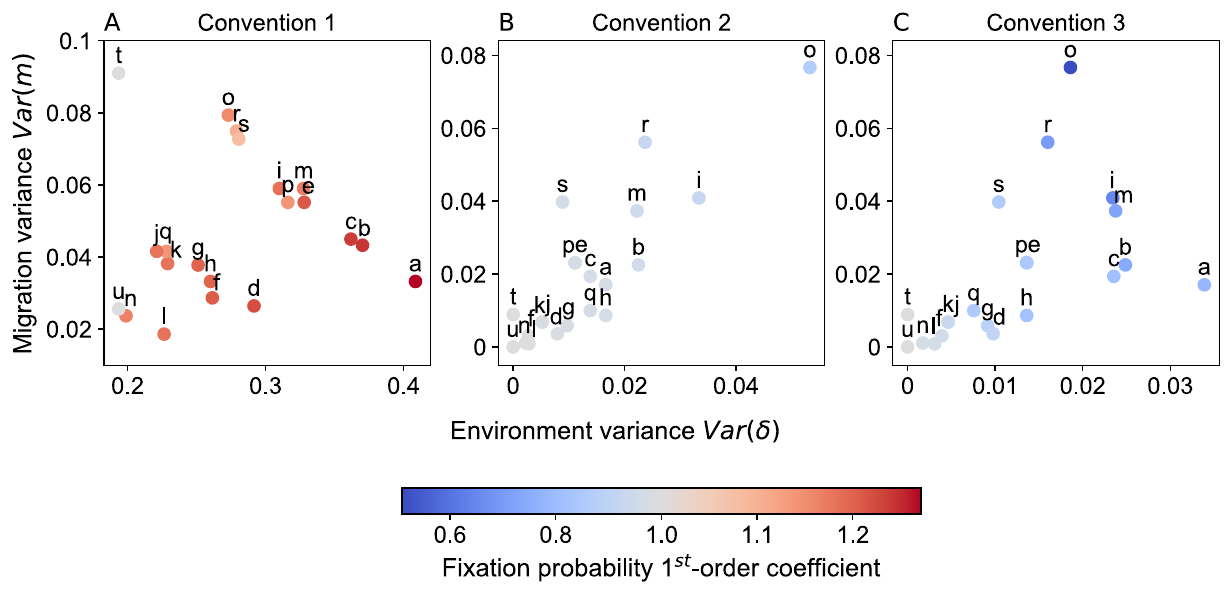}
\caption{\textbf{Fixation probability in all distinct connected graphs with 5 nodes, under 3 different conventions of migration and environment heterogeneities.} In each panel, each marker corresponds to one of the 21 distinct connected graphs with 5 nodes shown in Figure~\ref{fig:all-graphs}. Its color indicates the value of the first-order coefficient in the expansion of the fixation probability in $st$, computed numerically as described in Section~\ref{sec:numerical_pfix}, and is labeled using the same lowercase letter as in Figure~\ref{fig:all-graphs}. Here, $\langle \delta \rangle=0.5$, which entails that there is amplification of selection if the first-order coefficient is larger than $2\langle \delta \rangle=1$ (red) and suppression of selection if it is smaller than that (blue). Results are plotted versus the variance $\mathrm{Var}(m)$ of nonzero migration probabilities and the variance $\mathrm{Var}(\delta)$ of the mutant fitness advantage prefactor across demes in the graph considered. Each panel corresponds to a different convention for defining migration probabilities and mutant fitness advantage prefactors in each graph, as defined in the text (in particular, panel A corresponds to convention 1, which is represented in Figure~\ref{fig:all-graphs}). }
\label{fig:all-graphs-FO}
\end{figure}

Figure~\ref{fig:all-graphs-FO} shows the first-order coefficient in the expansion of the fixation probability in $st$ under the three different conventions defined above. Under convention 1, we observe that amplification of selection exists in all graphs considered, except the cycle and the clique, which have no effect (labeled t and u respectively, see Figure~\ref{fig:all-graphs}). Consistently, we showed above that, to first order in $st$, there is no amplification or suppression of selection in the clique and in the cycle, which are both circulations (see Sections~\ref{subs:pfix_clique} and~\ref{subs:pfix_cycle}). Conversely, under convention 2 and 3, we observe suppression of selection in all graphs except the cycle and the clique, and we note that this suppression tends to be stronger under convention 3 than under convention 2.

Under convention 1, the overall migration flow is directed from the demes where the mutant has the strongest fitness advantage to those where this advantage is smaller. Our results suggest that this can generically lead to amplification of selection. We further observe that this amplification of selection is stronger when the deme where mutants are most advantaged has degree 1, while other nodes are strongly connected together (graphs a, b, c, e), and next, similar graphs but where the deme that most advantages mutants has degree two (graphs d, f, g, h). Besides, the Pearson correlation between $\mathrm{Var}(\delta)$ and the first-order coefficient of the fixation probability is 0.71, illustrating the importance of environment heterogeneity for amplification. Note that a variant of convention 1, where migration probabilities are the same but the mutant fitness advantage prefactor in deme $i$ is set to $2^{-i}$ also leads to amplification of selection, albeit a bit more weakly (mean first-order coefficient 1.10 versus 1.17 under convention 1). Nevertheless, the results obtained under conventions 2 and 3 suggest that suppression of selection remains the most generic effect of spatial structure under frequent migrations in the branching process regime, as for homogeneous environments~\cite{Abbara__Frequent}. Specific situations, such as the overall migration flow going in the same direction as the environment gradient, are required to inverse that trend. Overall, these results show that our conclusions obtained with specific graphs extend to others.

\newpage

\section{Rare migration regime}
\label{sec:raremigr}
In this Section, we turn to the rare migration regime, that is to the case where fixation or extinction of a type within a deme happens on faster timescales than any migrations between demes. Recall that so far, all analytical calculations were made assuming frequent migrations, in the branching process framework. However, with rare migrations, another approach allows to obtain analytical results. Indeed, given the separation of timescales that exists in this regime, one can consider that each deme is either fully mutant or fully wild-type when migrations happen. One can then study mutant fixation using a Markov chain in a coarse-grained description, where migrations can result into a change of state of demes from fully mutant to fully wild-type, and vice-versa~\cite{Slatkin81,marrec_toward_2021}. Below, we consider the star and the line with rare migrations. We calculate their mutant fixation probabilities for heterogeneous environments, and we show that amplification of selection can be observed in this regime. Importantly, in this regime, we can address deleterious mutants as well as beneficial ones, and consider a broad range of fitness differences, while the branching process approach is restricted to beneficial mutants ($s>0$) satisfying $st\ll 1$ and $st\gg 1/K$.

\subsection{Star with heterogeneous environment}
\label{subs:RM-star}

\subsubsection{Mutant fixation probability}

Let us consider a star with $D$ demes, where migrations from each leaf to the center (resp.\ from the center to each leaf) occur with a rate per individual $m_I$ (resp.\ $m_O$). In the rare migration regime, the state of the system can be fully described by two numbers: (1) a binary number indicating whether the center is wild-type or mutant, and (2) the number $i$ of mutant leaves. 

Below, we determine the probability of fixation of a mutant in the star in the rare migration regime, starting from a fully mutant deme chosen uniformly at random. For this, we generalize the calculation of Ref.~\cite{marrec_toward_2021}, which assumed the same environment in each deme (see the Supplementary Material of that paper), to the case of heterogeneous environments. We focus on the case where the center has a different environment from the leaves. We assume that all leaves have the same environment, thereby preserving the symmetry of the star, where all leaves are equivalent. Thus, the key difference compared to the homogeneous case~\cite{marrec_toward_2021} is that here, the probability of fixation of a given type (mutant or wild-type) in the center is different from that in a leaf. In practice, here, for simplicity, we will assume that this arises from the fact that the mutant has a different fitness advantage in the center and in leaves.

Note that, compared to Ref.~\cite{marrec_toward_2021}, we do not take into account the fact that the equilibrium deme size (or bottleneck size) of a deme can differ depending on whether the deme is mutant or wild type. Indeed, here, we consider serial dilution models, which are similar to Wright-Fisher models, and where this bottleneck size is set by a regulation, assumed to be independent from composition. (In models with individual-level birth and death probabilities and logistic regulation, such as that of Ref.~\cite{marrec_toward_2021}, the steady-state deme size depends on fitness, and thus differs between mutants and wild-types, but this difference is small when fitnesses $f$ satisfy $f\gg g$, where $g$ represents death rate. The calculation below can easily be extended to take this difference into account if necessary~\cite{marrec_toward_2021}.)

Upon a given migration event, the probability that the mutant type fixes in the center, if the center is initially wild-type and $i$ leaves are mutant, reads
\begin{equation}
T_{(0,i)\rightarrow(1,i)}=\frac{m_Ii}{m_Ii+m_I(D-1-i)+m_ON_W(D-1)}\rho_M^C\,,
\end{equation} 
where $\rho_M^C$ is the probability of fixation of one mutant in a wild-type center. Indeed, this scenario happens if there is a migration from a mutant leaf to the center, and the mutant then fixes in the center. Similarly, the probability that the wild-type fixes in the center, if the center is initially mutant and $i$ leaves are mutant, reads
\begin{equation}
T_{(1,i)\rightarrow(0,i)}=\frac{m_I(D-1-i)}{m_Ii+m_I(D-1-i)+m_O(D-1)}\rho_W^C\,,
\end{equation} 
where $\rho_W^C$ is the probability of fixation of one wild-type individual in a mutant center. 
Then, the probability that the number of mutant leaves increases by 1 if the center is mutant is
\begin{equation}
T_{(1,i)\rightarrow(1,i+1)}=\frac{m_O(D-1-i)}{m_Ii+m_I(D-1-i)+m_O(D-1)}\rho_M^L\,,
\end{equation}  
where $\rho_M^L$ is the probability of fixation of one mutant in a wild-type leaf. 
Finally, the probability that the number of mutant leaves decreases by 1 if the center is wild-type is
\begin{equation}
T_{(0,i)\rightarrow(0,i-1)}=\frac{m_O i}{m_Ii+m_I(D-1-i)+m_O(D-1)}\rho_W^L\,,
\end{equation} 
where $\rho_W^L$ is the probability of fixation of one wild-type individual in a mutant leaf.

As in Ref.~\cite{marrec_toward_2021}, we call $\Phi_{0,i}^\textrm{star}$ (resp.\ $\Phi_{1,i}^\textrm{star}$) the fixation probability of the mutant type starting from $i$ fully mutant leaves and a wild-type center (resp.\ a mutant center). The fixation probabilities $\Phi_{0,i}^\textrm{star}$ and $\Phi_{1,i}^\textrm{star}$ satisfy the following recurrence relationship (see Refs.~\cite{broom_analysis_2008,marrec_toward_2021}):
\begin{equation}
\left\{
\begin{aligned}
\Phi_{0,0}^\textrm{star}=\,&\,0 \,,\\
\Phi_{1,i}^\textrm{star}=\,&\,T_{(1,i)\rightarrow(0,i)}\Phi_{0,i}^\textrm{star}+T_{(1,i)\rightarrow(1,i+1)}\Phi_{1,i+1}^\textrm{star}\\&+\left[1-T_{(1,i)\rightarrow(0,i)}-T_{(1,i)\rightarrow(1,i+1)}\right]\Phi_{1,i}^\textrm{star}\mbox{ for }0 \leq i \leq D-2\,,\\
\Phi_{0,i}^\textrm{star}=\,&\,T_{(0,i)\rightarrow(1,i)}\Phi_{1,i}^\textrm{star}+T_{(0,i)\rightarrow(0,i-1)}\Phi_{0,i-1}^\textrm{star}\\&+\left[1-T_{(0,i)\rightarrow(1,i)}-T_{(0,i)\rightarrow(0,i-1)}\right]\Phi_{0,i}^\textrm{star}\mbox{ for }1 \leq i \leq D-1\,,\\
\Phi_{1,D-1}^\textrm{star}=\,&\,1\mbox{ },
\end{aligned}
\right.
\label{Phi_star_1}
\end{equation} 
which yields:
\begin{equation}
\left\{
\begin{aligned}
&\Phi_{0,0}^\textrm{star}=0\,, \\
&\Phi_{1,i}^\textrm{star}=\frac{-1+\Gamma_1\left[-1+(1+\Gamma_0)\left(\frac{\Gamma_0(1+\Gamma_1)}{1+\Gamma_0}\right)^i\right]}{-1+\Gamma_1\left[-1+(1+\Gamma_0)\left(\frac{\Gamma_0(1+\Gamma_1)}{1+\Gamma_0}\right)^{D-1}\right]}\mbox{ for }0 \leq i \leq D-2\,,\\
&\Phi_{0,i}^\textrm{star}=\frac{(1+\Gamma_1)\left[-1+\left(\frac{\Gamma_0(1+\Gamma_1)}{1+\Gamma_0}\right)^i\right]}{-1+\Gamma_1\left[-1+(1+\Gamma_0)\left(\frac{\Gamma_0(1+\Gamma_1)}{1+\Gamma_0}\right)^{D-1}\right]}\mbox{ for }1 \leq i \leq D-1\,,\\
&\Phi_{1,D-1}^\textrm{star}=1\mbox{ },
\end{aligned}
\right.
\label{Phi_star}
\end{equation}
with 
\begin{align}
    \Gamma_1&=\alpha\frac{\rho_W^C}{\rho_M^L}\,,\\
    \Gamma_0&=\frac{1}{\alpha}\frac{\rho_W^L}{\rho_M^C}\,,
\end{align}
where we have used migration asymmetry 
\begin{equation}
\alpha=\frac{m_I}{m_O}\,.
\end{equation}

Let us now consider the probability $\rho_1^\textrm{star}$ that a single mutant, appearing uniformly at random in the population, fixes in the whole population. In the rare migration regime, it first fixes in the deme where it appeared, and then it spreads. Let us denote by $\rho_M^C$ and $\rho_M^L$, respectively, the fixation probability of a single mutant in the center and in a leaf. Note that they depend on whether we consider a Moran-like model (as in Ref.~\cite{marrec_toward_2021}) or a Wright-Fisher-like model (as our dilution model introduced in~\cite{Abbara__Frequent} and used here). For consistency with the rest of the paper, we employ the serial dilution model, yielding, for a single mutant placed in the center at a bottleneck:
 \begin{equation}
    \rho_M^C=\frac{1-e^{-2\delta_C st}}{1-e^{-2K\delta_C st}}\,\,\,\textrm{and}\,\,\,\rho_W^C=\frac{1-e^{2\delta_C st}}{1-e^{2K\delta_C st}}\,,
    \label{eq:Kimura-1deme-1}
 \end{equation}
and similarly for $\rho_M^L$, but with $\delta_L$ taking the place of $\delta_C$, if the mutant starts in a leaf. Here, we have denoted by $\delta_C st$ and $\delta_L st$ the mutant effective relative fitness advantages in the center and in a leaf, respectively.

Then, we have:
\begin{align}
\rho_1^\textrm{star}&=\frac{1}{D}\rho_M^C\Phi_{1,0}^\textrm{star}+\frac{D-1}{D}\rho_M^L\Phi_{0,1}^\textrm{star}\,,
\label{PhiStar_2}
\end{align}
with $\Phi_{1,0}^\textrm{star}$ and $\Phi_{0,1}^\textrm{star}$ given in Eq.~\ref{Phi_star}.

\subsubsection{Amplification of selection}
\label{sec:star-rare-ampli}

Figure~\ref{fig:star_rare} shows that the heterogeneous star in the rare migration regime is an amplifier of selection compared to a homogeneous circulation with the same mean mutant fitness when $\delta_C$ is substantially larger than $\delta_L$, i.e.\ when beneficial (resp.\ deleterious) mutants are substantially more advantaged (resp.\ disadvantaged) in the center than in leaves. We observe this amplification for negative and for small absolute baseline mutant fitness advantages, corresponding to weakly deleterious and weakly beneficial mutants, in a broad range of migration asymmetries $\alpha=m_I/m_O$. Recall that the homogeneous star is a suppressor of selection for $\alpha<1$ and an amplifier of selection for $\alpha>1$~\cite{marrec_toward_2021}, and that the amplification effect is observed under rare migrations for $\alpha>1$ when $st\lesssim 1/K$~\cite{Abbara__Frequent}. To better understand these results, let us consider different selection regimes, under rare migrations, and assuming $s>0$. 

First, if $1/K\ll 2\delta_C st\ll 1$ and $1/K\ll2\delta_L st\ll 1$, then the fixation probabilities in a single deme (see Eq.~\ref{eq:Kimura-1deme-1}) reduce to 
\begin{equation}
    \rho_M^C\approx 1-e^{-2\delta_C st}\approx 2\delta_C st- 2\delta_C^2 s^2t^2\,\,\,\textrm{and}\,\,\,\rho_W^C\approx 0\,,
\end{equation}
and similarly in the leaf (note that $\rho_W^C$ is in fact exponentially suppressed). In this regime, once a mutant has fixed in a deme, the wild type is unable to retake over that deme, and fixation in the whole population always ensues. Thus, the mutant fixation probability in the star, averaged over the deme where the mutant starts, reads:
\begin{equation}
    \rho\approx \frac{1}{D}\rho_M^C+\frac{D-1}{D}\rho_M^L\approx 2\langle\delta\rangle st- 2\langle\delta^2\rangle s^2t^2\,.
\end{equation}
Under our assumptions, we have $1/K\ll2\langle \delta\rangle st\ll 1$, and hence, we can express the difference of fixation probability between a heterogeneous and a homogeneous structure with the same $\langle\delta\rangle$ as:
\begin{equation}
    \rho-\rho^\mathrm{homo}\approx-2(st)^2\left(\langle\delta^2\rangle-\langle\delta\rangle^2\right)<0\,,
    \label{eq:comp-star-rare}
\end{equation}
which demonstrates that in this regime, environment heterogeneity leads to suppression of selection, and explains the observations of Figure~\ref{fig:star_rare}B for larger $st$.

Second, if $\delta_C>\delta_L$, there may be a regime where $1/K\ll 2\delta_C st\ll 1$ but $2\delta_L st\lesssim 1/K$. In this case, the situation is the same as before for a mutant starting in the center: if it fixes there, it should ultimately take over the whole population. Indeed, wild-type individuals migrating to the center cannot take over due to their substantial fitness disadvantage (their fixation probability is exponentially suppressed). However, the situation is different for a mutant starting in a leaf: once it has fixed there, it may be re-invaded by wild-type individuals from the center, before the mutant can fix in the center. In a homogeneous circulation with the same $\langle\delta\rangle$,  if $D$ is large enough, we will have $2\langle\delta\rangle st\lesssim 1/K$, as in a leaf, and thus the homogeneous circulation will not benefit from a ``safe'' deme for mutants: this leads to amplification in this regime. Moreover, the heterogeneous star structure with $\delta_C>\delta_L$ means that a mutant starting in a leaf will take over the whole structure if it just takes over the center, making this structure quite favorable to mutant fixation in this regime. Takeover of the center starting from a leaf is more likely if $\alpha$ is large, leading to more amplification in that case.

\subsection{Line with heterogeneous environment}
\label{subs:RM-line}

\subsubsection{Mutant fixation probability}

Here, we extend our derivation of mutant fixation probability in a line with rare migrations, made in the homogeneous case in Ref.~\cite{servajean_impact_2025}, to heterogeneous case.

\paragraph{Equation on the fixation probability.}
Let us consider a line graph made of $D$ demes in the rare migration regime. We denote by $m_L$ (respectively $m_R$) the migration rate to the left (respectively to the right). We introduce environment heterogeneity by considering $D_L$ demes on the left-hand side of the line where the mutant relative  fitness advantage is $\delta_L s$, and $D_R = D-D_L$ demes on the right-hand side where the mutant relative  fitness advantage is $\delta_R s$. Considering the serial dilution model, the fixation probability of one mutant (resp. wild-type) in a wild-type (resp. mutant) deme is
\begin{align}
    \rho_{M}^b = \frac{1-e^{-2\delta_b st}}{1-e^{-2 K\delta_b st}}  \qquad \text{and} \qquad \rho_{W}^b = \frac{1-e^{2\delta_b st}}{1-e^{2 K\delta_b st}},
    \label{eq:Kimura-1deme}
\end{align}
with $b \in\left\{L, R\right\}$.

As pointed out in Ref.~\cite{servajean_impact_2025}, in the rare migration regime, the fact that mutants have to spread from the deme where the mutant appeared to other ones in order to fix entails that, at all times, there is one contiguous cluster of demes of mutants. Let us denote by $\phi_{i, j}$ the probability that mutants fix in the line starting from a cluster of $j-i$ adjacent fully mutant demes. Here, $i\in[0,D]$ indicates the position of the first mutant deme along the line, while $j\in[0,D]$ is the position of the first wild-type deme along the line, with the constraint that $i\leq j$ (the case $j=D$ corresponds to a situation where there are no wild-type demes to the right of the mutant cluster). Note that exceptionally, in this derivation, demes are numbered from $0$ to $D-1$. 

\begin{figure}[h!]
\centering
\includegraphics[width=.3\linewidth]{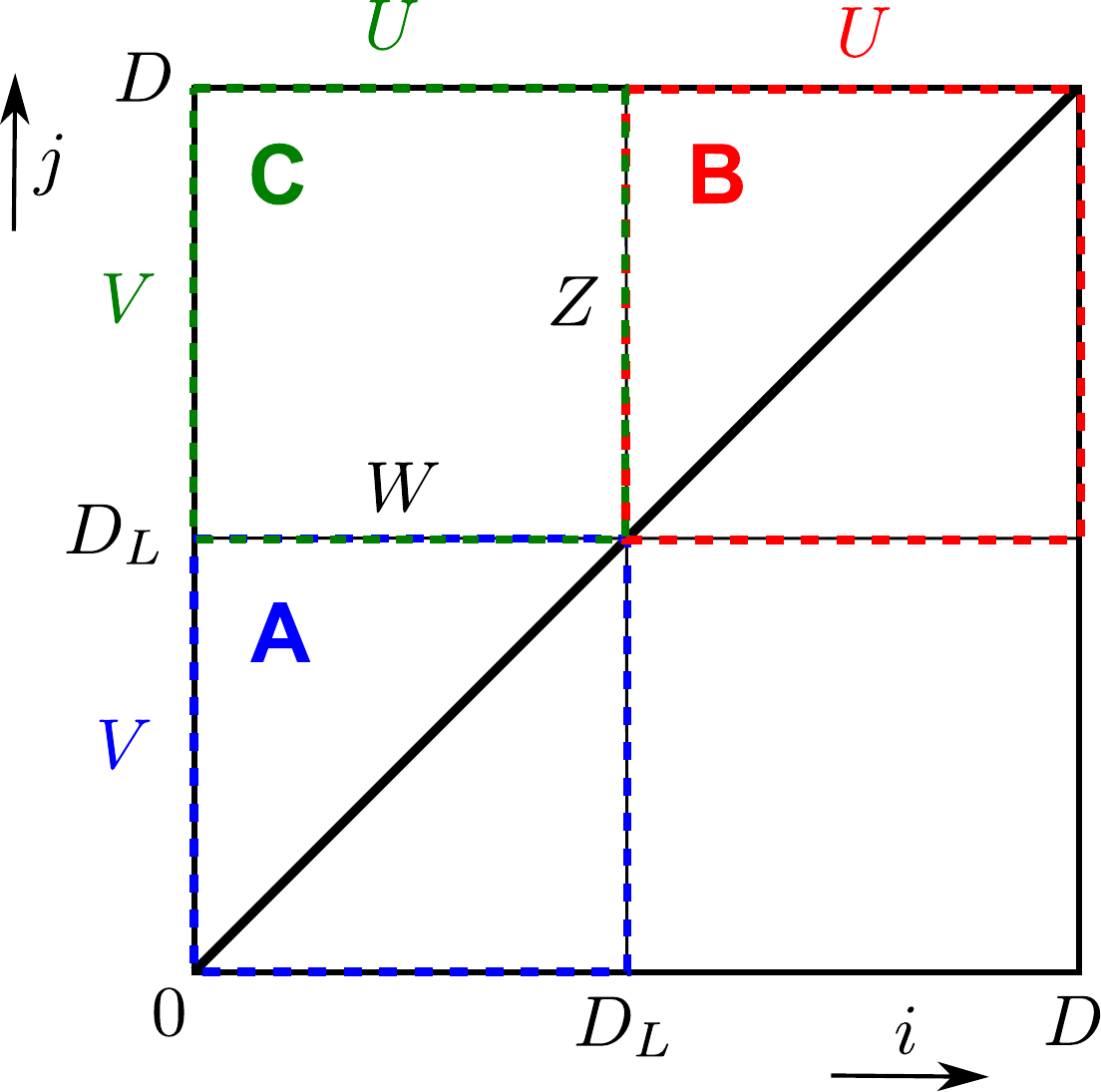}
\caption{\textbf{Schematic of the zones where the indices $i$ and $j$ can vary.} Each zone (A, B, C) is associated with specific values of the transition probabilities $p_1$, $p_2$, $p_3$, and $p_4$. Boundary conditions ($U$, $V$, $W$, $Z$) are shown. The black diagonal corresponds to mutant extinction, where $\phi_{i,i} = 0$ for all $i \in [0, D]$, while the upper-left corner corresponds to mutant fixation on the graph, with $\phi_{0,D} = 1$.  }
\label{fig:schema_square}
\end{figure}

The equation satisfied by the fixation probability $\phi_{i,j}$ is 
\begin{equation}
    \phi_{i,j} = p_1 \phi_{i,j+1}+p_2 \phi_{i-1,j}+p_3 \phi_{i,j-1}+p_4 \phi_{i+1,j}\,,
\end{equation}
where the indices $i, j$ correspond respectively to the position of the first mutant (type $M$) deme
and the position of the next first wild type (type $W$). Here, in contrast with the homogeneous case addressed in Ref.~\cite{servajean_impact_2025}, the transition probabilities $p_1, p_2, p_3, p_4$ depend on where the cluster of mutant demes is located, because of the environment heterogeneity. Specifically, they depend on the zones where the indices $i,j$ are located, represented in \cref{fig:schema_square}. 

\begin{figure}[ht!]
    \centering\includegraphics[width=.6\linewidth]{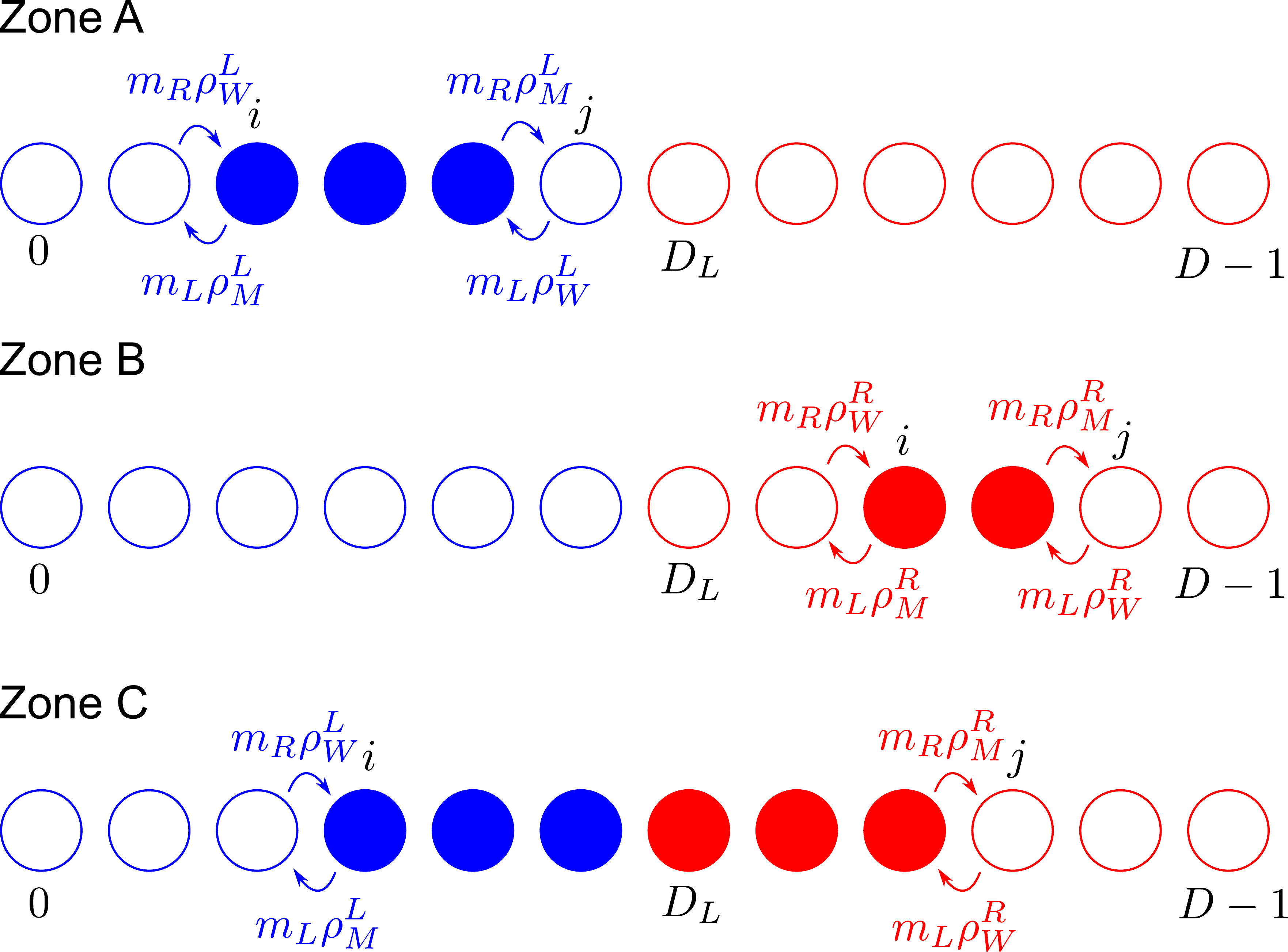}
    \caption{Example configurations for each region A, B and C.}
    \label{fig:schema_3_lines}
\end{figure}

Their expressions in these different zones, illustrated in \cref{fig:schema_3_lines}, are the following:
\begin{itemize}
    \item Zone A ($0 < i < j< D_L $):
    \begin{align}
        p_1^A = \frac{\alpha}{(1+\alpha)(1+\gamma_L)}, p_2^A = \frac{1}{(1+\alpha)(1+\gamma_L)}, \\ p_3^A = \frac{\gamma_L}{(1+\alpha)(1+\gamma_L)}, p_4^A = \frac{\alpha\gamma_L}{(1+\alpha)(1+\gamma_L)},
    \end{align}
    \item Zone B ($D_L<i < j<D$):
    \begin{align}
        p_1^B = \frac{\alpha}{(1+\alpha)(1+\gamma_R)}, p_2^B = \frac{1}{(1+\alpha)(1+\gamma_R)}, \\ p_3^B = \frac{\gamma_R}{(1+\alpha)(1+\gamma_R)}, p_4^B = \frac{\alpha\gamma_R}{(1+\alpha)(1+\gamma_R)},
    \end{align}
     \item Zone C ($i<D_L, j > D_L$):
    \begin{align}
        &p_1^C = \frac{\alpha\beta_M}{1+\gamma_R\beta_M + \alpha (\gamma_L+\beta_M)}, p_2^C = \frac{1}{1+\gamma_R\beta_M + \alpha (\gamma_L+\beta_M)}, \\&p_3^C = \frac{\gamma_R \beta_M}{1+\gamma_R\beta_M + \alpha (\gamma_L+\beta_M)}, p_4^C = \frac{\alpha \gamma_L}{1+\gamma_R\beta_M + \alpha (\gamma_L+\beta_M)},
    \end{align}
\end{itemize}
with migration asymmetry $\alpha = m_R/m_L$, and introducing $\beta_a = \rho_a^R / \rho_a^L$ ($a\in (W,M)$), and $\gamma_b = \rho_W^b/\rho_M^b$, ($b\in (L,R)$). 
The boundary conditions are denoted by $\phi_{i,i} = 0$, $\phi_{0,j} = V_j$, $\phi_{i,D} = U_i$, see \cref{fig:schema_square}, where $U_i$ and $V_j$ are expressed below.

\newpage

\paragraph{Boundary conditions.}
\begin{itemize}
    \item Expression of $V_j = \phi_{0,j}$:
\begin{equation} \left\{ \begin{array}{cc}
    & (\alpha +\gamma_L) V_j = \alpha V_{j+1} + \gamma_L V_{j-1}\text{ , if } j < D_L, \\
    & (\alpha +\gamma_R) V_j = \alpha V_{j+1} +\gamma_RV_{j-1}\text{ , if }  j> D_L,
\end{array} \right.
\end{equation}
and for $j = D_L$,
\begin{align}
    (\alpha \beta_M + \gamma_L) V_{D_L}=\alpha \beta_M V_{D_L+1} + \gamma_L V_{D_L-1} .
\end{align}
The solution is 
\begin{equation} V_j = \left\{ \begin{array}{cc}
    & A_L + B_L \left(\frac{\gamma_L}{\alpha} \right)^j \text{ , if } j < D_L ,\\
    &A_R + B_R \left(\frac{\gamma_R}{\alpha} \right)^j \text{ , if } j > D_L. \\
\end{array} \right.
\end{equation}
The boundary conditions are $A_L + B_L = 0$ and $A_R + B_R\left(\frac{\gamma_R}{\alpha} \right)^D  = 1$.
Branching left and right solutions imposes that
$A_L + B_L \left(\frac{\gamma_L}{\alpha} \right)^{D_L} = A_R + B_R\left(\frac{\gamma_R}{\alpha} \right)^{D_L} $. It yields
\begin{align}
A_L &= \frac{1}{Z_V} \beta_M \left(\frac{\gamma_R}{\gamma_L} \right)^{D_L} \left(1- \frac{\gamma_R }{\alpha} \right),\\
B_L &= -\frac{1}{Z_V} \beta_M \left(\frac{\gamma_R}{\gamma_L} \right)^{D_L} \left(1- \frac{\gamma_R }{\alpha}\right),\\
A_R &= \frac{1}{Z_V} \left[ \beta_M \left(\frac{\gamma_R}{\gamma_L} \right)^{D_L} \left(1- \frac{\gamma_R }{\alpha} \right) \left(1 - \left(\frac{\gamma_L}{\alpha } \right)^{D_L} \right) + \left(1- \frac{\gamma_L}{\alpha }  \right) \left(\frac{\gamma_R}{\alpha }\right)^{D_L}\right] ,\\
B_R &= \frac{1}{Z_V} \left(1- \frac{\gamma_L}{\alpha } \right),
\end{align}
with \begin{align}
Z_V = \beta_M \left(\dfrac{\gamma_R}{\gamma_L} \right)^{D_L}\left(1- \dfrac{\gamma_R }{\alpha} \right) \left(1 - \left(\dfrac{\gamma_L}{\alpha } \right)^{D_L} \right) + \left(1- \dfrac{\gamma_L}{\alpha }  \right) \left(\left(\dfrac{\gamma_R}{\alpha }\right)^{D_L}  - \left(\dfrac{\gamma_R}{\alpha }\right)^{D} \right)
.\end{align}
\item Expression of $U_i = \phi_{i,D}$:
\begin{equation} \left\{ \begin{array}{cc}
    & (1 + \alpha\gamma_L) U_i = \alpha\gamma_L U_{i+1} + U_{i-1}\text{ , if } i < D_L, \\
    & (1 + \alpha\gamma_R)  U_i = \alpha\gamma_R U_{i+1} + U_{i-1}\text{ , if } i > D_L,
\end{array} \right.
\end{equation}
and for $i = D_L$
\begin{equation}
    \left(1 +\alpha \beta_M\gamma_R \right) U_{D_L} = \alpha \beta_M\gamma_R U_{D_L+1} +  U_{D_L-1}.
\end{equation}
The general solution is
\begin{equation} U_i = \left\{ \begin{array}{cc}
    & E_L + \frac{F_L}{\left(\alpha\gamma_L\right)^i}   \text{ , if } i < D_L, \\
    &E_R +\frac{F_R}{\left(\alpha\gamma_R\right)^i} \text{ , if } i>D_L. \\
\end{array} \right.
\end{equation}
The boundary conditions are $E_L + F_L = 1$ and $E_R +  \dfrac{F_R}{\left(\alpha \gamma_R \right)^D} = 0$.
Branching both solutions imposes that
$E_L + \dfrac{F_L}{(\alpha \gamma_L)^{D_L}}  = E_R + \dfrac{F_R}{(\alpha \gamma_R)^{D_L}}  $. It yields
\begin{align}
    E_L &= \frac{1}{Z_U} \left[  -\dfrac{\beta_M}{\left(\alpha\gamma_L\right)^{D_L}}\left(1- \dfrac{1 }{\alpha \gamma_R} \right) + \left(\dfrac{\gamma_R}{\gamma_L} \right)^{D_L-1}\left(1- \dfrac{1}{\alpha \gamma_L}  \right) \left(\dfrac{1}{\left(\alpha \gamma_R \right)^{D_L} } - \dfrac{1}{\left(\alpha\gamma_R \right)^{D} }\right) \right] ,\\
    F_L& = \frac{1}{Z_U}  \beta_M \left(1- \dfrac{1 }{\alpha \gamma_R} \right) ,\\
     E_R &=- \frac{1}{Z_U}  \left(\dfrac{\gamma_R}{\gamma_L} \right)^{D_L-1}\left(1- \dfrac{1}{\alpha \gamma_L}  \right)  \dfrac{1}{\left(\alpha\gamma_R \right)^{D} } , \\
     F_R& = \frac{1}{Z_U}  \left(\dfrac{\gamma_R}{\gamma_L} \right)^{D_L-1}\left(1- \dfrac{1}{\alpha \gamma_L}  \right) ,
\end{align}
with 
\begin{align}
Z_U = \beta_M \left(1- \dfrac{1 }{\alpha \gamma_R} \right) \left(1 - \dfrac{1}{\left(\alpha\gamma_L\right)^{D_L}  } \right) + \left(\dfrac{\gamma_R}{\gamma_L} \right)^{D_L-1}\left(1- \dfrac{1}{\alpha \gamma_L}  \right) \left(\dfrac{1}{\left(\alpha \gamma_R \right)^{D_L} } - \dfrac{1}{\left(\alpha\gamma_R \right)^{D} }\right).
\end{align}
\end{itemize}

\paragraph{Expression of $\phi_{i,j}$ in the different zones.}
In each zone, the fixation probability is given by \cite{miller1994matrix} 
\begin{align}
        \phi_{i,j}^k & = \sum_{a= 1}^{L_x^k -1} \left[ p_1^k \phi_{a, L_y^k} T_{(a,L_y^k-1)i,j}^k +p_3^k \phi_{a, 0} T_{(a,1)i,j}^k \right] +\sum_{b= 1}^{L_y^k -1} \left[p_2^k \phi_{0, b} T_{(1,b)i,j}^k +p_4^k \phi_{L_x^k,a} T_{(L_x^k-1,b)i,j}^k  \right]
    \end{align}
with
\begin{align}
    T_{(a,b)i,j}^k & = \frac{4}{L_x^k L_y^k} \left( \frac{p_2^k}{p_4^k}\right)^{\frac{i-a}{2}}\left( \frac{p_3^k}{p_1^k}\right)^{\frac{j-b}{2}} \sum_{r = 1}^{L_x^k-1}\sum_{s = 1}^{L_y^k-1} \frac{\sin\left(\frac{i r \pi}{L_x^k} \right) \sin\left(\frac{a r \pi}{L_x^k} \right) \sin\left(\frac{b s \pi}{L_y^k} \right) \sin\left(\frac{j s \pi}{L_y^k} \right)}{1- 2 \sqrt{p_2^k p_4^k} \cos\left(\frac{r \pi}{L_x^k} \right) - 2 \sqrt{p_1^k p_3^k} \cos\left(\frac{s \pi}{L_y^k} \right)},
\end{align}
with $k \in(A,B,C)$, $i\in [1,...,L_x^k-1]$ and $j\in[1,...,L_y^k-1]$. 
\begin{itemize}
    \item Zone A ($0< i,j< D_L$): $L_x^A = L_y^A = D_L$
    \begin{align}
        \phi_{i,j}^A &= \sum_{a= 1}^{D_L -1} \left[ p_1^A \phi_{a, D_L} T_{(a,D_L-1)i,j}^A +p_2^A \phi_{0, a} T_{(1,a)i,j}^A +p_3^A \phi_{a, 0} T_{(a,1)i,j}^A +p_4^A \phi_{D_L,a} T_{(D_L-1,a)i,j}^A  \right] \nonumber \\
        &= \sum_{a= 1}^{D_L-1} \left[ p_1^A W_a T_{(a,D_L-1)i,j}^A +p_2^A V_a T_{(1,a)i,j}^A +p_3^A \phi_{a, 0} T_{(a,1)i,j}^A +p_4^A \phi_{D_L,a} T_{(D_L-1,a)i,j}^A  \right],
    \end{align}
where $V_a = \phi_{0,a}$ was computed previously, but $W_a = \phi_{a,D_L}$ is unknown for now.
We determine $\phi_{a, 0} $ and $\phi_{D_L,a}$ using the constraint
\begin{align}
    \phi_{i,i}^A &= 0 = \sum_{a= 1}^{D_L-1} \left[ p_1^A W_a T_{(a,D_L-1)i,i}^A +p_2^A V_a T_{(1,a)i,i}^A +p_3^A \phi_{a, 0} T_{(a,1)i,i}^A +p_4^A \phi_{D_L,a} T_{(D_L-1,a)i,i}^A  \right],
\end{align}
and the property $T_{(a,b)i,i}= \gamma_L^{a-b} T_{(b,a)i,i} $, which yields
\begin{align}
    0 = \sum_{a= 1}^{D_L-1} \left[  T_{(a,D_L-1)i,i}^A \left( p_1^A W_a + p_4^A \phi_{D_L,a} \gamma_L^{D_L-1-a}\right)+ T_{(a,1)i,i}^A \left(p_3^A \phi_{a, 0} + p_2^A V_a \gamma_L^{1-a} \right) \right].
\end{align}
Building on Ref.~\cite{broom_analysis_2008}, one can determine the solution: 
\begin{align}
    \phi_{D_L, a}& = - \gamma_L^{a-D_L}W_a ,\\
    \phi_{a,0}& = - \gamma_L^{-a}V_a .
\end{align}
Finally,
\begin{align}
        \phi_{i,j}^A &= \sum_{a= 1}^{D_L-1} p_1^A W_a \left[T_{(a,D_L-1)i,j}^A - \gamma_L^{a-D_L+1} T_{(D_L-1,a)i,j}^A \right] +p_2^A V_a \left[ T_{(1,a)i,j}^A  - \gamma_L^{1-a} T_{(a,1)i,j}^A\right].
    \end{align}
\item Zone B ($D_L< i,j < D$): $L_x^B = L_y^B = D-D_L = D_R$
\begin{align}
        \phi_{i+D_L,j+D_L}^B = \sum_{a= 1}^{D_R-1}& \left[ p_1^B \phi_{a+D_L, D} T_{(a,D_R-1)i,j}^B +p_2^B\phi_{D_L, a+D_L} T_{(1,a)i,j}^B \right. \nonumber\\& \left.+p_3^B\phi_{a+D_L, D_L} T_{(a,1)i,j}^B +p_4^B \phi_{D,a+D_L} T_{(D_R-1,a)i,j}^B  \right]\nonumber \\
     =  \sum_{a= 1}^{D_R-1} &\left[ p_1^B U_{a+D_L} T_{(a,D_R-1)i,j}^B +p_2^B Z_{a+D_L} T_{(1,a)i,j}^B\right. \nonumber\\& \left. +p_3^B\phi_{a+D_L, D_L} T_{(a,1)i,j}^B +p_4^B \phi_{D,a+D_L} T_{(D_R-1,a)i,j}^B \right],
\end{align}
where $Z_{a+D_L} = \phi_{D_L,a+D_L}$, which is unknown.
Using the same technique as for Zone A, we find
\begin{align}
    \phi_{D,a+D_L} &= - \gamma_R^{a-D_R} U_{a+D_L},\\
    \phi_{a+D_L,D_L} &= - \gamma_R^{-a} Z_{a+D_L}.
\end{align}
Finally,
\begin{align}
        \phi_{i+D_L,j+D_L}^B = \sum_{a= 1}^{D_R-1} &p_1^B U_{a+D_L} \left[T_{(a,D_R-1)i,j}^B - \gamma_R^{a-D_R+1} T_{(D_R-1,a)i,j}^B \right] \nonumber\\&+p_2^B Z_{a+D_L} \left[ T_{(1,a)i,j}^B  - \gamma_R^{1-a} T_{(a,1)i,j}^B\right].
    \end{align}
\item Zone C ($0<i< D_L$, $D_L< j < D$): $L_x^C = D_L$, $L_y^C = D_R$
\begin{align}
        \phi_{i,j+D_L}^C = &\sum_{a= 1}^{D_L-1} \left[ p_1^C \phi_{a, D} T_{(a,D_R-1)i,j}^C +p_3^C\phi_{a, D_L} T_{(a,1)i,j}^C \right] \nonumber\\& +\sum_{b= 1}^{D_R-1} \left[ p_2^C\phi_{0, b+D_L} T_{(1,b)i,j}^C  +p_4^C \phi_{D_L,b+D_L} T_{(D_L-1,b)i,j}^C  \right] \nonumber\\ 
 = &\sum_{a= 1}^{D_L-1} \left[ p_1^C U_a T_{(a,D_R-1)i,j}^C +p_3^C W_a T_{(a,1)i,j}^C \right] \nonumber\\& +\sum_{b= 1}^{D_R-1} \left[ p_2^C V_{b+D_L} T_{(1,b)i,j}^C  +p_4^C Z_{b+D_L} T_{(D_L-1,b)i,j}^C  \right].
\end{align}
\end{itemize}

\begin{figure}[h!]
    \centering
\includegraphics[width=.6\linewidth]{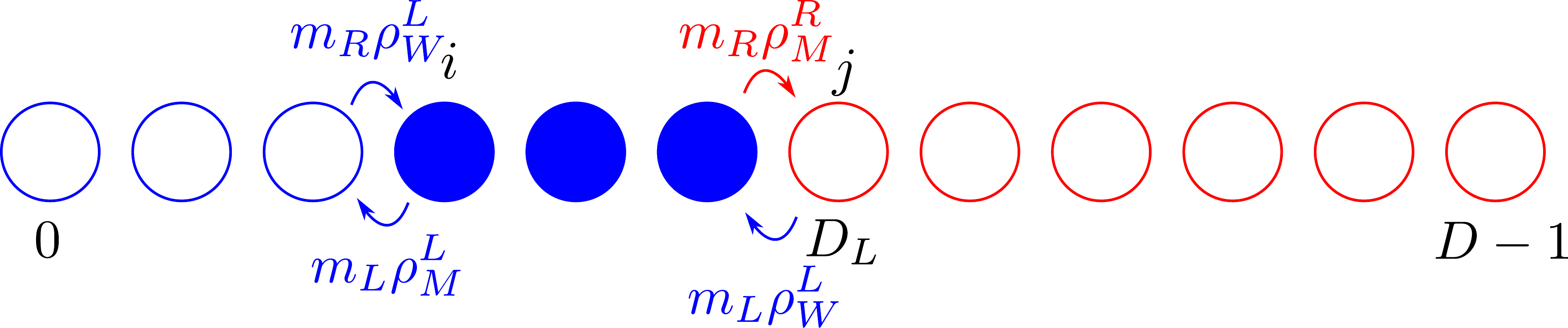}
    \caption{Example configuration in the specific case where  $j=D_L$.}
    \label{fig:schema_W}
\end{figure}

\paragraph{Equations on $W$ and $Z$.}
We would like to determine the remaining unknown boundary conditions $W$ and $Z$. First, the equation on $W$ (see \cref{fig:schema_W}) is, for all $i\in\left[ 1, D_L-1\right]$:
\begin{equation}
    W_i = \phi_{i, D_L} = p_1^W \phi_{i, D_L+1} + p_2^W \phi_{i-1, D_L}+p_3^W \phi_{i, D_L-1}+p_4^W \phi_{i+1, D_L} ,\label{eq:Eq_W}
\end{equation}
with 
\begin{align}
        &p_1^W = \frac{\alpha\beta_M }{1+ \gamma_L + \alpha (\beta_M+\gamma_L)}, p_2^W = \frac{1}{1+ \gamma_L + \alpha (\beta_M+\gamma_L)}, \\&p_3^W = \frac{\gamma_L}{1+ \gamma_L + \alpha (\beta_M+\gamma_L)}, p_4^W = \frac{\alpha \gamma_L}{1+ \gamma_L + \alpha (\beta_M+\gamma_L)},
    \end{align}
where $W_0 = V_{D_L}$ and $W_{D_L} =0$. 

\begin{figure}[h!]
    \centering
\includegraphics[width=.6\linewidth]{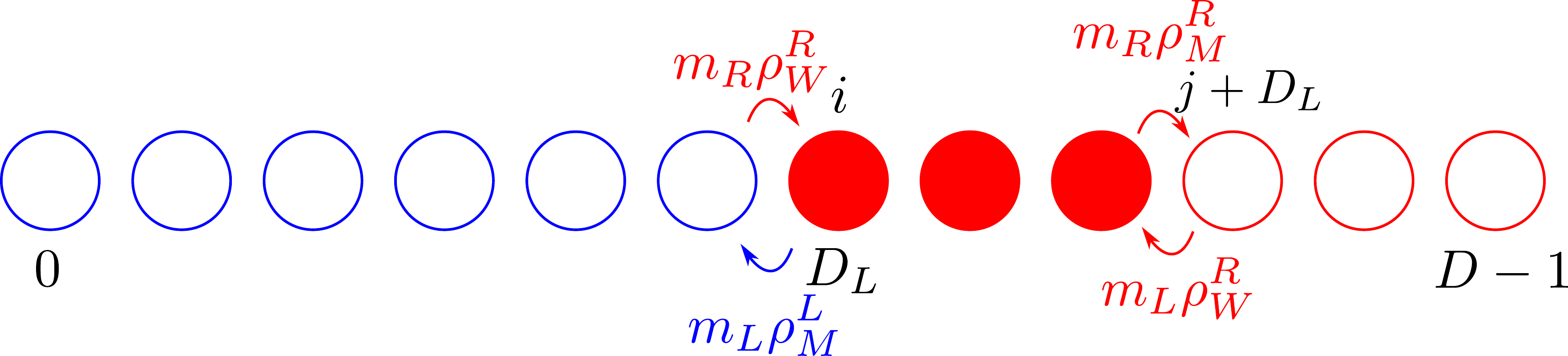}
    \caption{Example configuration in the specific case where  $i=D_L$.}
    \label{fig:schema_Z}
\end{figure}

Eq.~\ref{eq:Eq_W} can be rewritten as:
\begin{align}
    W_i=\  & p_1^W \phi_{i, D_L+1}^C + p_2^W W_{i-1} + p_3^W \phi_{i, D_L-1}^A + p_4^W W_{i+1} \nonumber \\
     =\ & p_1^W \Bigg\{ \sum_{a= 1}^{D_L-1} \left[ p_1^C U_a T_{(a,D_R-1)i,1}^C +p_3^C W_a T_{(a,1)i,1}^C \right] \Bigg.\nonumber \\ & \Bigg.+\sum_{b= 1}^{D_R-1} \left[ p_2^C V_{b+D_L} T_{(1,b)i,1}^C  +p_4^C Z_{b+D_L} T_{(D_L-1,b)i,1}^C  \right] \Bigg\} \nonumber \\&+ p_3^W \Bigg\{\sum_{a= 1}^{D_L-1} p_1^A W_a \left[T_{(a,D_L-1)i,D_L-1}^A - \gamma_L^{a-D_L+1} T_{(D_L-1,a)i,D_L-1}^A \right] \Bigg.\nonumber \\ & \Bigg.+p_2^A V_a \left[ T_{(1,a)i,D_L-1}^A  - \gamma_L^{1-a} T_{(a,1)i,D_L-1}^A\right] \Bigg\}\nonumber  \\&+ p_2^W W_{i-1} +p_4^W W_{i+1} .\label{eq:Eq_W2}
\end{align}
Similarly, the equation on $Z$ (see \cref{fig:schema_Z}) is, for all $j\in\left[ 1, D_R-1\right]$:
\begin{equation}
    Z_{j+D_L} = \phi_{D_L, j+D_L} = p_1^Z \phi_{D_L, j+D_L+1} + p_2^Z \phi_{D_L-1, j+D_L}+p_3^Z \phi_{D_L, j+D_L-1} +p_4^Z \phi_{D_L+1, j+D_L},
\end{equation}
with 
\begin{align}
        &p_1^Z = \frac{\alpha}{\gamma_R+ 1 / \beta_M + \alpha (1+\gamma_R)}, \qquad p_2^Z = \frac{1 / \beta_M}{\gamma_R+ 1 / \beta_M + \alpha (1+\gamma_R)}, \\&p_3^Z = \frac{\gamma_R}{\gamma_R+ 1 / \beta_M + \alpha (1+\gamma_R)}, \qquad  p_4^Z = \frac{\alpha \gamma_R}{\gamma_R+ 1 / \beta_M + \alpha (1+\gamma_R)}.
    \end{align}
    
Again, we find
\begin{align}
    Z_{j+D_L}  =\ & p_1^Z Z_{j+D_L+1} + p_2^Z \phi_{D_L-1,j+D_L}^C + p_3^Z Z_{j+D_L-1} + p_4^Z \phi_{D_L+1, j+D_L}^B \nonumber\\
      =\ & p_2^Z \Bigg\{ \sum_{a= 1}^{D_L-1} \left[ p_1^C U_a T_{(a,D_R-1)D_L-1,j}^C +p_3^C W_a T_{(a,1)D_L-1,j}^C \right] \Bigg.\nonumber \\ & \Bigg. +\sum_{b= 1}^{D_R-1} \left[ p_2^C V_{b+D_L} T_{(1,b)D_L-1,j}^C  +p_4^C Z_{b+D_L} T_{(D_L-1,b)D_L-1,j}^C  \right] \Bigg\}\nonumber\\&+ p_4^Z \Bigg\{ \sum_{a= 1}^{D_R-1} p_1^B U_{a+D_L} \left[T_{(a,D_R-1)1,j}^B - \gamma_R^{a-D_R+1} T_{(D_R-1,a)1,j}^B \right] \Bigg.\nonumber \\ & \Bigg.+p_2^B Z_{a+D_L} \left[ T_{(1,a)1,j}^B  - \gamma_R^{1-a} T_{(a,1)1,j}^B\right] \Bigg\} \nonumber\\&+ p_1^Z Z_{j+D_L+1} +p_3^Z Z_{j+D_L-1} .\label{eq:Eq_Z2}
\end{align}
It is convenient to rewrite Eqs.~\ref{eq:Eq_W2} and \ref{eq:Eq_Z2} in the form of a linear system:
\begin{align}
    \sum_{k=0}^{D_L} A_{i,k}W_k + \sum_{\ell = 0}^{D_R} B_{i,\ell} Z_{\ell + D_L} &= E_i \qquad(0\leq i\leq D_L), \label{linsyst1}\\
    \sum_{k=0}^{D_L} C_{j,k}W_k + \sum_{\ell = 0}^{D_R} D_{j,\ell} Z_{\ell + D_L} &= F_j \qquad(0\leq j\leq D_R). \label{linsyst2}
\end{align}
where for all $0 < i < D_L$ and $0 < j < D_R$:
\begin{align}
    A_{i,k}&= p_3^W p_1^A \left[T_{(k,D_L-1)i,D_L-1}^A- \gamma_L^{k-D_L+1} T_{(D_L-1,k)i,D_L-1}^A\right] \nonumber \\ &+p_1^W p_3^C T_{(k,1)i,1}^C +p_2^W \delta_{k,i-1} +p_4^W \delta_{k,i+1}- \delta_{k,i}, \\
    B_{i,\ell} &=  p_1^W p_4^C T_{(D_L-1,\ell)i,1}^C,\\
    C_{j,k}&= p_2^Z p_3^C T_{(k,1)D_L-1,j}^C , \\
    D_{j,\ell} &= p_4^Z p_2^B \left[T_{(1, \ell) 1,j}^B- \gamma_R^{1-\ell} T_{(\ell,1) 1,j}^B\right] \nonumber\\& +p_2^Z p_4^C T_{(D_L-1, \ell)D_L-1,j}^C +p_1^Z \delta_{\ell,j+1} +p_3^Z \delta_{\ell,j-1}- \delta_{\ell, j} , \\
    E_i &= - p_1^W  p_1^C \sum_{a= 1}^{D_L-1}   U_a T_{(a,D_R-1)i,1}^C  - p_1^W p_2^C \sum_{b= 1}^{D_R-1}   V_{b+D_L} T_{(1,b)i,1}^C \nonumber\\&- p_3^W p_2^A \sum_{a= 1}^{D_L-1} V_a \left[ T_{(1,a)i,D_L-1}^A  - \gamma_L^{1-a} T_{(a,1)i,D_L-1}^A\right],  \\
   F_j &= - p_2^Z p_1^C \sum_{a= 1}^{D_L-1} U_a T_{(a,D_R-1)D_L-1,j}^C  - p_2^Z p_2^C \sum_{b= 1}^{D_R-1}  V_{b+D_L} T_{(1,b)D_L-1,j}^C \nonumber \\&- p_4^Z p_1^B \sum_{a= 1}^{D_R-1}  U_{a+D_L} \left[T_{(a,D_R-1)1,j}^B - \gamma_R^{a-D_R+1} T_{(D_R-1,a)1,j}^B \right] ,
    \end{align}
    and 
\begin{align}
    A_{0,k} &= \delta_{0,k}, & \quad A_{D_L, k} &= \delta_{D_L,k}, & \quad D_{0,\ell} &= \delta_{0,\ell}, & \quad D_{D_R, \ell} &= \delta_{D_R, \ell}, \\
    B_{0,\ell} &= 0, & \quad B_{D_L,\ell} &= 0, & \quad C_{0,k} &= 0, & \quad C_{D_R, k} &= 0, \\
    E_0 &= V_{D_L}, & \quad E_{D_L} &= 0, & \quad F_0 &= 0, & \quad F_{D_R} &= U_{D_L}.
\end{align}

\paragraph{Conclusion.}
Numerically solving the linear system in Eqs.~\ref{linsyst1} and~\ref{linsyst2} (using matrix inversion) provides the values of $W$ and $Z$. Then, the fixation probability of mutants, starting from one mutant placed uniformly at random in the line graph is
\begin{align}
    \rho_1^{\text{line}} = \frac{1}{D } \left\{\rho_M^L \left[ V_1 + \sum_{i=1}^{D_L-2}\phi_{i,i+1}^A + W_{D_L-1} \right] + \rho_M^R \left[ Z_{D_L+1} + \sum_{i=D_L+1}^{D-2}\phi_{i,i+1}^B + U_{D-1} \right] \right\}, \label{eq:p_mutant_line}
\end{align}
with
\begin{align}
        \phi_{i,i+1}^A &= \sum_{a= 1}^{D_L-1}\  p_1^A W_a \left[T_{(a,D_L-1)i,i+1}^A - \gamma_L^{a-D_L+1} T_{(D_L-1,a)i,i+1}^A \right] \nonumber \\&+ \sum_{a= 1}^{D_L-1}p_2^A V_a \left[ T_{(1,a)i,i+1}^A  - \gamma_L^{1-a} T_{(a,1)i,i+1}^A\right],\\
        \phi_{i+D_L,i+D_L+1}^B& = \sum_{a= 1}^{D_R-1} p_1^B U_{a+D_L} \left[T_{(a,D_R-1)i,i+1}^B - \gamma_R^{a-D_R+1} T_{(D_R-1,a)i,i+1}^B \right] \nonumber\\&+ \sum_{a= 1}^{D_R-1}p_2^B Z_{a+D_L} \left[ T_{(1,a)i,i+1}^B  - \gamma_R^{1-a} T_{(a,1)i,i+1}^B\right].
\end{align}

\subsubsection{Amplification of selection}

\begin{figure}[ht!]
    \centering
\includegraphics[width=.75\linewidth]{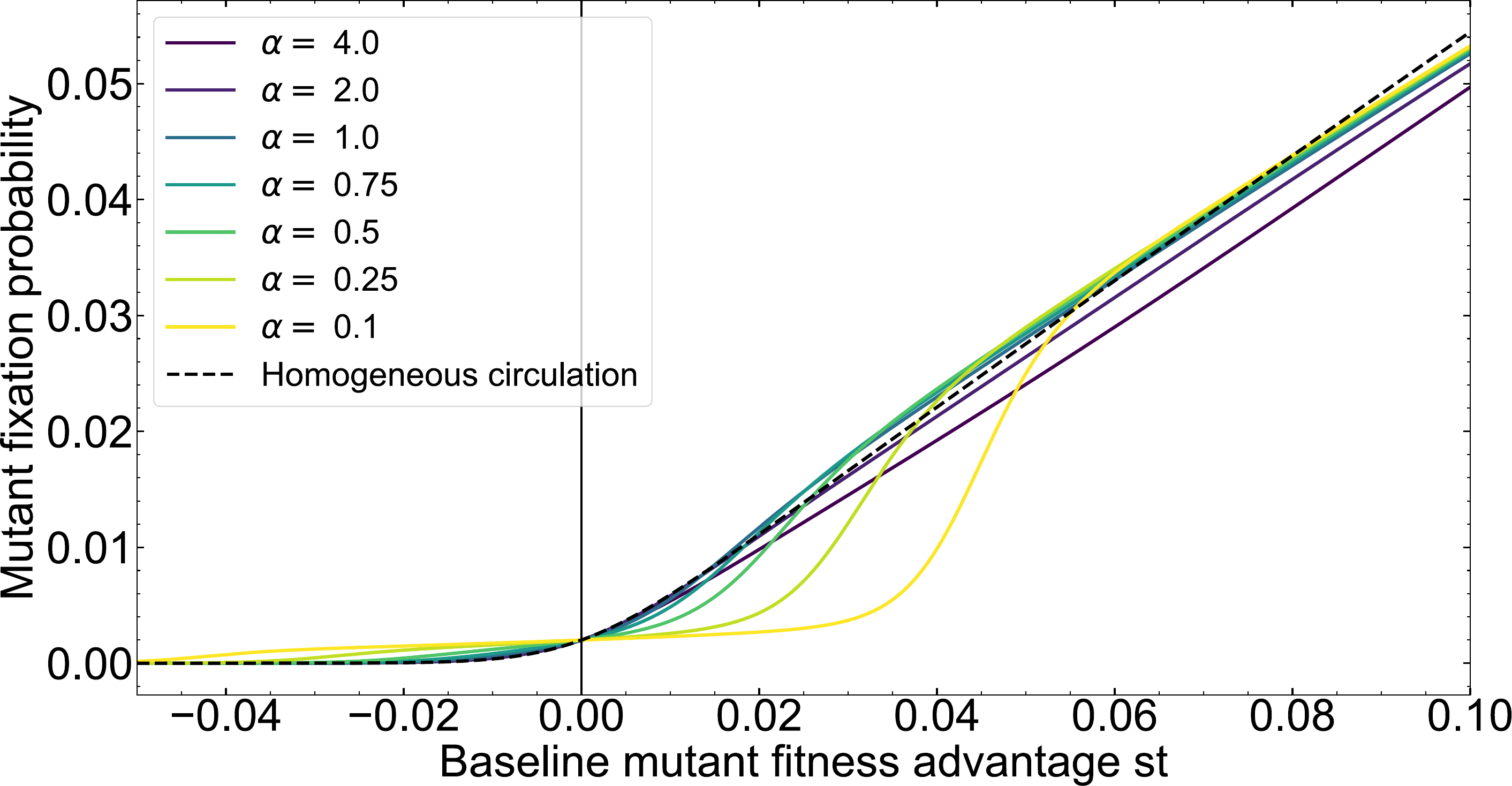}
    \caption{\textbf{Fixation probability in the heterogeneous line with rare migrations.} We plot the fixation probability starting from one mutant appearing uniformly at random in the line graph, given by \cref{eq:p_mutant_line}, as a function of the baseline mutant fitness advantage $st$. Parameters: $D= 5$, $D_L = 1$, $\delta_L = 1$, $\delta_R = 0.1$, $K = 100$. The case of a homogeneous circulation with the same average mutant relative fitness advantage is shown for reference. }
    \label{fig:p_mutant_line}
\end{figure}

In \cref{fig:p_mutant_line}, the fixation probability obtained in Eq.~\ref{eq:p_mutant_line}, for a mutant in the heterogeneous line with rare migrations, is plotted as a function of the baseline relative mutant fitness advantage $s t$. Several distinct regimes emerge as $st$ is increased while $s t \ll 1$. First, for very small $st$, selection is suppressed compared to the homogeneous circulation. Then, for slightly larger $s t$, we observe a small amplification of selection, provided the migration asymmetry is less than 1. Next, suppression is again observed. While these results are complex, due to the interplay of migration and environment heterogeneity and to different regimes of selection, they show that the line can amplify selection in the rare migration regime when the environment is heterogeneous.

A qualitative analysis can be made along the lines of Section~\ref{sec:star-rare-ampli}. Here, for simplicity, and as in Figure~\ref{fig:p_mutant_line}, we assume that only one deme on the left has $\delta_L$ and the other ones have $\delta_R$. 

First, if $1/K\ll 2\delta_L st\ll 1$ and $1/K\ll2\delta_R st\ll 1$,  the difference of fixation probability between a heterogeneous and a homogeneous structure with the same $\langle\delta\rangle$ can be expressed as for the star, see Eq.~\ref{eq:comp-star-rare}, and thus, environment heterogeneity leads to suppression of selection, explaining the observations of Figure~\ref{fig:p_mutant_line} for larger $st$.

Second, if $\delta_L>\delta_R$, there may be a regime where $1/K\ll 2\delta_L st\ll 1$ but $2\delta_R st\lesssim 1/K$. Following the same reasoning as for the star, we conclude that a mutant starting in the left-most deme will take over in the whole population if it fixes there. Meanwhile, a mutant starting elsewhere that fixes in its original deme may be re-invaded by wild-type individuals. This is more likely in the line than in our heterogeneous star, because there are more ways to get re-invaded before full fixation. Nevertheless, the benefit of having one deme where the mutant is substantially advantaged and safe from re-invasions makes the heterogeneous line slightly amplify selection in this regime, compared to a homogeneous circulation.

Finally, when both $\delta_L$ and $\delta_R$ are very small, the benefit from having a deme where the mutant is fitter is minimal, and the heterogeneous line behaves as a suppressor of selection, as the homogeneous line~\cite{servajean_impact_2025}.

\end{document}